Ben-Gurion University of the Negev

Faculty of Engineering Sciences

Department of Mechanical Engineering

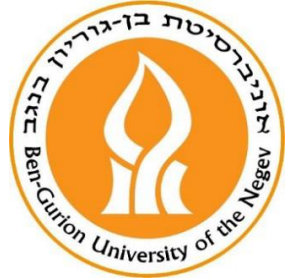

# Development of methods for thermo-hydraulic simulation of nuclear reactors and similar systems in normal working conditions and in transient processes

Thesis submitted in partial fulfillment of the requirements for the M.Sc. degree

Submitted by:                                  Dmitry Zviaga

Supervised by:                              Dr. Yuri Feldman

Author: 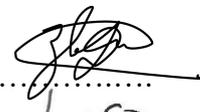 ………………………………………            Date: 30/09/21

Supervisor: 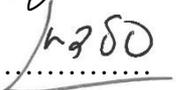 ………………………………            Date: 30/09/21

Chairman of Graduate Studies Committee: ………..……           Date: 30/09/21

September 2021

# Abstract

The goal of this report is to present the final project conducted in order to fulfill the requirements of the M.Sc. degree at the Department of Mechanical Engineering, Ben-Gurion University (BGU) of the Negev. The project comprises theoretical research investigating natural convection compressible flow with high temperature differences and with complex geometries. The research motivation comes from long-term research investigating and simulating the steady state and transient multiphase flow regimes existing in the reactor core, that was established by the Soreq Nuclear Center.

The main objective of this project is to develop a comprehensive numerical methodology that is capable of theoretical modeling of natural convection compressible flow with high temperature differences and with complex geometries, using standard techniques of computational fluid dynamics (CFD) – pressure-based solution algorithms and immersed boundary methods.

This report contains:

- A comprehensive literature review surveying methods for the simulation of natural convection flow and immersed boundary methods.
- An extended outline of the objectives of the performed research.
- A comprehensive physical model, including the governing equations, definitions, constitutive laws, and dimensional analysis.
- A verification study by favorable comparison with corresponding independent numerical data available in the literature for incompressible, and non-Bossinesq compressible flows, without complex geometry.
- A comparison between results obtained in the present study and results from previous studies for configurations with low temperature difference and complex geometry.
- A solution and analysis of the configurations with high temperature differences and complex geometry.
- A summary, conclusions, and recommendations for possible future work.

# Acknowledgements

I would like to take this opportunity to express my immense gratitude to all those who have given their invaluable support and assistance. In particular, I am profoundly indebted to my supervisor, Dr. Yuri Feldman, for his professional and knowledgeable support in the physical, computational, and numerical fields. His great contribution to this research, from his laboratory computing equipment, through the collaboration process with the Soreq Nucelar Center, and to the countless hours of meetings and advice at any time, is remarkable. Especially, I would like to thank him for his close supervision during the writing of a scientific manuscript.

Additionally, I would like to thank Dr. Ido Silberman who generously contributed his time and assisted me with his theoretical background.

The research for this thesis was financially supported by Ministry of Energy and Infrastructure under grant contract number 218-11-038, in addition to a student scholarship provided by the Faculty of Engineering Sciences of Ben-Gurion University of the Negev. These sources of support contributed much to the success of the study.

In this section, I would like to include a mention of my colleagues from the Group of Computational Physics Lab founded by my supervisor, Dr. Yuri Feldman, for their discussions and technical support.

It is also appropriate to thank for spiritual support, so I would also like to thank my family and friends for their interest and encouragement along the way.

# Contents

# List of Figures

# List of Tables

# Nomenclature

| Initials | Description | Units |
|---|---|---|
| $C_p$ | Specific heat capacity at constant pressure | $J/kgK$ |
| $C_v$ | Specific heat capacity at constant volume | $J/kgK$ |
| $e$ | Internal energy | $J/kg$ |
| $\vec{f}^\Gamma(f_x^\Gamma, f_y^\Gamma)$ | Immersed forces on Eulerian grid points | $N$ |
| $\vec{F}_K(F_{K,x}, F_{K,y})$ | Immersed forces on Lagrangian grid points | $N$ |
| $g$ | Gravity acceleration | $m/s^2$ |
| $h$ | Enthalpy | $J/kg$ |
| $H$ | Domain height | $m$ |
| $k$ | Thermal conductivity | $W/mK$ |
| $L$ | Domain length | $m$ |
| $L(\vec{u})$ | Diffusion fluxes of the dimensionless momentum equations (linear term) | $-$ |
| $\dot{m}$ | Mass flow rate | $kg/s$ |
| $\Delta\dot{m}$ | Mass flow imbalance after the prediction step | $kg/s$ |
| $n$ | Direction | $-$ |
| $N(\vec{u})$ | Convection fluxes of the dimensionless momentum equations (non-linear term) | $-$ |
| $p$ | Pressure | $Pa$ |
| $q^\Gamma$ | Immersed heat fluxes on Eulerian grid points | $W/m^2$ |
| $Q_K$ | Immersed heat fluxes on Lagrangian grid points | $W/m^2$ |
| $R$ | Specific gas context | $J/kgK$ |
| $t$ | Time | $s$ |
| $\Delta t$ | Time interval | $s$ |
| $S_k$ | Sutherland temperature for thermal conduction | $K$ |
| $S_\mu$ | Sutherland temperature for viscosity | $K$ |
| $T$ | Temperature | $K$ |
| $\vec{u}$ | Velocity | $m/s$ |
| $V$ | Volume | $m^3$ |
| $\vec{x}(x,y)$ | Coordinates on Eulerian grid | $m$ |
| $\vec{X}_K(X_{K,x}, Y_{K,y})$ | Coordinates of Lagrangian grid points | $m$ |

# Greek Letters

| Initials | Description | Units |
|---|---|---|
| $\alpha$ | Thermal diffusivity | $m^2/s$ |
| $\delta$ | Discrete delta function | $m$ |
| $\varepsilon$ | Normalized temperature difference parameter | — |
| $\kappa$ | Ratio of specific heat capacities | — |
| $\mu$ | Viscosity | $kg/m \cdot s$ |
| $\rho$ | Density | $kg/m^3$ |
| $\tau$ | Stress tensor | $Pa$ |

# Dimensionless variables

| Symbol | Description | Formulation |
|---|---|---|
| $C_k$ | Sutherland temperature for thermal conductivity | $\dfrac{S_k}{T_0}$ |
| $C_\mu$ | Sutherland temperature for viscosity | $\dfrac{S_\mu}{T_0}$ |
| $C_\rho$ | Density coefficient | $\dfrac{1}{T}$ |
| $M_0$ | Mach number | $\dfrac{\alpha_0/L_0}{\sqrt{\kappa R T_0}}$ |
| $Pr$ | Prandtl number | $\dfrac{\mu_0}{\rho_0 \alpha_0}$ |
| $Ra$ | Rayleigh number | $\dfrac{\rho_0 g (T_h - T_c) L_0^3}{\mu_0 \alpha_0 T_0}$ |

# Subscripts

| Symbol | Description | Symbol | Description |
|---|---|---|---|
| 0 | Characteristic value | ∞ | Value at long distance from the immersed body |
| $h$ | Hot surface | $c$ | Cold surface |
| $x$ | Horizontal component | $y$ | Vertical component |
| $i, j$ | Eulerian grid | $K$ | Lagrangian grid |
| $ns$ | Normal to surface | $ss$ | Tangential to surface |
| $avg$ | Average value | $w$ | Wall |

# Superscripts

| Symbol | Description | Symbol | Description |
|---|---|---|---|
| ~ | Dimensional value | → | Vector |
| ∧ | Unit vector | = | Matrix/Tensor |
| $n$ | Current time step | $n+1$ | Next time step |
| $n-1$ | Previous time step | $m$ | Current iteration |
| $m-1$ | Previous iteration | ∗ | Corrected value after first correction |
| ∗∗ | Corrected value after second correction | ′ | First correction value |
| ″ | Second correction value | $d$ | Desired value |

# Operators

| Symbol | Description | Symbol | Description |
|---|---|---|---|
| + | Addition | − | Subtraction |
| · | Scalar multiplication | × | Vector multiplication |
| ⊗ | Tensor multiplication | $\partial/\partial j$ | Derivative by variable $j$ |
| ∇ | Gradient | ∇· | Divergence |

# 1. Introduction and literature survey

## 1.1 Motivation of the study

The thermal-hydraulic behaviour of nuclear reactors and similar systems, both in normal working conditions and in transient operation, is a sophisticated physical process for which in-depth investigation is essential in order to promote further scientific research and engineering development. Different regimes in the reactor core host several coupled phenomena. The most important of them is the multiphase flow coupled with nuclear reactions and neutron distribution in the reactor core. Physical modelling of this phenomenon is essential for enhancing the reliability and the engineering maintenance of nuclear reactors. For this reason, a long-term research project has been established, that aims to develop a state-of-the-art solver capable of the simulation of the steady state and transient multiphase flow regimes existing in the reactor core. These flow regimes may contain both incompressible and compressible phases, involve heat and mass transfer and host different phase change phenomena, such as boiling and condensation. The final version of this advanced solver should employ state-of-the-art numerical methods to handle fully coupled steady state and transient continuity, momentum, and energy equations by simultaneously employing the equation of state, and thermodynamic correlations necessary to obtain thermo-physical properties of the flow. In addition, the solver should be capable of addressing the complex geometry of the reactor core, as well as the moving liquid-vapor interface in loss-of-coolant-accidents (LOCA). The current study constitutes a first, but very important, step within the long-term research, while focussing on the development of a single-phase compressible solver for the simulation of non-Boussinesq flows in the presence of complex geometries.

## 1.2 Computational approaches

Numerical approaches typically utilized in Computational Fluid Dynamics can be classified into two general families, namely, density-based and pressure-based approaches, both aimed to ensure continuity constraints of the simulated flows. The decision regarding the choice of specific approach should be made by carefully, considering the physical characteristics of the specific problem. Traditionally, pressure-based solvers were developed for low-speed incompressible flows in which the energy equation (if it exists) is typically written solely in terms of temperature, while all the thermo-physical flow properties are assumed to be constant (i.e., there is no need for the equation of state). In this case, the pressure field plays the role of a distributed Lagrange multiplier that aims to ensure the incompressibility constraint of the flow. The density-based schemes were typically used to solve high-speed compressible flows, in which the viscous effects can be neglected, and the non-viscid momentum (Euler) equations are derived from the full Navier-Stokes equations.

The pressure-based algorithms use velocity components and pressure (or pressure-correction) as their primary variables. In the first step, the pressure field is taken as known from the previous time step and the momentum equations are solved by yielding the velocity field, which does not satisfy the continuity equation. In the second stage, the pressure-correction equation (which is derived from the continuity equation) is solved. Afterwards the pressure and the velocity components are corrected to meet the continuity constraint. If the solver is compressible, then, after the solution of the pressure-correction equation the energy equation should be solved, followed by updating the density and the viscosity fields by relying on the equations of state and Sutherland equations. Typically, outer iterations are needed to achieve convergence of the velocity, the temperature, and the density fields and to satisfy the continuity constraint. In density-based algorithms, the density and all the velocity component fields play the role of primary variables. Continuity, momentum, and all other transport equations are first solved in a fully coupled manner, and then the pressure field is derived from the equation of state. Both pressure-based and density-based formulations have been extensively used to simulate a broad range of flows, but the density-based formulation is typically expected to be more accurate for simulating high-speed compressible flows, while the pressure-based formulation is advantageous for simulating low-speed viscous compressible flows.

The main criteria that characterize flow regimes in the nuclear reactor core are: (i) low Mach numbers, (ii) high temperature differences and, therefore, large deviations in density, viscosity, and thermal conductivity, (iii) the presence and significance of gravity and (iv) an enclosed expanse. As a result, in the framework of the current study the choice was made in favor of employing the pressure-based approach. We next briefly survey the state-of-the-art of the numerical techniques that were adopted in the current study when analyzing isothermal and thermally driven non-Boussinesq flows.

### 1.2.1 Pressure-based algorithms

Pressure-based algorithms belong to the family of pressure-velocity segregated approaches, the vast majority of which are either based on the SIMPLE method or on its derivatives (e.g., the SIMPLER, SIMPLEC, SIMPLEST, SIMPLEX, SIMPLEM, PISO and PRIME approaches) as reviewed in [1], or on the fractional-step method (FSM) [2], [3], [4], [5], [6], [7]. The two families can be classified as marching methods, consisting of four major steps: (i) predictor step, constituting solution of the momentum equations to obtain intermediate velocity using velocity and pressure fields from the previous iteration/time step, (ii) solution of the Poisson equation for pressure-correction, (iii) correction-projection step, constituting correction of the pressure and projection of velocity fields to fulfill the continuity constraint and (iv) solution of other transport equations, if required. Depending on the time discretization (fully- or semi-implicit) of non-linear terms, numerical solution of the governing equations may (or may not) contain outer iterations [8]. It is also noteworthy that in FSMs the velocity correction only appears in the transient term, while in the SIMPLE and related methods the corrections appear in all the velocity entries contributing to the main diagonal of the transient, convection and diffusion terms. The theoretical basis of all state-of-the-art pressure-correction approaches from the SIMPLE and the FSM families is given in the monograph [8]. This book can also be consulted for gaining in-depth understanding of the ways of extending the SIMPLE and the FSM approaches for the simulation of compressible flows, including the properties of the pressure-correction equation and implementation of specific boundary conditions (e.g., prescribed total pressure, total temperature, static pressure for outflow boundary and supersonic outflow boundary).

We next briefly review the major developments of the pressure-based algorithms for the SIMPLE and the FSM families, in chronological order.

*The FSM approach*

A numerical solution of the incompressible Navier-Stokes equations based on the fractional-step approach was presented in the seminal study of Kim and Moin in [2]. The solution was performed on a staggered grid by calculating the pressure correction in the corrector rather than in the predictor step. The study was then extended to collocated grids and curvilinear coordinates [3]. Both works employed an approximate factorization ADI scheme in the predictor step. The importance of these formulations is the implicit treatment of the viscous (or turbulent) terms, which enabled avoiding the time-step limitations typical of their explicit discretization. According to Gresho [4], the FSM can be classified into P1, P2 and P3 methods. In the P1 method, the pressure field is set to zero in the momentum equation, whose solution is further used as a predictor for the velocity field. The Poisson equation is then solved to obtain the pressure and to project non-solenoidal velocity on the divergence free subspace. Following this classification, it can be concluded that studies [2] and [3] adopted the P1 method. In the P2 method, the pressure term is present in the momentum equation, where its value is taken from the previous time step. The pressure-correction Poisson equation is then solved, followed by a standard correction-projection step. The P3 method is very similar to the P2 method with the only difference being that the pressure used in the momentum equation is extrapolated with a second order of accuracy by utilizing the pressure field values from two previous time steps.

Further contributions in the assessment of the numerical accuracy of different formulations of FSM is due to the works of [9], [10], [5], [6], [7], [11]. It was shown that for both staggered [9] and collocated [10] grids the solution of the Navier-Stokes equations obtained by employing a simplified formulation of the pressure-correction equation not including elliptic pressure coupling is superior in terms of computational time, and is characterized by the same accuracy compared to its counterpart obtained by employing the full pressure-correction equation. High efficiency of the methods based on the solution of a simplified formulation of the pressure-correction equation, not requiring outer iterations, was further demonstrated in [5] where the second order accuracy of the results was achieved by modifying boundary conditions for the predicted velocity. In their next study [6], the authors confirmed that the standard P1 and P2 methods provide the first and the second order accuracy in time of the pressure field, respectively. It was also shown that the second order accuracy in time could be achieved for the P1 method by utilizing the full pressure correction. Third order accuracy in time of the P3 method was reported in [7]. In the same study, the authors also proposed a new approach, which they called the pressure method. The key idea was to solve the Poisson pressure correction equation prior to the solution of the momentum equations. Both alternatives were shown to reduce the overall error and to increase the efficiency as compared to the standard P2 method.

### *The SIMPLE and related approaches*

The SIMPLE approach was first formulated by Patankar and Spalding in [12] as a marching procedure to calculate the transport processes in three-dimensional incompressible flows. The algorithm consists of three major steps: (i) implicit solution of the momentum equations to obtain the intermediate velocity field based on the estimated pressure field, (ii) obtaining the pressure correction by solution of the Poisson equation and calculation of the corrected pressure and velocity fields to fulfill the continuity constraint and (iii) solution of other transport equations, if required. If the continuity constraint is not satisfied up to a given precision, the pressure and velocity fields should be updated and steps (i) and (ii) are reiterated. In the SIMPLE algorithm, velocity corrections of the convection and the diffusion terms that are not in the main diagonal of the coefficient matrix in the corrector step are neglected. As stated by the authors of [12], the SIMPLE algorithm can overestimate the pressure term, which may negatively affect the convergence rate. For this reason, Patankar recommended to apply underrelaxation for pressure-correction calculation in order to improve the convergence rate and to avoid divergence [13], [14]. In an attempt to improve the convergence rate, a SIMPLER algorithm [13], [14] was developed. A key idea was not to correct the velocity field after the pressure field had been corrected. Instead, the momentum equations are solved again, but this time by using an updated pressure field. Then, a second pressure corrector is employed, and pressure and velocity fields are updated. The time step is completed by proceeding to the solution of other transport equations, if required. Similarly to the SIMPLE algorithm, the SIMPLER algorithm only takes into account the velocity corrections entering the main diagonal of transient, convection and diffusion terms. Van Doormal and Raithby introduced a SIMPLEC method [15] as a different development of SIMPLE, in which the definitions of SIMPLE's coefficients in the pressure-correction equation are modified so that there is no need for underrelaxation of pressure. Maliska and Raithby presented the PRIME algorithm [16], in which the momentum equations in the predictor step are solved explicitly, and then the solution procedure continues as in SIMPLE. Issa presented the PISO method [17], which, similarly to the SIMPLER algorithm, is based on two corrector steps. The differences are that velocity corrections are omitted in the convective and diffusive terms of the second corrector, and the energy equation is solved twice, namely, after the first and the second correctors. We finish this chapter by mentioning the developments in the field of implementation of the boundary conditions for incompressible and compressible flows in the variety of pressure-based solvers based on the SIMPLE algorithm, which can be found in Moulkalled et al. [18].

### 1.2.2 Isothermal compressible single-phase flow

When developing a novel solver for the compressible Navier Stokes equations, it is common practice to verify its capability to successfully handle incompressible (or slightly compressible) isothermal flows. In recent decades, a significant number of studies have been dedicated to developing pressure-based solvers, adopting various prediction-projection algorithms for the simulation of both incompressible and compressible flows typical of a broad range of configurations. Armaly et al. [19] utilized an iterative finite-difference numerical scheme developed in [20] for the investigation of laminar, transitional, and turbulent flows of air in an extended channel for a Reynolds number's range of $70 < Re < 8000$. The authors assumed isothermal flow when solving compressible equations of the conservation of mass and momentum. An algorithm capable of the simulation of unsteady/transient viscous stratified compressible flows, characterized by a broad range of subsonic Mach numbers, including nearly incompressible flows, was developed by Mary et al. [21]. In their work the authors solved the fully compressible Navier-Stokes equations and verified their results by comparison with the corresponding data reported in [19] and in [22]. To remove the stiffness of the numerical problem due to the large disparity between the flow and the acoustic wave speeds at low Mach numbers, an approximate Newton method, based on artificial compressibility, was used by the authors in [21]. Hauke et al. [23] developed computational techniques employing segregated stabilized methods with standard pressure boundary conditions capable of the simulation of both incompressible and isothermal compressible flows. Their goal was to establish robust segregated methods characterized by superior computational performance compared to more common coupled methods typically used for the simulation of compressible flows. In their next work [24] the authors successfully extended their methodology to non-isothermal compressible flows. Frehse et al. [25] addressed isothermal compressible flow from the mathematical point of view. They validated the existence of a weak solution for the case of mixed boundary conditions by the means of the stream function analysis.

### 1.2.3 Thermally driven compressible single-phase flow

Another important step in the development of a general compressible solver is to prove its capability to handle compressible natural convection flows characterized by low values of Mach number. For years, natural convection flows were addressed by applying the Boussinesq approximation, i.e., by assuming incompressible flow, while introducing the buoyancy effects in the form of a temperature dependent source in the momentum equations. However, the Boussinesq approximation is only valid for small temperature differences (not more than 30 °C). To address the configurations characterized by high temperature differences, fully compressible Navier Stokes equations must be solved. An extensive overview of both compressible and incompressible approximations when simulating natural convection flows was given by Mayeli and Sheard in [26]. Regarding incompressible approximations, the Oberbeck-Boussinesq, the thermodynamic Boussinesq and the Gay-Lussac approximations are typically utilized in numerical simulations. Regarding compressible approximations, the fully compressible and the weakly compressible approaches are typically employed.

*Weakly compressible approximations*

The majority of weakly compressible approximations developed for the simulation of low Mach number compressible flows are based on the asymptotic model, in which the total pressure is split into the thermodynamic and the hydrodynamic pressures. In the literature, this model is often referred to as a "classical low Mach number model". The asymptotic model for the simulation of thermally driven natural convection flow was introduced for the first time by Rehm and Baum [27]. The key idea was to split the pressure field into the large, time-dependent thermodynamic part, and the stationary part including extremely small spatial deviations. Such a decomposition was found to be applicable for the simulation of low Mach number thermally driven flows, and allows the pressure terms to be of the same order of magnitude as other terms in the momentum and the energy equations. Further progress in this field is due to the work of Paolucci [28], who formulated and solved anelastic transient equations, allowing to avoid the appearance of acoustic waves in natural convection non-Boussinesq flows. The study presented in [29] extended the methodology described in [28] by applying it to the simulation of two-dimensional compressible natural convection flow in a vertical slot with large horizontal temperature differences. Le Quere et al. [30] approached the non-Boussinesq natural convection confined flow by extending a pseudo-spectral algorithm from [31] that was originally developed by them for Boussinesq natural convection flow. The algorithm developed in [29] was further successfully extended for three-dimensional configurations, as presented in [32]. Paillere et al. [33] investigated two numerical methods to solve low Mach number compressible flows by simulating natural convection flow in a differentially heated cavity with a high temperature difference. Elmo and Cioni [34] compared the Boussinesq approximation and quasi-compressible model, based on a classical low Mach number model, when applying it to the simulation of the pebble bed reactor. Schall et al. [35] studied steady state and transient thermal compressible flow by applying Turkel preconditioned Roe splitting, which is more likely to apply to the fully compressible hyperbolic solvers (see below), and to flows modelled by the classical low Mach number model. Le Maitre et al. [36] developed a stochastic projection method (SPM) for the quantitative propagation of uncertainty in compressible low Mach number flows (which they referred to as "zero Mach number flows") under non-Boussinesq conditions. Beccantini et al. [37] investigated a transient injection flow in a low Mach number regime, employing three different approaches where two of them are based on asymptotic models of the Navier-Stokes equations. Reddy et al. [38] studied conjugate natural convection in a vertical annulus with a centrally located vertical heat generating rod, employing compressible fluid transport equations and an asymptotic model for low Mach numbers. Lappa [39] employed low Mach number asymptotics and developed a flexible and modular solver that takes into account all the molecular (translational, rotational and vibrational) degrees of freedom and their effective excitation, and guarantees adequate interplay between molecular and macroscopic-level entities and processes. Armengol et al. [40] studied the effects of air variable properties in the transient case of the classical differentially heated square cavity problem, employing the SIMPLE algorithm, a low Mach number approximation and the finite volume method.

Some authors approached the weakly compressible approximation and addressed thermal systems with mixed natural convection-radiation thermal mechanisms, considering compressible flow. Darbandi and Abrar [41] developed a hybrid incompressible-compressible method to solve the combined natural convection-radiation heat transfer in a participating medium at steady state without addressing the Boussinesq approximation or employing the low Mach number asymptotic assumption. Their study showed that compressibility effects become more dominant in combined natural convection-radiation problems than in the pure natural-convection problem. Parmananda et al. [42] presented a computational framework for non-Oberbeck-Boussinesq buoyancy driven turbulent transient convection coupled with thermal radiation at large temperature differences. They used a low Mach number model based on the Favre-averaged (Navier-Stokes and energy) equations, with the standard $k - \varepsilon$ model presented using an unstructured finite volume method. The same research group as in [42] continued their work, and in [43] presented the development of a non-Boussinesq flow solver employing low Mach number asymptotics and a fractional-step method to simulate combined radiative-convective heat transfer, which is suitable for arbitrary polygonal meshes. They also continued their work in [44], where they investigated three different algorithms for the numerical simulation of non-Boussinesq convection with thermal radiative heat transfer based on a low Mach number formulation. The first algorithm (Algorithm A) uses conservation of mass and energy equations to compute density and temperature. The other two algorithms (Algorithm B and Algorithm C) calculate temperature and density from the equation of state, respectively, and solve a conservative form of the continuity and energy equations to obtain density and temperature, respectively.

*Fully compressible approximation*

Compared to the weakly compressible approximations, fully compressible approximations for low Mach number flows adopted several different approaches. Some fully compressible approximations for low Mach number compressible flows employ density-based solvers with splitting schemes and preconditioning. Paillere et al. [33] employed a hyperbolic solver based on the resolution of the compressible Euler equations, with a Roe's flux difference splitting scheme and Turkel preconditioning. Beccantini et al. [37] used the fully compressible Navier-Stokes equations, employing a density-based solver, and using a HLLC computation scheme and a ILUTP preconditioner. Fu et al. [45] studied natural convection in a channel under a high temperature difference with a fully compressible approximation using Roe preconditioning and dual time stepping. In order to resolve reflections induced by acoustic waves at the boundaries of the channel, non-reflection conditions at the boundaries of the channel were derived. El-Gendi and Aly in [46] simulated natural convection compressible flow using a Roe scheme and a dual time method, in square and sinusoidal cavities without applying the Boussinesq approximation. Li in [47] studied the laminar-turbulent transition induced by natural convection with high temperature differences, employing compressible transport equations, Roe preconditioning and an implicit large eddy simulation.

Other fully compressible approximations employ pressure-based solvers. Sewall and Tafti in [48] developed a variable property algorithm for time-dependent resolution of flows with large temperature differences without using the low Mach number assumption; in this algorithm, the momentum and energy equations are integrated in time using an implicit Crank-Nicolson method, the Helmholtz equation for pressure was solved at each inner iteration, and local density changes were coupled with variations of both local temperature and pressure fields, whereas other properties are only coupled with temperature variations. Barrios-Pina et al. [49] adapted the method from [48] to conduct thermodynamic analysis to determine the contribution of each term in the total energy equation.

Another fully compressible approximation employed a mesoscopic computational method based on a model Boltzmann equation. Wen et al. [50], [51] recovered the fully compressible Navier–Stokes equations by employing the Boltzmann equation with the Bhatnagar–Gross–Krook (BGK) model used by Guo et al. [52]. A standard differentially heated cavity with a large temperature difference was chosen as a testbed for validating the mesoscopic method when applied to the simulation of Boussinesq natural convection flow, as well as for addressing the non-Boussinesq flow at near-turbulent and turbulent steady states [50] and for transient [51] flows.

### 1.2.4 Immersed boundary method for thermal compressible flow

In order to acquire the ability to simulate thermal compressible flows in the presence of complex boundaries it is common to extend the existing solvers by incorporating capabilities of the immersed boundary method (IBM) within them. IBM, initially developed by Peskin [53], can simulate flow in the presence of complex, movable, and deformable boundaries. The simulations take advantage of solvers utilizing compact and simple stencils of discretized differential operators that can be efficiently employed on structured grids. The boundary conditions of no-slip and prescribed values of temperature (or heat flux) on each immersed boundary are enforced by introducing forces and heat fluxes as additional unknowns of the problem. A closure of the overall system is achieved by including additional equations in the form of kinematic constraints for all the unknowns.

In the last decade IBM has been widely utilized for investigating natural convection within enclosures with embedded discrete thermally active sources (or sinks) of various geometries. Interest in this field has been motivated by its addressing a broad spectrum of engineering applications based on gas-solid heat exchangers and a fundamental understanding of the instability of highly separated confined flows. It is worth mentioning in this context the works of [54], [55], [56], [57] and the studies of [58], [59], [60], [61], [62], which addressed natural convection confined flows in the presence of complex two-dimensional and three-dimensional geometries, respectively.

Studies utilizing IBM for the analysis of thermal compressible flows are relatively scarce. Most of the works in this field addressed high Mach number compressible flows, focusing on transonic/supersonic transitions, or comparing between characteristics of subsonic and supersonic flows. For high Mach number flows the impact of viscosity and the thermal behavior of the flow is negligible compared to the compressibility effects, and thus in these studies both phenomena are typically neglected. In contrast, when simulating low Mach number thermal flows, which is the focus of the current study, both effects play a significant role and should be carefully addressed.

It should be noted that accurate implementation of IBM forcing in low Mach number compressible flows is still the subject of active research. In a recently developed IBM scheme [63] the authors introduced a novel pressure-based correction of the IBM forcing (in addition to the classical one based on the time derivative of velocity)[1] and applied it to the analysis of three-dimensional low and high Mach number pressure driven flows. Comparison of the obtained results with the corresponding data obtained by body-fitted DNS revealed that pressure-correction of the IBM forcing significantly improves the accuracy of the IBM procedure for low Mach number flows. An additional contribution to the application of IBM for the simulation of compressible thermally driven confined flow is due to the work of Kumar and Natarajan [64]. The authors developed a diffuse immersed boundary approach for thermally driven non-Boussinesq flows, which, however, relies on a quasi-incompressible formulation of the governing equations and therefore cannot be considered to be a fully compressible approach.

## 1.3  Objectives of the study

The present study aims to develop and extensively verify a general transient pressure-based solver for the simulation of thermally driven non-Boussinesq flows within complex geometries typical of modern nuclear reactors. A set of governing equations, including the continuity, the momentum and the energy equations along with the equation of state, are solved, while the thermophysical properties of the fluid are determined at each time step. The solver employs backward second-order finite difference and standard second-order finite volume methods for the temporal and the spatial discretizations, respectively. A semi-implicit fractional-step method for all the flow speeds is implemented for the pressure-velocity coupling. The IBM formulation that has been successfully applied for compressible pressure driven flows [63] was chosen, and further extended for the simulation of thermally driven non-Boussinesq flows.

The present study is accomplished in three stages: (i) the development of a generic solver for the simulation of non-Boussinesq thermally driven flows in rectangular containers, (ii) the development of a novel formulation of the IBM suitable for non-Boussinesq thermally driven flows and (iii) implementation of the developed IBM by incorporating it into the generic solver developed in stage (i).

 In the first stage, the results reported in [19] and [22] are carefully replicated with regard to the simulation of isothermal compressible flow. The methodology utilized in the current study

---

[1] We note in passing that pressure-correction of the direct forcing IBM must not be confused with a pressure-correction equation of the fractional-step and the SIMPLE related methods

differs from that reported in [19] and [22] in the sense that: (i) the flow is assumed to be compressible a priori, which implies solution of the compressible continuity, momentum and energy equations, (ii) the governing equations are unsteady and therefore time integration is performed until the flow converges to its steady state up to the predefined convergence criterion. We next carefully replicate the results reported in [29] with respect to both Bousinesq and non-Bousinesq flows in differentially heated square cavities. Finally, the current study extends the IBM formulation reported in [63] to the non-Boussinesq thermally driven flows and presents in-depth analysis of the characteristics of non-Boussinesq natural convection flow typical of the configuration consisting of a hot cylinder placed within a square cold cavity.

## 2. Theoretical background

Natural convection flow is ubiquitous in a vast variety of engineering applications. The flow is governed by the system of continuity, Navier-Stokes, and energy equations, which, due to their non-linearity, must be solved numerically. The methods developed over the years for numerical simulation of natural convection phenomena are classified into coupled and segregated approaches to provide full pressure-velocity coupling. The current study utilizes the segregated pressure–velocity coupling approach, namely the fractional-step method. When utilized in the context of the Boussinesq approximation, (i.e., flow is assumed incompressible) the method consists of a number of basic steps: (i) predictor that aims to accurately estimate the velocity field by utilizing the pressure field from the previous time step, (ii) corrector that aims to obtain the pressure correction for the current time step, (iii) projection using the pressure-correction values to update the pressure field and to project the velocity field on the divergence free subspace and (iv) solution of the energy equation. For non-Boussinesq natural convection flows (i.e., the flow is assumed to be compressible), the procedure is more sophisticated, since in this case the pressure is a thermodynamic rather than simply a hydrodynamic property. An additional difficulty is that for compressible flow the density, the viscosity and the thermal conductivity fields are not constant anymore. Rather, the density is coupled with the pressure by the equation of state, while the viscosity and thermal conductivity are coupled with the temperature field by the Sutherland laws. To resolve this, step (iii) is modified. Namely, at step (iii) the pressure corrections are further utilized to update both the pressure and the velocity fields at the current time step and to calculate intermediate density by utilizing the equation of state. At step (iv) the energy equation is solved, and the current time-step temperature is obtained. At the final step, step (v), the viscosity and thermal conductivity are updated by utilizing the Sutherland equations. Note that steps (iii) – (v) are repeated within the outer iteration until the continuity equation is satisfied up to a given precision.

Most engineering applications of fluid dynamics and heat transfer involve complex geometries whose modeling by standard body conformed grids is often a challenge. The problem exists regardless of the heat transfer mechanism governing the system. There are several strategies to overcome this problem, including stepwise approximation of orthogonal grids, overlapping grids and boundary-fitted non-orthogonal grids, and utilizing IBM.

IBM utilizes regular (typically Cartesian) grids and resolves the boundary irregularity by locally inducing additional volumetric sources and heat fluxes to impose kinematic constraints on the immersed surface. Details on the immersed boundary formulation utilized in the current study are given in the following.

The current study focuses on the solution of compressible natural convection flows in the presence of complex geometries by combining implicit fractional-step and immersed boundary methods. A detailed description of the methodology and the numerical formulation utilized in the current study is presented below.

## 2.1 Governing equations
### 2.1.1 Continuity, momentum, and energy equations

The dimensional form of the continuity, momentum, and total energy equations is given by:

$$\frac{\partial \tilde{\rho}}{\partial \tilde{t}} + \tilde{\nabla} \cdot (\tilde{\rho}\tilde{\vec{u}}) = 0, \tag{2.1}$$

$$\frac{\partial (\tilde{\rho}\tilde{\vec{u}})}{\partial \tilde{t}} + \tilde{\nabla} \cdot (\tilde{\rho}\tilde{\vec{u}} \otimes \tilde{\vec{u}}) = \tilde{\rho}\vec{g} - \tilde{\nabla}\tilde{p} + \tilde{\nabla} \cdot \left(\tilde{\mu}\left[(\tilde{\nabla}\tilde{\vec{u}} + \tilde{\nabla}\tilde{\vec{u}}^T) - \frac{2}{3}(\tilde{\nabla} \cdot \tilde{\vec{u}}) \cdot \bar{\bar{I}}\right]\right), \tag{2.2}$$

$$\frac{\partial \left(\tilde{\rho}\left[\tilde{e} + |\tilde{\vec{u}}|^2/2\right]\right)}{\partial \tilde{t}} + \tilde{\nabla} \cdot \left(\tilde{\rho}\tilde{\vec{u}}\left[\tilde{e} + |\tilde{\vec{u}}|^2/2\right]\right) =$$
$$= \tilde{\nabla} \cdot (\tilde{k}\tilde{\nabla}\tilde{T}) + \tilde{\nabla} \cdot \left(\tilde{\mu}\left[(\tilde{\nabla}\tilde{\vec{u}} + \tilde{\nabla}\tilde{\vec{u}}^T) - \frac{2}{3}(\tilde{\nabla} \cdot \tilde{\vec{u}}) \cdot \bar{\bar{I}}\right] \cdot \tilde{\vec{u}}\right) + \tilde{\rho}\vec{g}\tilde{\vec{u}}, \tag{2.3}$$

where $\tilde{\vec{u}}$ is the velocity, $\tilde{\rho}$ is the density, $\tilde{p}$ is the pressure, $\tilde{e}$ is the internal energy, $\tilde{\mu}$ is the dynamic viscosity, $\tilde{k}$ is the thermal conductivity, $\vec{g}$ is the standard gravity, and $\tilde{t}$ is time. The generic subscript symbol $\sim$ corresponds to a dimensional variable.

The energy equation can be modified slightly. First, the kinetic and potential energies are relatively small in natural convection flows. Second, the dissipation term is typically neglected. Finally, the internal energy, $\tilde{e}$, is a thermodynamic state variable that is rarely used in practical engineering applications, while the more commonly used quantity is the enthalpy, $\tilde{h}$, that is related to the internal energy. For ideal gases, enthalpy is equal to the temperature multiplied by the specific heat capacity at constant pressure. Then, the energy equation is rendered as:

$$\frac{\partial (\tilde{\rho}\widetilde{C_p}\tilde{T})}{\partial \tilde{t}} + \tilde{\nabla} \cdot (\tilde{\rho}\tilde{\vec{u}}\widetilde{C_p}\tilde{T}) = \tilde{\nabla} \cdot (\tilde{k}\tilde{\nabla}\tilde{T}) + \frac{\partial \tilde{p}}{\partial \tilde{t}} + \tilde{\nabla} \cdot (\tilde{p}\tilde{\vec{u}}) - \tilde{p}(\tilde{\nabla} \cdot \tilde{\vec{u}}), \tag{2.4}$$

where $\widetilde{C_p}, \widetilde{T}$ are the specific heat capacity at constant pressure and temperature, respectively.

For an ideal gas, the equation of state is:

$$\tilde{p} = \tilde{\rho} R \tilde{T}, \tag{2.5}$$

where $R$ is the specific gas constant.

The governing equations were rendered dimensionless by utilizing the scaling proposed in [29]:

$$\vec{x} = \frac{\tilde{\vec{x}}}{L_0}, \quad t = \frac{\tilde{t}}{L_0^2/\alpha_0}, \quad \vec{u} = \frac{\tilde{\vec{u}}}{\alpha_0/L_0}, \quad p = \frac{\tilde{p}}{p_0}, \quad \rho = \frac{\tilde{\rho}}{\rho_0}$$
$$T = \frac{\tilde{T}}{T_0}, \quad C_p = \frac{\widetilde{C_p}}{C_{p_0}}, \quad \mu = \frac{\tilde{\mu}}{\mu_0}, \quad k = \frac{\tilde{k}}{k_0}, \tag{2.6}$$

where $\vec{x}$ are geometric coordinates, $L_0$ is a geometric characteristic length, $\alpha_0$ is the thermal diffusivity, $p_0$ is the characteristic pressure, $\rho_0$ is the characteristic density, $T_0$ is the characteristic temperature, $C_{p_0}$ is the characteristic specific heat capacity for constant pressure, $\mu_0$ is the characteristic dynamic viscosity and $k_0$ is the characteristic thermal conductivity.

The above scaling yields the following non-dimensional groups governing the natural convection flow:

$$Ra = \frac{\rho_0 g (T_h - T_c) L_0^3}{\mu_0 \alpha_0 T_0}, \quad Pr = \frac{\mu_0}{\rho_0 \alpha_0}, \quad M_0^2 = \frac{\alpha_0^2/L_0^2}{\kappa R T_0}, \quad \varepsilon = \frac{T_h - T_c}{2 T_0}, \tag{2.7}$$

where $Ra, Pr$, and $M_0$ are the Rayleigh, Prandtl and Mach numbers, respectively, $\kappa$ is the ratio of specific heat capacities, i.e., $\kappa = C_{p_0}/C_{v_0}$, $\varepsilon$ is a normalized temperature difference parameter, $T_h, T_c$ are the maximal and minimal temperatures, respectively, whereas the relation between these temperatures and the characteristic temperature is $T_0 = 0.5(T_h + T_c)$.

The non-dimensional mass, momentum and energy equations are next formulated as:

$$\frac{\partial \rho}{\partial t} + \nabla \cdot (\rho \vec{u}) = 0, \tag{2.8}$$

$$\frac{\partial (\rho \vec{u})}{\partial t} + \nabla \cdot (\rho \vec{u} \otimes \vec{u}) = \frac{Ra \, Pr}{2\varepsilon} \rho \hat{n}_g - \frac{1}{\kappa M_0^2} \nabla p + Pr \nabla \cdot \left( \mu \left[ \nabla \vec{u} + \nabla \vec{u}^T - \frac{2}{3} (\nabla \cdot \vec{u}) \cdot \bar{\bar{I}} \right] \right), \tag{2.9}$$

$$\frac{\partial (\rho C_p T)}{\partial t} + \nabla \cdot (\rho \vec{u} C_p T) = \nabla \cdot (k \nabla T) + \frac{\kappa - 1}{\kappa} \left( \frac{\partial p}{\partial t} + \vec{u} \cdot \nabla p \right). \tag{2.10}$$

The term $(\rho_0 R T_0/p_0)$ appears to be equal to unity, and therefore the non-dimensional equation of state is:

$$\rho = \frac{p}{T} = C_\rho p, \qquad (2.11)$$

where $C_\rho$ is a density coefficient equal to $1/T$.

### 2.1.2 Determining viscosity and thermal conductivity

#### 2.1.2.1 Dynamic viscosity

For compressible flow, viscosity must be determined using specific correlation. For ideal gases, the Sutherland equation is typically used:

$$\tilde{\mu} = \mu_0 \left(\frac{\tilde{T}}{T_0}\right)^{\frac{3}{2}} \frac{T_0 + S_\mu}{\tilde{T} + S_\mu}, \qquad (2.12)$$

where $S_\mu$ is the dimensional Sutherland temperature for viscosity.

Eq. (2.12) is rendered dimensionless as:

$$\mu = \frac{1 + C_\mu}{T + C_\mu} T^{\frac{3}{2}}, \qquad (2.13)$$

where $C_\mu = \frac{S_\mu}{T_0}$ is the dimensionless Sutherland temperature for viscosity.

#### 2.1.2.2 Thermal conductivity

For ideal gases, Sutherland's law can also be applied for thermal conductivity:

$$\tilde{k} = k_0 \left(\frac{\tilde{T}}{T_0}\right)^{\frac{3}{2}} \frac{T_0 + S_k}{\tilde{T} + S_k}, \qquad (2.14)$$

where $S_k$ is the dimensional Sutherland temperature for thermal conductivity.

Eq. (2.14) is rendered dimensionless as:

$$k = \frac{1 + C_k}{T + C_k} T^{\frac{3}{2}}, \qquad (2.15)$$

where $C_k = \frac{S_k}{T_0}$ is the dimensionless Sutherland temperature for thermal conductivity.

## 2.2 Semi-implicit fractional-step method for thermally driven compressible flow

### 2.2.1 Predictor step

When solving the system of continuity, energy and Navier-Stokes equations, outer iterations are used to ensure the mass conservation of the flow. Specifically, at the $m^{\text{th}}$ iteration of the current time step the momentum equation is first solved in order to predict the velocity field. The 2$^{\text{nd}}$-order, three-time-level scheme was utilized for the time integration, while the diffusion term was taken in implicit/explicit form, and convective and pressure terms were taken explicitly from the previous calculations:

$$\frac{3\rho^{m-1}\vec{u}^*}{2\Delta t} - L(\vec{u})^{*/n} = \frac{RaPr}{2\varepsilon}\rho^{m-1}\hat{n}_g - N(\vec{u}^n) - \frac{1}{\kappa M_0^2}\nabla p^{m-1} + \frac{4(\rho\vec{u})^n - (\rho\vec{u})^{n-1}}{2\Delta t}, \quad (2.16)$$

where $\vec{u}^*$ is the predictor field for $\vec{u}^m$, $N(\vec{u})$ is the convection fluxes (non-linear term) and $L(\vec{u})$ is the diffusion fluxes (linear term).

### 2.2.2 Momentum corrector

After obtaining the predicted velocity field, the pressure-correction equation developed from the continuity equation is solved as follows:

$$\frac{3}{2\Delta t}C_\rho p' - \nabla \cdot \left(\frac{2\Delta t}{3\kappa M_0^2}\nabla p'\right) - \nabla \cdot (C_\rho p' \vec{u}^*) = -\Delta\dot{m}, \quad (2.17)$$

where $\Delta\dot{m}$ is a mass flow imbalance that exists because the predicted velocity does not necessarily satisfy the continuity equation:

$$\Delta\dot{m} = \frac{\partial \rho^{m-1}}{\partial t} + \nabla \cdot (\rho^{m-1}\vec{u}^*). \quad (2.18)$$

### 2.2.3 Projection step for the pressure, velocity, and density fields

The solution of Eq. (2.17) is followed by the projection step, at which pressure, density, and velocity are updated:

$$p^m = p^{m-1} + p', \quad \vec{u}^m = \vec{u}^* + \vec{u}', \quad \rho^* = \rho^{m-1} + \rho', \quad (2.19)$$

where:

$$\vec{u}' = -\frac{2\Delta t}{3\rho^{m-1}\kappa M_0^2}\nabla p', \quad \rho' = \frac{\partial\rho}{\partial p}p' = \frac{1}{T}p' = C_\rho p'. \tag{2.20}$$

### 2.2.4 Solution of the energy equation

The procedure sequence is finalized by solution of the energy equation to obtain the new temperature field:

$$\frac{\partial(\rho^* C_p T^m)}{\partial t} + \nabla\cdot(\rho^*\vec{u}^m C_p T^m) - \nabla\cdot(k^{m-1}\nabla T^m) = \frac{\kappa-1}{\kappa}\left(\frac{\partial p^m}{\partial t} + \vec{u}^m\cdot\nabla p^m\right), \tag{2.21}$$

where $C_p$ is a non-dimensional specific heat capacity which is equal to unity.

### 2.2.5 Updating thermophysical properties

After the energy equation has been solved, the viscosity, $\mu_{ij}^m$, and thermal conductivity, $k_{ij}^m$, are updated using the Sutherland equations given by Eqs. (2.13) and (2.15). The $C_\rho$ coefficient and density field are updated by using the equation of state:

$$C_\rho = \frac{1}{T^m}, \quad \rho^m = C_\rho p^m. \tag{2.22}$$

### 2.2.6 Summary of the correction loop

Eqs. (2.17), (2.19)-(2.22) constitute an outer iteration loop, which is run at each time step. At the end of the iteration, after updating the thermophysical properties, the solver performs a mass conservation check by restricting the L-infinity norm to a value of $10^{-6}$. The fields of $p$, $\rho$ and $\vec{u}$ are updated until predefined values of convergence criteria are achieved for each field. After a sufficient number of iterations has been performed, corrections become negligible, the flow can be considered as mass-conservative and the solution can proceed to the next time step by setting $\vec{u}^{n+1} = \vec{u}^m$, $p^{n+1} = p^m$, $T^{n+1} = T^m$ and $\rho^{n+1} = \rho^m$. A graphical representation of all the steps that are performed in each time step is shown in Fig 2.1.

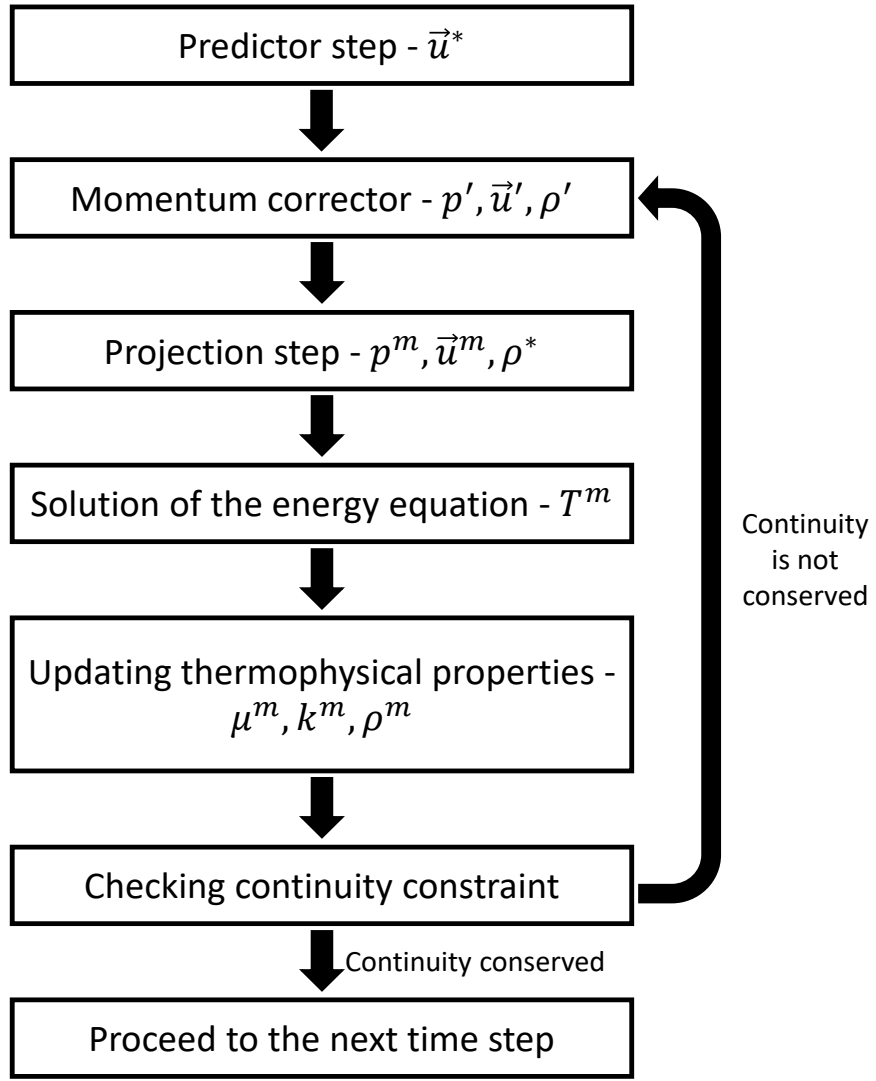

Figure 2.1: Graphical representation of the calculation steps at each time step for semi-implicit FSM for thermally driven compressible flow

## 2.3 Immersed boundary formulation

### 2.3.1 Immersed boundary method for compressible flows

The immersed boundary formulation utilized in the framework of the current study is briefly introduced first, and the strategies for its application in the analysis of compressible natural convection flow are next explained. Fig. 2.2 shows the setup of a typical spatial discretization implemented on a staggered grid. The grid is characterized by offset velocity, temperature, pressure, density, viscosity, and thermal conductivity fields. An arbitrary immersed object, $B$, within a computational domain, $D$, (whose geometry does not, in general, have to conform to the underlying spatial grid), is represented by the surface, $\partial B$, determined by a set of Lagrangian points, $\vec{X}_K$. At the Lagrangian points, appropriate surface forces, $\vec{F}_K$, and heat sources, $Q_K$, are applied to enforce the non-slip velocity and the Dirichlet temperature boundary conditions along $\partial B$.

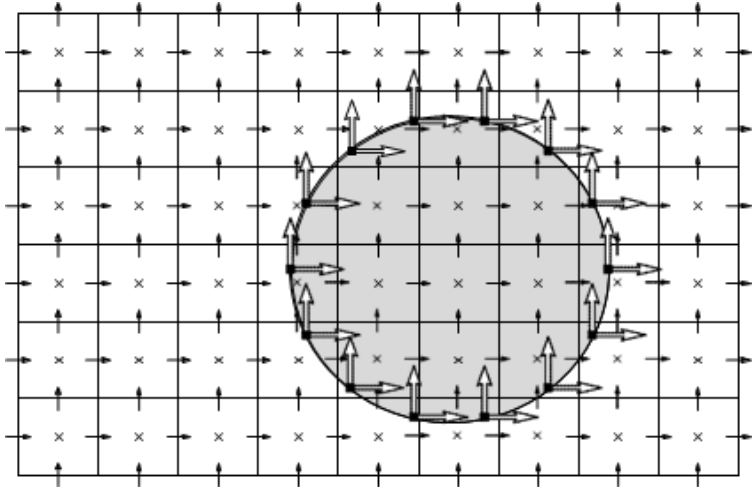

Figure 2.2: Staggered grid discretization of a two-dimensional computational domain $D$ and immersed boundary formulation for a body $B$, depicted by a shaded object. The horizontal and vertical arrows ($\rightarrow, \uparrow$) represent the discrete $u_j$ and $v_j$ velocity locations, respectively. Pressure $p_j$, temperature $T_j$, density $\rho_j$, viscosity $\mu_j$ and thermal conductivity $k_j$ are applied at the center of each cell ($\times$). Lagrangian points $\vec{X}_K(X_K, Y_K)$ along $\partial B$ are shown as filled squares ($\blacksquare$) where boundary forces $\vec{F}_K\left(F_{K_x}, F_{K_y}\right)$ or boundary heat sources $Q_K$ are applied ($\Rightarrow, \Uparrow$).

These forces and heat fluxes appear as additional unknown variables, whose values – along with those for the pressure, temperature, and velocity fields – are provided by solving the Navier Stokes and the energy equations and are directly accounted for in the overall balance, enabling direct calculation of the Nusselt number. Since the location of the Lagrangian boundary points does not necessarily coincide with the underlying spatial discretization, regularization and interpolation operators must be defined to convey information in both directions of the body. The regularization operator smears singularly acting surface forces, $\vec{F}_K$, and heat sources, $Q_K$, on the nearby computational domain, while the interpolation operator acts in the opposite direction and imposes non-slip/thermal boundary conditions on the points located on the body surface. Equations governing the thermally driven, compressible natural convection flow, along with the embedded immersed boundary formulation can be written as:

$$\frac{\partial \rho}{\partial t} + \nabla \cdot (\rho \vec{u}) = 0, \qquad (2.23)$$

$$\frac{\partial (\rho \vec{u})}{\partial t} + \nabla \cdot (\rho \vec{u} \otimes \vec{u}) =$$
$$= \frac{RaPr}{2\varepsilon} \rho \hat{n}_g - \frac{1}{\kappa M_0^2} \nabla p + Pr \nabla \cdot \left(\mu \left[\nabla \vec{u} + \nabla \vec{u}^T - \frac{2}{3}(\nabla \cdot \vec{u}) \cdot \bar{\bar{I}}\right]\right) + \vec{f}^\Gamma, \qquad (2.24)$$

$$\frac{\partial (\rho C_p T)}{\partial t} + \nabla \cdot (\rho \vec{u} C_p T) = \nabla \cdot (k \nabla T) + \frac{\kappa - 1}{\kappa}\left(\frac{\partial p}{\partial t} + \vec{u} \cdot \nabla p\right) + q^\Gamma, \qquad (2.25)$$

where the volumetric force $\vec{f}^\Gamma$ and volumetric heat source $q^\Gamma$ reflect the impact of the immersed body on the surrounding flow.

The immersed boundary formulation of the compressible flow utilized in the current study was first introduced by Riahi et al. [63] and is currently extended for the simulation of thermally driven compressible flows. This method originates from the works of Uhlmann [65] and Pinelli et al. [66], combining advantages of both continuous [67] and discrete forcing methods [68].

### 2.3.2 Communication between the Eulerian and Lagrangian systems

Communication between the Lagrangian and Eulerian grids is performed in three steps, as described below.

#### 2.3.2.1 Interpolation step

In the interpolation step, physical quantities in the Eulerian mesh are interpolated on the boundary $\partial B$. We next give a specific example for the interpolation of $(\rho \vec{u})$, $(\rho C_p T)$, $(\nabla p)$ quantities discretized on the Eulerian mesh. The values of $(\rho \vec{U})_K$ and $(\rho C_p T)_K$ interpolated on any Lagrangian point $K$ are calculated by employing the interpolation operator $I$:

$$I[\rho \vec{u}]_{\vec{X}_K} = (\rho \vec{U})_K = \sum_{\substack{i \in N_x \\ j \in N_y}} (\rho \vec{u})_{i,j}^{N_x, N_y} \delta(x_i - X_K) \delta(y_j - Y_K) \Delta x \Delta y, \tag{2.26}$$

$$I[\nabla p]_{\vec{X}_K} = (\nabla P)_K = \sum_{\substack{i \in N_x \\ j \in N_y}} (\nabla P)_{i,j}^{N_x, N_y} \delta(x_i - X_K) \delta(y_j - Y_K) \Delta x \Delta y, \tag{2.27}$$

$$I[\rho C_p T]_{\vec{X}_K} = (\rho C_p T)_K = \sum_{\substack{i \in N_x \\ j \in N_y}} (\rho C_p T)_{i,j}^{N_x, N_y} \delta(x_i - X_K) \delta(y_j - Y_K) \Delta x \Delta y, \tag{2.28}$$

where $N_x \times N_y$ is a dimension of the Eulerian grid and $\delta$ is a discrete Dirac delta function defined in Eq. (2.29).

Convolutions with the Dirac delta function $\delta$ are used to facilitate the exchange of information from the Eulerian to the Lagrangian grid. Among the variety of discrete delta functions available, we chose the function described by Roma et al. [69], which was specifically designed for use on staggered grids, where even/odd de-coupling does not occur:

$$\delta(r) = \begin{cases} \frac{1}{3}\left(1 + \sqrt{-3r^2 + 1}\right), & 0 \leq r \leq 0.5, \\ \frac{1}{6}\left(5 - 3r - \sqrt{-3(1-r)^2 + 1}\right), & 0.5 \leq r \leq 1.5, \\ 0, & r > 1.5. \end{cases} \tag{2.29}$$

The chosen discrete delta function is supported only over three cells, which is advantageous for computational efficiency. As observed by Colonius and Taira [70], no significant differences in the results are expected when using the alternative discrete delta functions used in previous works.

#### 2.3.2.2 Calculation of direct forcing sources on Lagrangian grid

Following the interpolation step, the Lagrangian volumetric forces and heat sources are calculated as was suggested in [63]:

$$\vec{F}_K = \frac{1}{\Delta t}\left[\left(\rho\vec{U}\right)_K^d - \left(\rho\vec{U}\right)_K\right] - \left[\nabla P_K^d - \nabla P_K\right]\cdot\hat{n}_{ns}, \tag{2.30}$$

$$Q_K = \frac{1}{\Delta t}\left[\left(\rho C_p T\right)_K^d - \left(\rho C_p T\right)_K\right], \tag{2.31}$$

where superscript $d$ applies for the desired boundary condition of the immersed body.

#### 2.3.2.3 Regularization step

In the regularization step, the values of volumetric sources previously calculated on the Lagrangian grid are regularized (spread) back to the Eulerian grid. This backward communication is implemented by utilizing the same delta functions as for the interpolation step. The values of the volumetric force and heat source terms evaluated on the Eulerian mesh are given by:

$$\vec{f}^{\Gamma}(x,y) = \sum_{K\in N_K} \vec{F}_K\,\delta(x_i - X_K)\,\delta(y_j - Y_K)\Delta x \Delta y, \tag{2.32}$$

$$q^{\Gamma}(x,y) = \sum_{K\in N_K} Q_K\,\delta(x_i - X_K)\,\delta(y_j - Y_K)\Delta x \Delta y. \tag{2.33}$$

## 2.4 Semi-implicit fractional-step method for thermally driven compressible flow with implemented IBM

### 2.4.1 Predictor step

The solution of the momentum equation to predict the velocity field without taking into account the presence of an immersed body is given by:

$$\frac{3\rho^{m-1}\vec{u}^*}{2\Delta t} - L(\vec{u})^{*/n} = \frac{RaPr}{2\varepsilon}\rho^{m-1}\hat{n}_g - N(\vec{u}^n) - \frac{1}{\kappa M_0^2}\nabla p^{m-1} + \frac{4(\rho\vec{u})^n - (\rho\vec{u})^{n-1}}{2\Delta t}. \quad (2.34)$$

### 2.4.2 First momentum corrector

After obtaining the predicted velocity, the pressure-correction equation, which has been developed from the continuity equation, is solved again not taking into account the presence of the immersed body:

$$\frac{3}{2\Delta t}C_\rho p' - \nabla \cdot \left(\frac{2\Delta t}{3\kappa M_0^2}\nabla p'\right) - \nabla \cdot (C_\rho p'\vec{u}^*) = -\Delta \dot{m}^*, \quad (2.35)$$

where $\Delta \dot{m}^*$ is a mass flow imbalance that exists because the predicted velocity does not necessarily satisfy the continuity equation:

$$\Delta \dot{m}^* = \frac{\partial \rho^{m-1}}{\partial t} + \nabla \cdot (\rho^{m-1}\vec{u}^*). \quad (2.36)$$

Following calculation of the first pressure-correction, the intermediate pressure is calculated, and is used later when employing immersed boundary functionality:

$$p^* = p^n + p'. \quad (2.37)$$

### 2.4.3 Applying IBM for velocity to enforce a given velocity on the surface of the immersed body

After acquiring an intermediate pressure, IBM for velocity is implemented via Eqs. (2.26), (2.27), (2.29), (2.30) and (2.32). Note that the term $\vec{f}^\Gamma$ is not recalculated during the correction loop and is determined only once at the beginning of the time step. As a result, the volume force from the immersed body is added to the right-hand side of the momentum equation.

### 2.4.4 Solution of momentum equation with the impact of the immersed body

After the volumetric force has been calculated, the momentum equation with an updated right-hand side is solved to determine the new velocity field that takes into account the impact of the immersed body:

$$\frac{3\rho^{m-1}\vec{u}^m}{2\Delta t} - L(\vec{u})^{m/n} =$$
$$= \frac{RaPr}{2\varepsilon}\rho^{m-1}\hat{n}_g - N(\vec{u}^n) - \frac{1}{\kappa M_0^2}\nabla p^* + \frac{4(\rho\vec{u})^n - (\rho\vec{u})^{n-1}}{2\Delta t} + \vec{f}^{\Gamma}. \quad (2.38)$$

### 2.4.5 Second momentum corrector

At this stage, the pressure-correction equation is solved with an updated right-hand side:

$$\frac{3}{2\Delta t}C_\rho p'' - \nabla \cdot \left(\frac{2\Delta t}{3\kappa M_0^2}\nabla p''\right) - \nabla \cdot (C_\rho p''\vec{u}^*) = -\Delta \dot{m}^m, \quad (2.39)$$

where $\Delta \dot{m}^m$ is a mass flow imbalance that exists because the predicted velocity did not necessarily satisfy the continuity equation:

$$\Delta \dot{m}^m = \frac{\partial \rho^{m-1}}{\partial t} + \nabla \cdot (\rho^{m-1}\vec{u}^m). \quad (2.40)$$

Next, the new pressure and intermediate density are acquired:

$$p^m = p^* + p'', \quad \rho^* = \rho^n + C_\rho(p' + p''). \quad (2.41)$$

### 2.4.6 Solution of the energy equation without the impact of the immersed body

We next solve the energy equation without taking into account the existence of an immersed body to obtain the intermediate temperature:

$$\frac{\partial(\rho^* C_p T^*)}{\partial t} + \nabla \cdot (\rho^*\vec{u}^m C_p T^*) - \nabla \cdot (k^{m-1}\nabla T^*) = \frac{\kappa-1}{\kappa}\left(\frac{\partial p^m}{\partial t} + \vec{u}^m \cdot \nabla p^m\right). \quad (2.42)$$

### 2.4.7 Applying IBM to enforce a given temperature on the surface of the immersed body

At each time step in the first correction iteration, after acquiring an intermediate temperature, IBM for temperature is implemented via Eqs. (2.28), (2.29), (2.31) and (2.33). Note the term $q^{\Gamma}$ is not recalculated during the FSM loop and is determined only once at the beginning of the time step. As a result, the calculated volumetric heat source is added to the right-hand side of the energy equation.

### 2.4.8 Solution of the energy equation taking into account the impact of the immersed body

$$\frac{\partial(\rho^* C_p T^m)}{\partial t} + \nabla \cdot (\rho^* \vec{u}^m C_p T^m) - \nabla \cdot (k^{m-1} \nabla T^m) = \\ = \frac{\kappa - 1}{\kappa} \left(\frac{\partial p^m}{\partial t} + \vec{u}^m \cdot \nabla p^m\right) + q^\Gamma. \tag{2.43}$$

### 2.4.9 Updating thermophysical properties

After the energy equation has been solved, viscosity $\mu_{ij}^m$, thermal conductivity $k_{ij}^m$, the $C_\rho$ coefficient and density $\rho_{ij}^m$, are updated by using the Sutherland equations and the equation of state via Eqs. (2.13), (2.15) and (2.22).

### 2.4.10 Summary of the correction loop

Similarly to the general formulation of the compressible semi-implicit fractional-step method presented in section 2.2, in the semi-implicit fractional-step method combined with IBM, Eqs. (2.35), (2.37)-(2.39) and (2.41)-(2.43) constitute an outer iteration loop, which is run at each time step. At the end of the iteration, after updating the thermophysical properties, the solver performs a mass conservation check by restricting the L-infinity norm to a value of $10^{-6}$. The fields of $p, \rho$ and $\vec{u}$ are updated until predefined values of convergence criteria are achieved for each field. After performing a sufficient number of iterations, corrections become negligible and the flow can be considered to be mass-conservative. The solution can then proceed to the next time step by setting $\vec{u}^{n+1} = \vec{u}^m, p^{n+1} = p^m, T^{n+1} = T^m$ and $\rho^{n+1} = \rho^m$. The graphical representation of all the steps that are performed in each time step are shown in Fig 2.3.

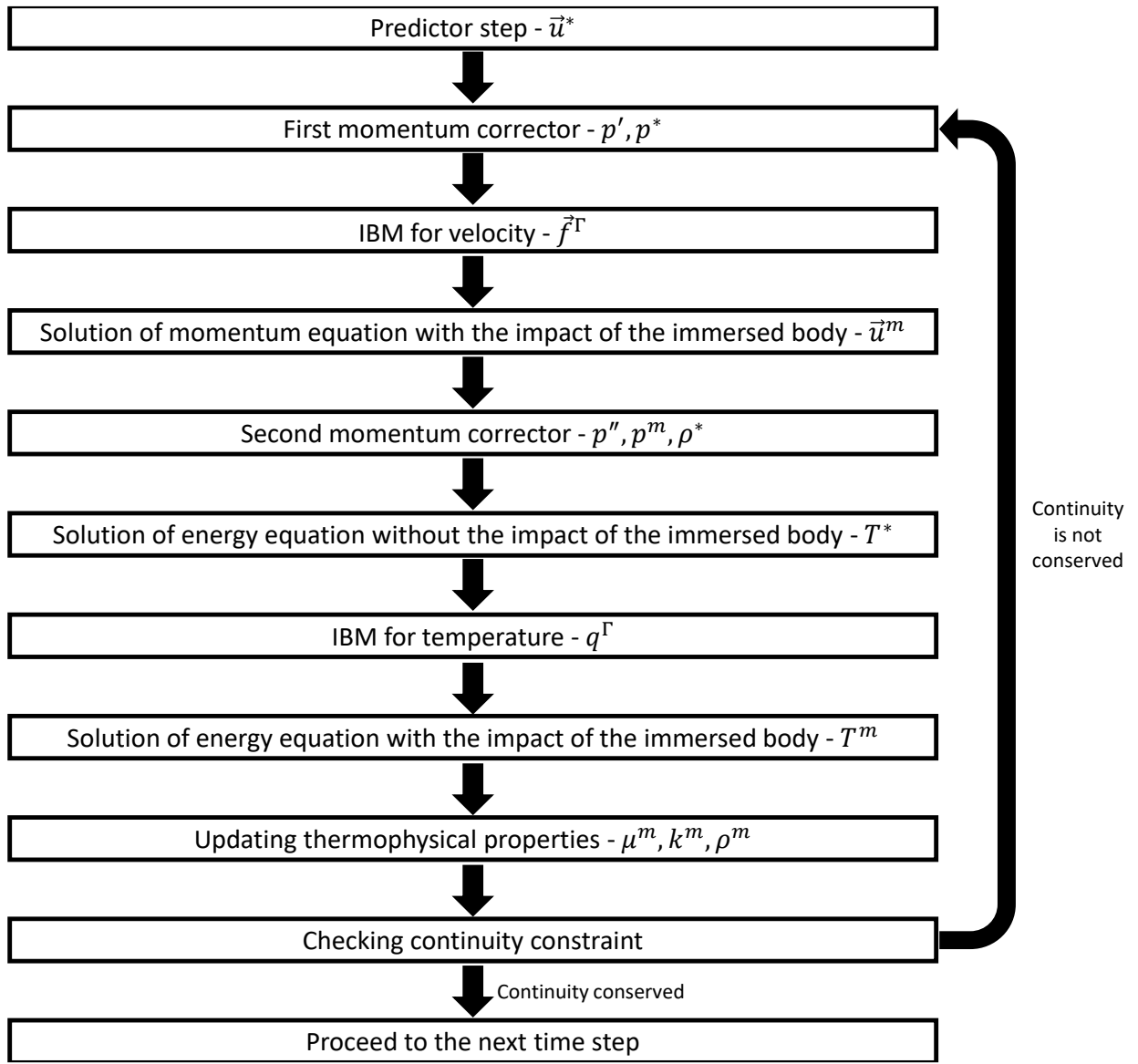

Figure 2.3: Graphical representation of the calculation steps at each time step for semi-implicit FSM for thermally driven compressible flow with implemented IBM

### 2.4.11 Boundary conditions

The numerical methodology applied in the current study was implemented by utilizing primitive variables, i.e., velocity components, pressure, and temperature. The reason for the above choice is the simplicity of formulating the boundary conditions, at least for standard configurations like tube/channel flow or natural convection flow within confinements. Determining proper boundary conditions for the velocity vector is a simple process of choosing between a few variants: no-slip at a wall ($u_{ns} = 0, u_{ss} = u_w$), slip at a wall ($u_{ns} = 0, \partial u_{ss}/\partial n = 0$), constant inlet profile, constant outlet profile and zero gradient at the outlet (if the pressure at the outlet is known). This is also correct with respect to determining boundary conditions for temperature which can be either constant temperature ($T = 0$) or constant heat flux $\left(\partial T/\partial n = f(\vec{x})\right)$ at a specified boundary. Determining boundary conditions for pressure is not as straightforward. In confined enclosures it is common to apply a zero gradient boundary condition normal to the wall direction. However, such an approximation requires further clarifications. The subject was investigated thoroughly by Gresho and Sani [71] with respect to applying physically correct boundary conditions for pressure for incompressible flow in enclosures. The key idea was to employ locally the momentum equation near the specified boundary, i.e., wall. In the current study we followed the same principle when simulating compressible flow, i.e., locally employing the momentum equation Eq. (2.9). As a result of the no-slip constraint, all the velocity components at the cavity walls are equal to zero, resulting in $\partial(\rho \vec{u})/\partial t = 0$ and $\nabla \cdot (\rho \vec{u} \otimes \vec{u}) = 0$. The momentum equation at the wall thus can be reduced to:

$$\nabla p = \kappa M_0^2 \left\{ \frac{RaPr}{2\varepsilon} \rho \hat{n}_g + Pr \nabla \cdot \left( \mu \left[ \nabla \vec{u} + \nabla \vec{u}^T - \frac{2}{3} (\nabla \cdot \vec{u}) \cdot \bar{\bar{I}} \right] \right) \right\}. \tag{2.44}$$

At this point the order of every term entering Eq. (2.44) can be assessed. In the first term the maximal value of $Ra$ is $10^6$, the $Pr$ is constant and equal to 0.71, and the values of $\varepsilon$ and density are in the range of $0.4 \leq \rho \leq 1.6$. Therefore, the maximal value of the first term is of order $10^7$. The second term multiplies the values of $Pr$, the dynamic viscosity and the second derivative of the velocity. As already mentioned, the value of $Pr$ is equal to 0.71; the values of viscosity lie in the range of $0.6 \leq \mu \leq 1.6$ (for the specific values of $\varepsilon$), which means that the order of the second term is determined by the second derivative of velocity at the wall. Both terms are multiplied by $M_0^2$, which is of order $10^{-12}$. Assuming that the order of the second derivative of velocity is significantly lower than the order of $M_0^2$, the boundary condition of zero pressure gradient normal to the wall direction can be justified. After the simulations were performed, the above assumption was further successfully verified by calculating the second derivative of velocity in the vicinity of all boundaries whose value was of order $10^6$, which bounds the normal pressure gradient at the cavity walls by $10^{-6}$.

# 3. Verification study

The methodology described in chapter 2 was verified by applying it to the solution of two benchmark problems - simulating incompressible, and thermally driven compressible flows. The flow was driven by two different mechanisms: isothermal flow in a long rectangular narrow channel driven by a pressure gradient, and natural convection flow within a differentially heated square cavity driven by a temperature gradient. The results acquired for both configurations were compared with data available in the literature.

## 3.1 Test case I - Isothermal compressible flow in narrow channel

### 3.1.1 Test case overview

This test case focusses on proving the capability to address incompressible isothermal flow by applying the currently developed methodology for compressible flow. The flow configuration presented in [22] simulates flow within an extended channel to minimize the effect of the outflow boundary on the upstream recirculation zones. A schematic of the geometric properties and boundary conditions of the considered configuration are shown in Fig. 3.1.

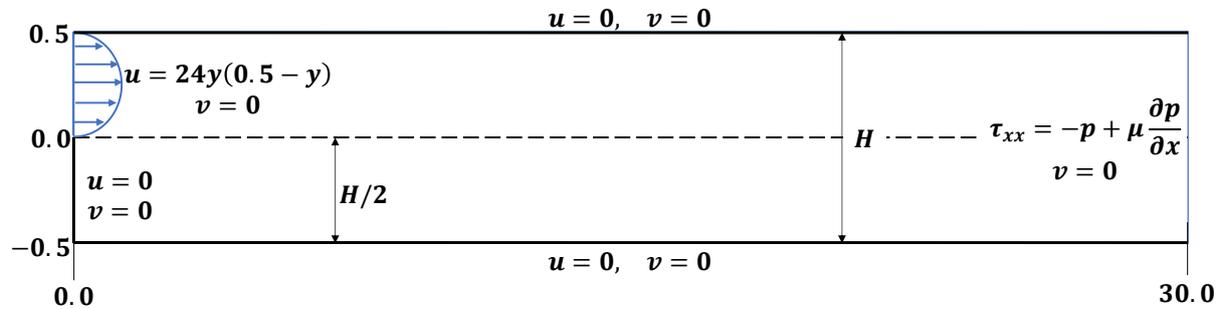

Figure 3.1: Schematic of the flow within an extended channel

The fluid enters the domain at the upper half of the left side, proceeds through the channel and exits at the right side. No-slip velocity and zero-gradient pressure boundary conditions were applied at all stationary walls. At the inlet, the vertical component of the velocity was equal to zero, a zero-gradient boundary condition was applied to the pressure field and a parabolic distribution with maximal and average values equal to $u_{max} = 1.5$ and $u_{avg} = 1$, respectively, were assigned to the horizontal velocity component. At the outflow, zero values were set to the normal stress, $\tau_{xx} = -p + \mu\, \partial u/\partial x$, and to the vertical velocity component. Considering that the pressure field at the outlet is known and set to zero, the gradient of the horizontal velocity component is also equal to zero.

The Navier-Stokes equations presented in subsection 2.1.1 governing the isothermal flow within an extended channel were rendered dimensionless by using the scales defined in Eq. (3.1):

$$\vec{x} = \frac{\tilde{\vec{x}}}{L_0}, \quad \vec{u} = \frac{\tilde{\vec{u}}}{U_0}, \quad t = \frac{\tilde{t}}{L_0/U_0}, \quad p = \frac{\tilde{p}}{\rho_0 U_0^2}, \quad \rho = \frac{\tilde{\rho}}{\rho_0}, \quad Re = \frac{\rho_0 u_{avg} H}{\mu}, \quad M_0^2 = \frac{U_0^2}{\kappa R T_0}, \quad (3.1)$$

where $U_0$, is the characteristic velocity and $Re$ is the Reynolds number, and all other quantities have been defined in subsection 2.1.1. The Reynolds number was set to be 800. When calculating the normal stress, in the postprocessing stage the viscosity was adjusted to meet the above value of the Reynolds number.

Non-dimensional continuity and momentum equations and the equation of state are formulated as:

$$\frac{\partial \rho}{\partial t} + \nabla \cdot (\rho \vec{u}) = 0, \qquad (3.2)$$

$$\frac{\partial (\rho \vec{u})}{\partial t} + \nabla \cdot (\rho \vec{u} \otimes \vec{u}) = -\nabla p + \frac{1}{Re} \nabla \cdot \left( \mu \left[ \nabla \vec{u} + \nabla \vec{u}^T - \frac{2}{3} (\nabla \cdot \vec{u}) \cdot \bar{\bar{I}} \right] \right), \qquad (3.3)$$

$$\rho = \frac{\kappa M_0^2}{T_0} p. \qquad (3.4)$$

### 3.1.2 Test results and comparison to the benchmark

The results obtained by the developed methodology were compared both qualitatively and quantitively with the corresponding data provided in [22]. To prove grid independence of the obtained results the simulations were performed on five uniform grids: $600 \times 20$, $1200 \times 40$, $1500 \times 50$, $2400 \times 80$ and $3000 \times 100$. The results that are presented below were achieved with the grid of $3000 \times 100$ cells. Tables 3.1-3.4 summarize the results of the grid independence study in terms of the values obtained for the $u$ and $v$ velocity components along two vertical lines passing through the $x = 7$ and $x = 15$ locations. It can be seen that all the values obtained on the two finest grids coincide up to four significant digits, which successfully proves grid independence of the obtained results.

*Table 3.1: Values of u velocity component currently obtained on different grids along vertical centerline passing through $x = 7$*

| $y$ | Grid $600 \times 20$ | Grid $1200 \times 40$ | Grid $1500 \times 50$ | Grid $2400 \times 80$ | Grid $3000 \times 100$ |
|---|---|---|---|---|---|
| 0.5 | 0.0000 | 0.0000 | 0.0000 | 0.0000 | 0.0000 |
| 0.45 | -0.0538 | -0.0411 | -0.0407 | -0.0387 | -0.0387 |
| 0.4 | -0.0720 | -0.0541 | -0.0525 | -0.0498 | -0.0498 |
| 0.35 | -0.0540 | -0.0380 | -0.0369 | -0.0325 | -0.0325 |
| 0.3 | -0.0005 | 0.0078 | 0.0092 | 0.0143 | 0.0143 |
| 0.25 | 0.0877 | 0.0847 | 0.0851 | 0.0923 | 0.0923 |
| 0.2 | 0.2112 | 0.1950 | 0.1954 | 0.2038 | 0.2038 |
| 0.15 | 0.3705 | 0.3399 | 0.3387 | 0.3496 | 0.3496 |
| 0.1 | 0.5595 | 0.5151 | 0.5130 | 0.5242 | 0.5242 |
| 0.05 | 0.7586 | 0.7054 | 0.7021 | 0.7121 | 0.7121 |
| 0 | 0.9365 | 0.8852 | 0.8810 | 0.8886 | 0.8886 |
| -0.05 | 1.0633 | 1.0267 | 1.0242 | 1.0270 | 1.0270 |
| -0.1 | 1.1226 | 1.1085 | 1.1065 | 1.1068 | 1.1068 |
| -0.15 | 1.1120 | 1.1209 | 1.1223 | 1.1176 | 1.1176 |
| -0.2 | 1.0402 | 1.0656 | 1.0673 | 1.0606 | 1.0606 |
| -0.25 | 0.9214 | 0.9538 | 0.9569 | 0.9460 | 0.9460 |
| -0.3 | 0.7712 | 0.8017 | 0.8035 | 0.7903 | 0.7903 |
| -0.35 | 0.6022 | 0.6272 | 0.6280 | 0.6123 | 0.6123 |
| -0.4 | 0.4189 | 0.4425 | 0.4425 | 0.4276 | 0.4276 |
| -0.45 | 0.2170 | 0.2399 | 0.2413 | 0.2320 | 0.2320 |
| -0.5 | 0.0000 | 0.0000 | 0.0000 | 0.0000 | 0.0000 |

*Table 3.2: Values of u velocity component currently obtained on different grids along vertical centerline passing through $x = 15$*

| y | Grid $600 \times 20$ | Grid $1200 \times 40$ | Grid $1500 \times 50$ | Grid $2400 \times 80$ | Grid $3000 \times 100$ |
|---|---|---|---|---|---|
| 0.5 | 0.0000 | 0.0000 | 0.0000 | 0.0000 | 0.0000 |
| 0.45 | 0.1090 | 0.1026 | 0.1018 | 0.1016 | 0.1016 |
| 0.4 | 0.2177 | 0.2050 | 0.2034 | 0.2030 | 0.2030 |
| 0.35 | 0.3267 | 0.3084 | 0.3061 | 0.3055 | 0.3055 |
| 0.3 | 0.4359 | 0.4132 | 0.4104 | 0.4095 | 0.4095 |
| 0.25 | 0.5430 | 0.5180 | 0.5150 | 0.5139 | 0.5139 |
| 0.2 | 0.6436 | 0.6188 | 0.6158 | 0.6147 | 0.6147 |
| 0.15 | 0.7316 | 0.7098 | 0.7075 | 0.7061 | 0.7061 |
| 0.1 | 0.7998 | 0.7837 | 0.7816 | 0.7807 | 0.7807 |
| 0.05 | 0.8418 | 0.8334 | 0.8329 | 0.8316 | 0.8316 |
| 0 | 0.8533 | 0.8539 | 0.8538 | 0.8535 | 0.8535 |
| -0.05 | 0.8331 | 0.8427 | 0.8443 | 0.8437 | 0.8437 |
| -0.1 | 0.7835 | 0.8007 | 0.8027 | 0.8032 | 0.8032 |
| -0.15 | 0.7097 | 0.7320 | 0.7351 | 0.7355 | 0.7355 |
| -0.2 | 0.6187 | 0.6430 | 0.6460 | 0.6471 | 0.6471 |
| -0.25 | 0.5178 | 0.5411 | 0.5442 | 0.5451 | 0.5451 |
| -0.3 | 0.4133 | 0.4330 | 0.4356 | 0.4366 | 0.4366 |
| -0.35 | 0.3090 | 0.3237 | 0.3257 | 0.3264 | 0.3264 |
| -0.4 | 0.2058 | 0.2153 | 0.2166 | 0.2170 | 0.2170 |
| -0.45 | 0.1033 | 0.1078 | 0.1084 | 0.1086 | 0.1086 |
| -0.5 | 0.0000 | 0.0000 | 0.0000 | 0.0000 | 0.0000 |

Table 3.3: Values of $v \times 10^{-3}$ velocity component currently obtained on different grids along vertical centerline passing through $x = 7$

| y | Grid $600 \times 20$ | Grid $1200 \times 40$ | Grid $1500 \times 50$ | Grid $2400 \times 80$ | Grid $3000 \times 100$ |
|---|---|---|---|---|---|
| 0.5 | 0.0000 | 0.0000 | 0.0000 | 0.0000 | 0.0000 |
| 0.45 | -0.6722 | -0.3348 | -0.3172 | -0.2812 | -0.2812 |
| 0.4 | -1.6185 | -0.9749 | -0.9269 | -0.8801 | -0.8801 |
| 0.35 | -1.8780 | -1.4744 | -1.4367 | -1.4769 | -1.4769 |
| 0.3 | -0.7839 | -1.5135 | -1.5818 | -1.8697 | -1.8697 |
| 0.25 | 1.9609 | -0.9820 | -1.2525 | -2.0565 | -2.0565 |
| 0.2 | 6.2750 | -0.0290 | -0.6444 | -2.2721 | -2.2721 |
| 0.15 | 11.6588 | 0.9508 | -0.1422 | -2.9272 | -2.9272 |
| 0.1 | 17.1599 | 1.4425 | -0.1892 | -4.4183 | -4.4183 |
| 0.05 | 21.6296 | 1.0645 | -1.1432 | -6.8694 | -6.8694 |
| 0 | 24.3636 | -0.1940 | -2.8775 | -9.9872 | -9.9872 |
| -0.05 | 25.2470 | -1.9869 | -5.0369 | -13.1653 | -13.1653 |
| -0.1 | 24.4974 | -3.8318 | -7.0856 | -15.7391 | -15.7391 |
| -0.15 | 22.4985 | -5.2920 | -8.5360 | -17.1865 | -17.1865 |
| -0.2 | 19.6939 | -6.0502 | -9.1280 | -17.2092 | -17.2092 |
| -0.25 | 16.5009 | -5.9358 | -8.6368 | -15.7420 | -15.7420 |
| -0.3 | 13.2518 | -4.9144 | -7.1458 | -12.9113 | -12.9113 |
| -0.35 | 10.0899 | -3.0621 | -4.6952 | -8.9450 | -8.9450 |
| -0.4 | 6.6974 | -0.8907 | -1.8688 | -4.4132 | -4.4132 |
| -0.45 | 2.7300 | 0.2389 | -0.1379 | -0.9633 | -0.9633 |
| -0.5 | 0.0000 | 0.0000 | 0.0000 | 0.0000 | 0.0000 |

Table 3.4: Values of $v \times 10^{-3}$ velocity component currently obtained on different grids along vertical centerline passing through $x = 15$

| $y$ | Grid $600 \times 20$ | Grid $1200 \times 40$ | Grid $1500 \times 50$ | Grid $2400 \times 80$ | Grid $3000 \times 100$ |
|---|---|---|---|---|---|
| 0.5 | 0.0000 | 0.0000 | 0.0000 | 0.0000 | 0.0000 |
| 0.45 | 0.1878 | 0.1988 | 0.2081 | 0.2033 | 0.2033 |
| 0.4 | 0.5894 | 0.6822 | 0.6998 | 0.7098 | 0.7098 |
| 0.35 | 1.0486 | 1.2980 | 1.3398 | 1.3659 | 1.3659 |
| 0.3 | 1.4253 | 1.8968 | 1.9710 | 2.0184 | 2.0184 |
| 0.25 | 1.6079 | 2.3424 | 2.4480 | 2.5253 | 2.5253 |
| 0.2 | 1.5224 | 2.5290 | 2.6774 | 2.7747 | 2.7747 |
| 0.15 | 1.1422 | 2.3988 | 2.5758 | 2.7031 | 2.7031 |
| 0.1 | 0.4947 | 1.9522 | 2.1651 | 2.3055 | 2.3055 |
| 0.05 | -0.3403 | 1.2455 | 1.4752 | 1.6332 | 1.6332 |
| 0 | -1.2451 | 0.3786 | 0.6202 | 0.7817 | 0.7817 |
| -0.05 | -2.0861 | -0.5233 | -0.2850 | -0.1264 | -0.1264 |
| -0.1 | -2.7379 | -1.3290 | -1.1108 | -0.9605 | -0.9605 |
| -0.15 | -3.1027 | -1.9216 | -1.7272 | -1.6007 | -1.6007 |
| -0.2 | -3.1273 | -2.2164 | -2.0655 | -1.9570 | -1.9570 |
| -0.25 | -2.8152 | -2.1797 | -2.0630 | -1.9881 | -1.9881 |
| -0.3 | -2.2329 | -1.8415 | -1.7692 | -1.7151 | -1.7151 |
| -0.35 | -1.5026 | -1.2976 | -1.2556 | -1.2258 | -1.2258 |
| -0.4 | -0.7805 | -0.6948 | -0.6766 | -0.6621 | -0.6621 |
| -0.45 | -0.2294 | -0.2047 | -0.2050 | -0.1949 | -0.1949 |
| -0.5 | 0.0000 | 0.0000 | 0.0000 | 0.0000 | 0.0000 |

Figs. 3.2-3.5 present a qualitative comparison between contours of the currently obtained flow fields and the corresponding results reported in [22]. As can be seen from Figs. 3.3-3.5 the values of all the currently obtained quantities lie in the same range compared to the corresponding data reported in [22]. The distribution of streamlines, shown in Fig. 3.2, along with the distributions of vorticity and velocity magnitude fields, shown in Figs. 3.4 and 3.5, respectively, confirm the presence of the staggered low-speed vortices adjacent to the upper and lower walls. The pressure field, shown in Fig. 3.3 confirms the presence of a "pressure

pocket" adjacent to the bottom wall between $x = 6$ and $x = 7$. As can be clearly seen from Fig. 3.6, the density variations over the domain are insignificant and the flow can be safely considered to be incompressible.

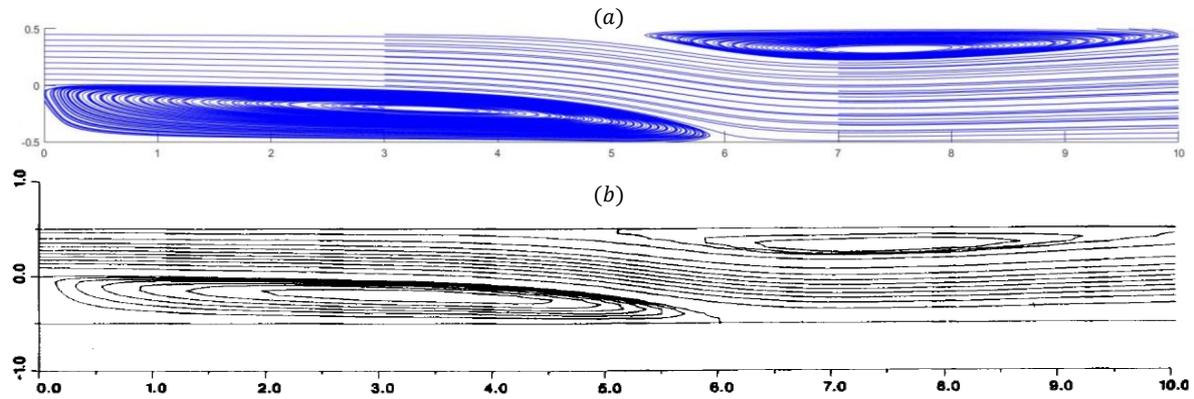

Figure 3.2: Comparison between the distribution of streamlines: (a) the current study; (b) the independent study [22]. Grid $3000 \times 100$.

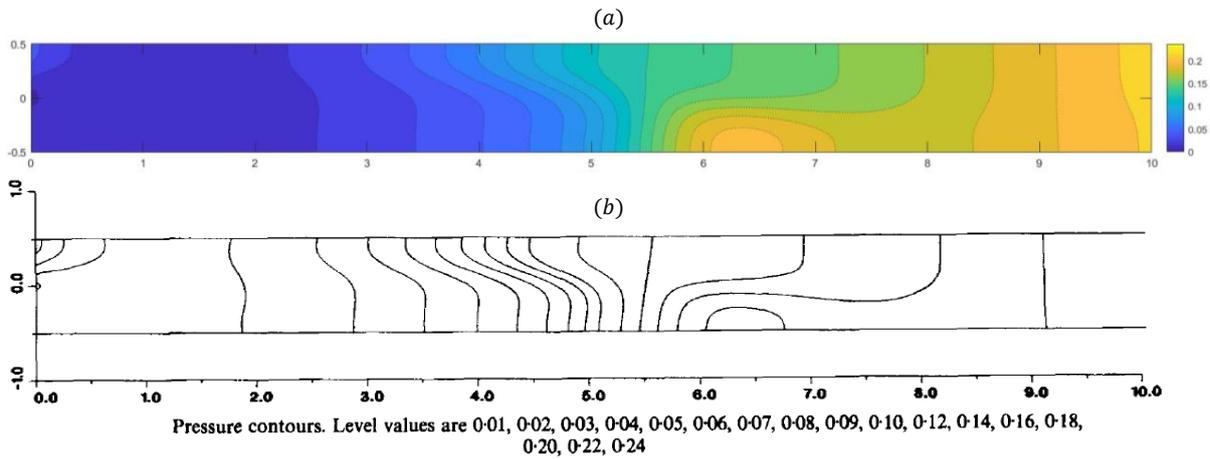

Figure 3.3: Comparison between the distribution of pressure: (a) the current study; (b) the independent study [22]. Grid $3000 \times 100$.

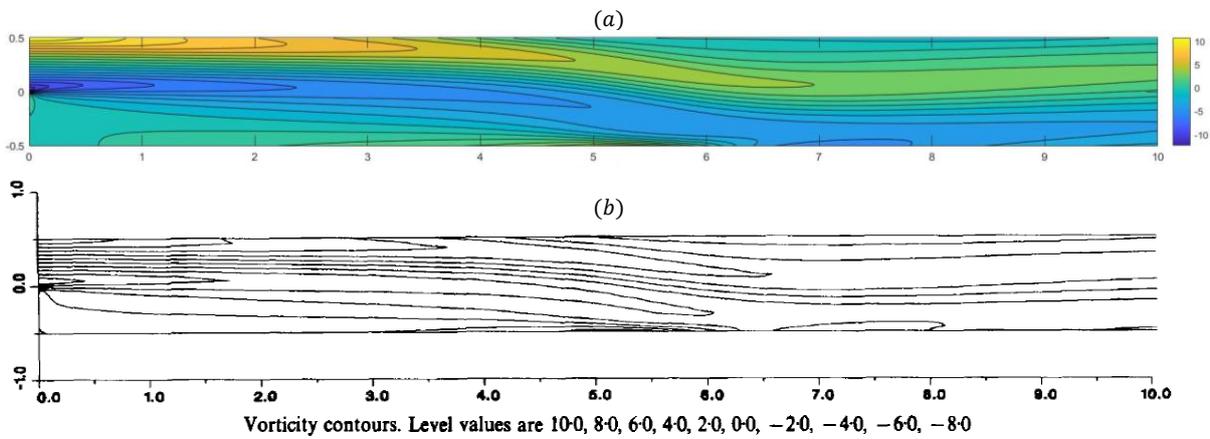

Figure 3.4: Comparison between the distribution of vorticity: (a) the current study; (b) the independent study [22]. Grid $3000 \times 100$.

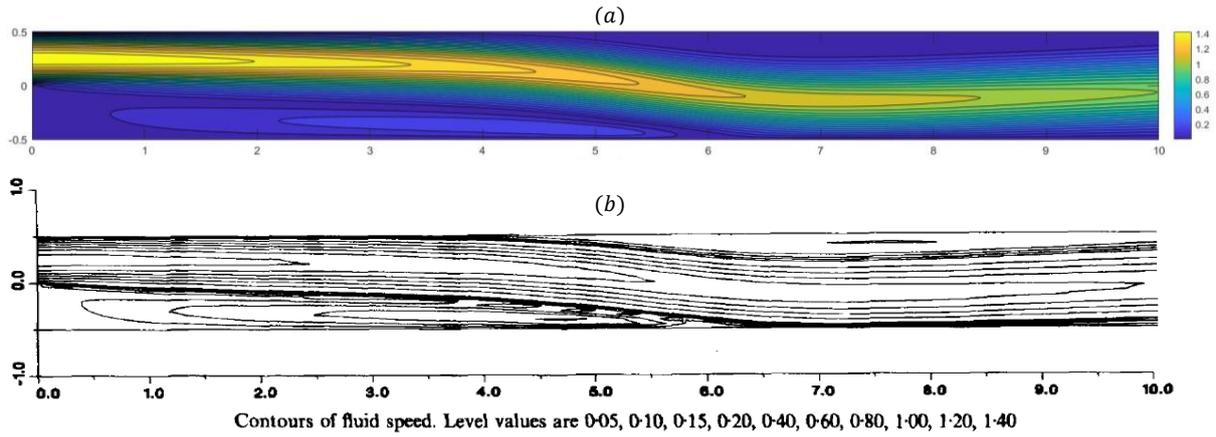

Figure 3.5: Comparison between distribution of the velocity magnitude: (a) the current study; (b) the independent study [22]. Grid $3000 \times 100$.

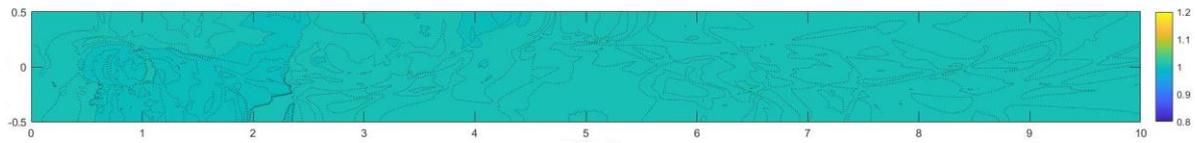

Figure 3.6: Distribution of the density field. Grid $3000 \times 100$.

Figs. 3.7-3.8 present quantitative comparisons between the currently obtained flow characteristics and the corresponding data reported in [22]. A comparison of the pressure and the shear stress distributions along the upper and lower walls of the channel is shown in Figs. 3.7a and 3.7b, respectively. A comparison between the currently obtained and the independent distributions of horizontal and vertical velocity components, pressure, vorticity, horizontal velocity gradient and normal stress along two verticals passing through the $x = 7$ and $x = 15$ locations is presented in Fig. 3.8. The figures demonstrate the same trends observed for all the flow characteristics obtained in the current and in the independent studies.

Figs. 3.7a, 3.8c and 3.8f compare between the corresponding pressure and the normal stress fields. A systematic offset between the corresponding fields observed in all the distributions can be explained by the fact that pressure is defined by up to a constant in incompressible flow. As such, the offset can be eliminated by simply adding a constant to the pressure field and does not affect the precision of the obtained results.

The maximal relative deviation between the distributions of all the flow characteristics in Figs. 3.7-3.8 is bounded by 10 percent, except for the cases where large gradients were observed or where the values of the flow characteristics are close to zero. In these cases, absolute deviation was calculated for each quantity, and its value was found to be below 8 percent of the maximum value of the corresponding quantity reported in the reference study.

In summary, acceptable qualitative and quantitative agreement exists between the currently obtained and the independent results for the entire range of flow characteristics, which successfully verifies the currently developed numerical methodology when applied for simulation of almost incompressible flows.

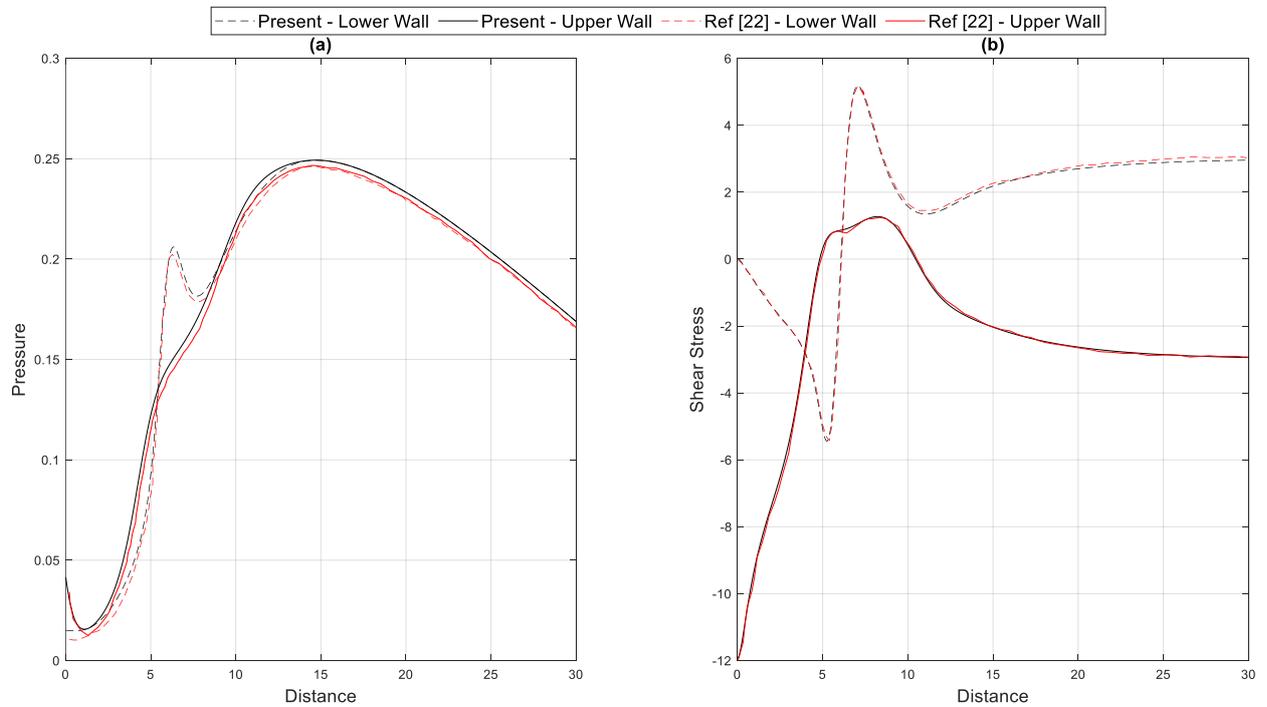

Figure 3.7: Distribution of: (a) pressure and (b) shear stress fields along upper and lower channel walls. Grid $3000 \times 100$.

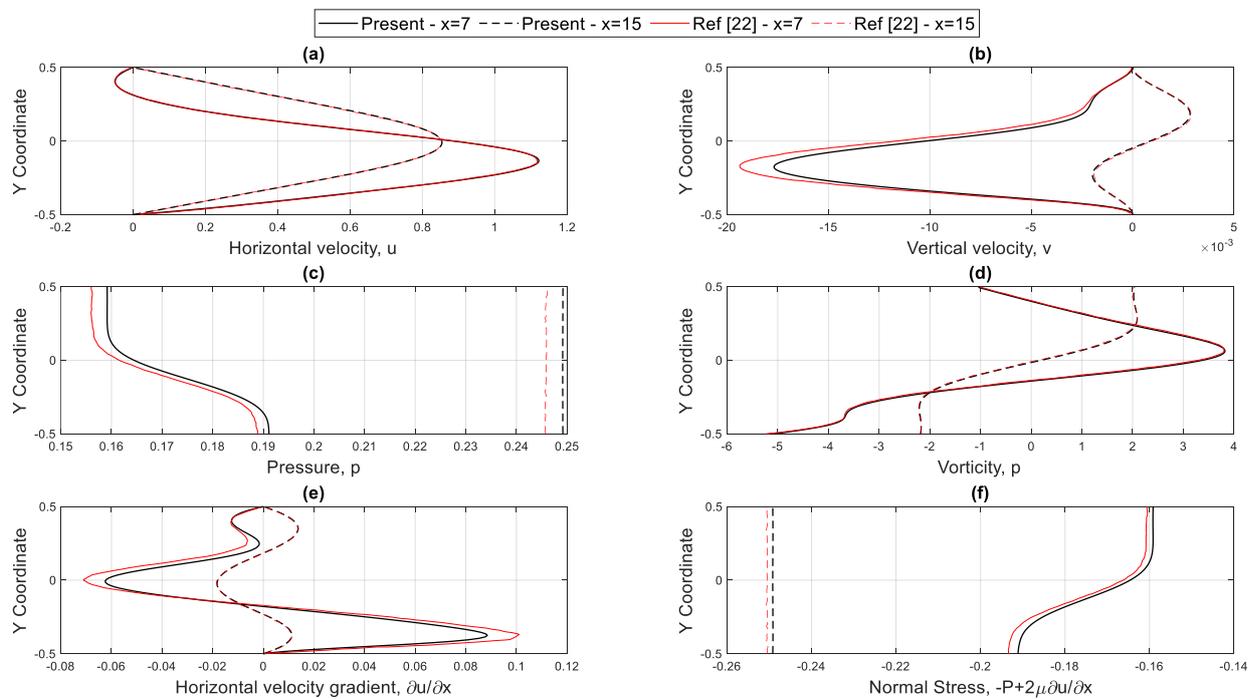

Figure 3.8: Distribution of: (a) horizontal velocity component; (b) vertical velocity component; (c) pressure; (d) vorticity; (e) horizontal velocity gradient; (f) normal stress at along vertical lines passing through $x = 7$ and $x = 15$ locations. Grid $3000 \times 100$.

## 3.2 Test case II - Natural convection flow in a differentially heated cavity

### 3.2.1 Test case overview

The results presented in this section were obtained by applying the developed methodology to the simulation of compressible natural convection flow developing within a differentially heated square cavity. The flow is driven by the temperature gradient between the two vertical walls in the presence of gravity. The obtained flow characteristics were compared with the corresponding independent data presented in [29]. The governing equations, characteristic values and non-dimensional groups of this flow configuration are presented in sections 2.1 and 2.2.

The hot and cold walls are maintained at constant temperature values equal to $T_h = 1 + \varepsilon$, and $T_c = 1 - \varepsilon$, respectively. Both horizontal walls of the cavity are thermally insulated. No-slip and zero gradient boundary conditions are applied for all the velocity components and pressure, respectively, at all the cavity walls. A schematic summarizing the geometry and boundary conditions of this flow configuration is shown in Fig. 3.9.

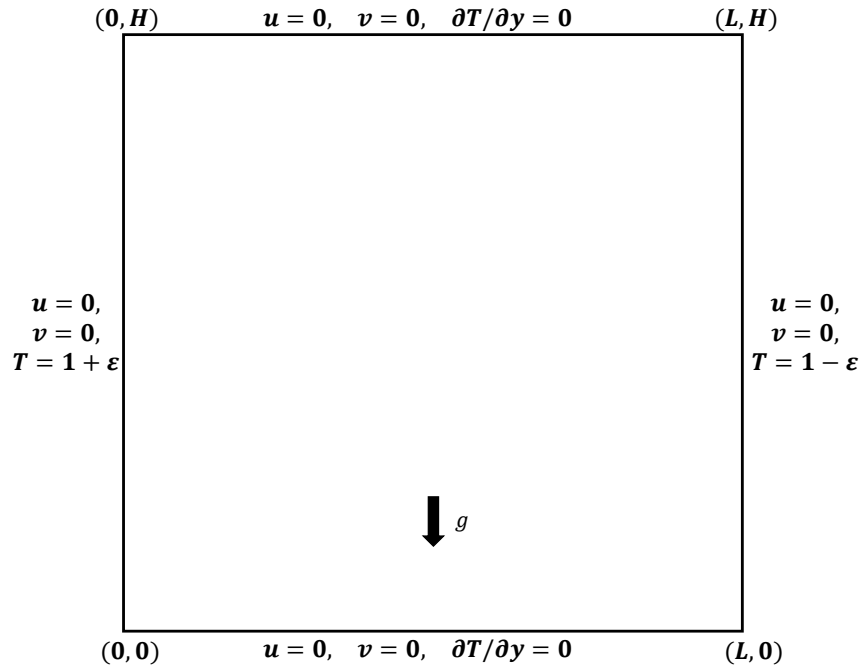

Figure 3.9: The differentially heated cavity: geometry and boundary conditions

### 3.2.2 Test results and comparison to the benchmark

The results obtained in the current study were compared qualitatively and quantitively with the corresponding independent data reported in [29]. The qualitative comparison was focused on the distributions of the velocity magnitude and temperature. The quantitative data was compared in terms of the values of the vertical temperature stratification parameter and the Nusselt number defined in the following and determined by Eqs. (3.5)-(3.9). In accordance with [29] the vertical temperature stratification parameter $\theta A$ was calculated by:

$$\theta A = \frac{AR}{2\varepsilon} \left(\frac{dT}{dy}\right)_{x=\frac{1}{2}, y=\frac{1}{2}}, \tag{3.5}$$

where $AR = H/L$ is the aspect ratio.

The Nusselt number was obtained by employing a balance between the conductive and the convective heat transfer from the surface by applying non-dimensional analysis of Fourier's law of thermal conduction and Newton's law of cooling:

$$q_s = -\tilde{k}\tilde{A}\tilde{\nabla}\tilde{T} = \tilde{h}\tilde{A}(\widetilde{T_h} - \widetilde{T_c}), \tag{3.6}$$

where $q_s$ is the amount of heat transferred from the surface and $h$ is the convection heat transfer coefficient.

By utilizing the characteristic scales determined in subsection 2.1.1, Eq. (3.6) can be written as:

$$-\frac{k_0 T_0}{L_0} k \nabla T = \tilde{h} T_0 [(1+\varepsilon) - (1-\varepsilon)] \tag{3.7}$$

where $\nabla T$ is the temperature gradient at the cavity wall. We next determine the local Nusselt number, $Nu_{L_0}$ as:

$$Nu_{L_0} = \frac{\tilde{h} L_0}{k_0} = -\frac{k \nabla T}{2\varepsilon}, \tag{3.8}$$

and the average Nusselt number as:

$$\overline{Nu} = -\frac{1}{S} \int_S \frac{k \nabla T}{2\varepsilon} dS, \tag{3.9}$$

where $S$ is the non-dimensional surface area.

In the current study, the Nusselt number was calculated on the hot wall of the cavity. The calculations were performed for the range of $Ra \in [10^3, 10^7]$ and $\varepsilon = 0.005, 0.2, 0.4$ and $0.6$, In all the simulations the value of the aspect ratio $AR = H/L$ was equal to unity. Figs. 3.10-3.13 summarize the qualitative comparison between the velocity and the temperature fields obtained for the value of $Ra = 10^5$ and the entire range of $\varepsilon$ values. Good agreement between the currently obtained and the independent results can clearly be recognized for all the performed simulations. Note that the temperature and the velocity distributions vary with $\varepsilon$. In fact, for the lowest value of $\varepsilon = 0.005$ both velocity and temperature distributions are almost skew-symmetric relatively to the cavity center, and resemble those typical of incompressible flows obtained by applying the Boussinesq approximation. With an increase in the $\varepsilon$ value the flow loses its skew-symmetry as a result of the dependence of its conductivity, $k$, and viscosity, $\mu$, values on temperature.

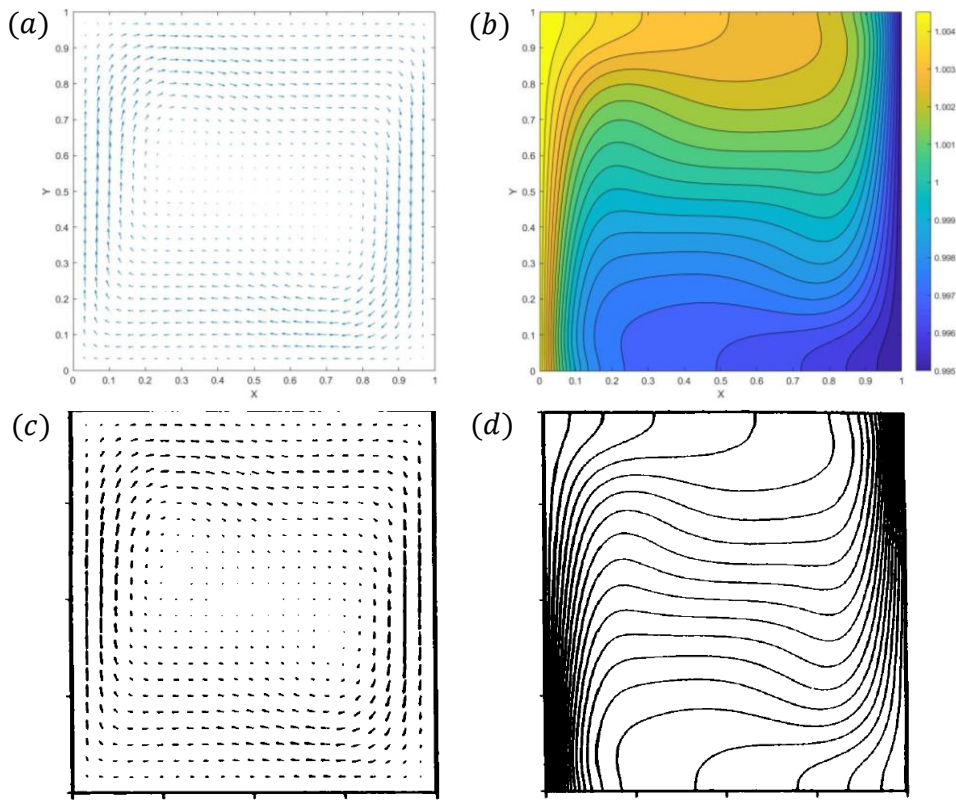

Figure 3.10: Comparison of contours for $\varepsilon = 0.005, Ra = 10^5$: (a) velocity, current study; (b) temperature, current study; (c) velocity, independent study [29]; (d) temperature, independent study [29]

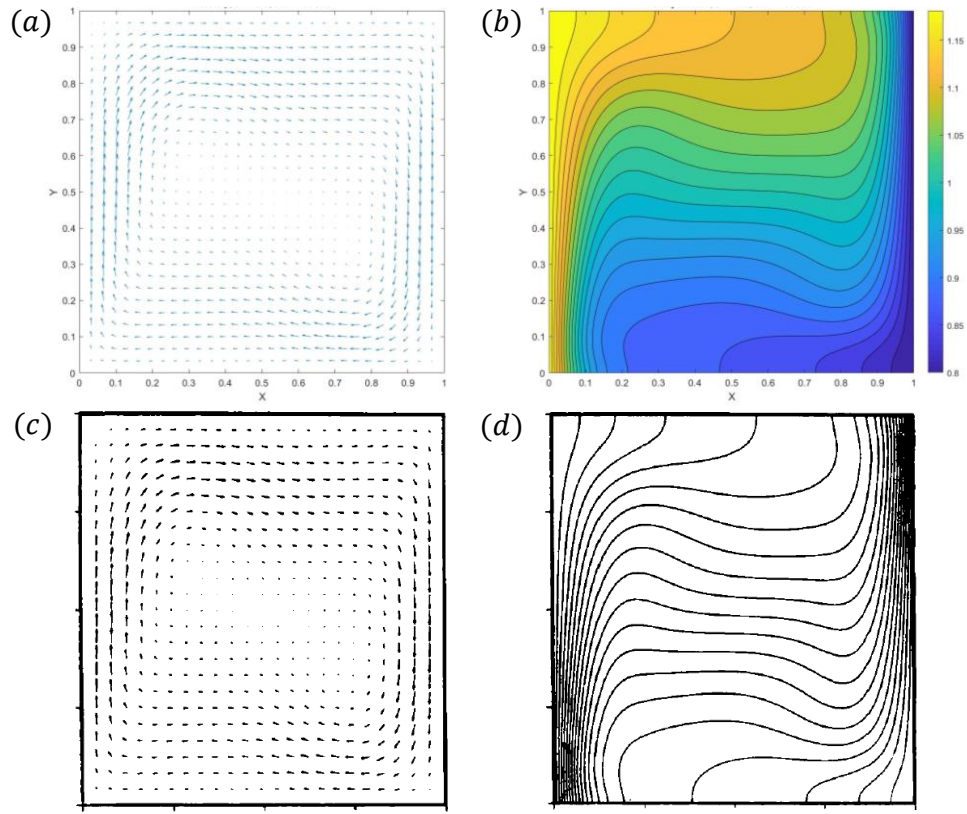

Figure 3.11: Comparison of contours for $\varepsilon = 0.2, Ra = 10^5$: (a) velocity, current study; (b) temperature, current study; (c) velocity, independent study [29]; (d) temperature, independent study [29]

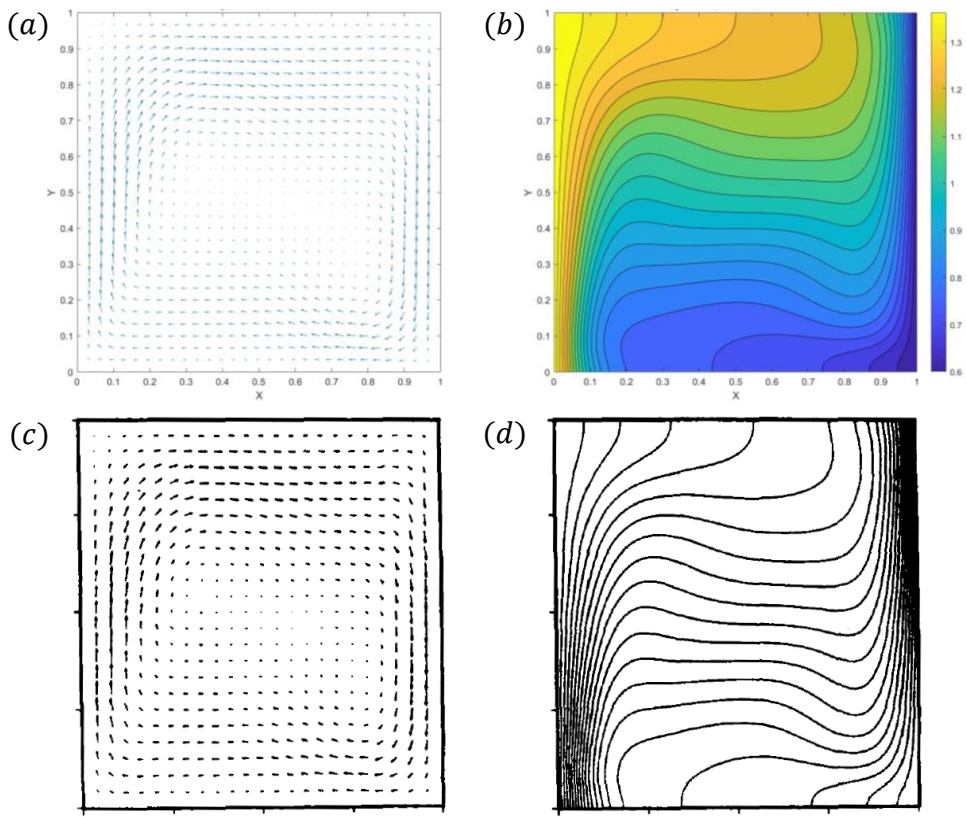

Figure 3.12: Comparison of contours for $\varepsilon = 0.4, Ra = 10^5$: (a) velocity, current study; (b) temperature, current study; (c) velocity, independent study [29]; (d) temperature, independent study [29]

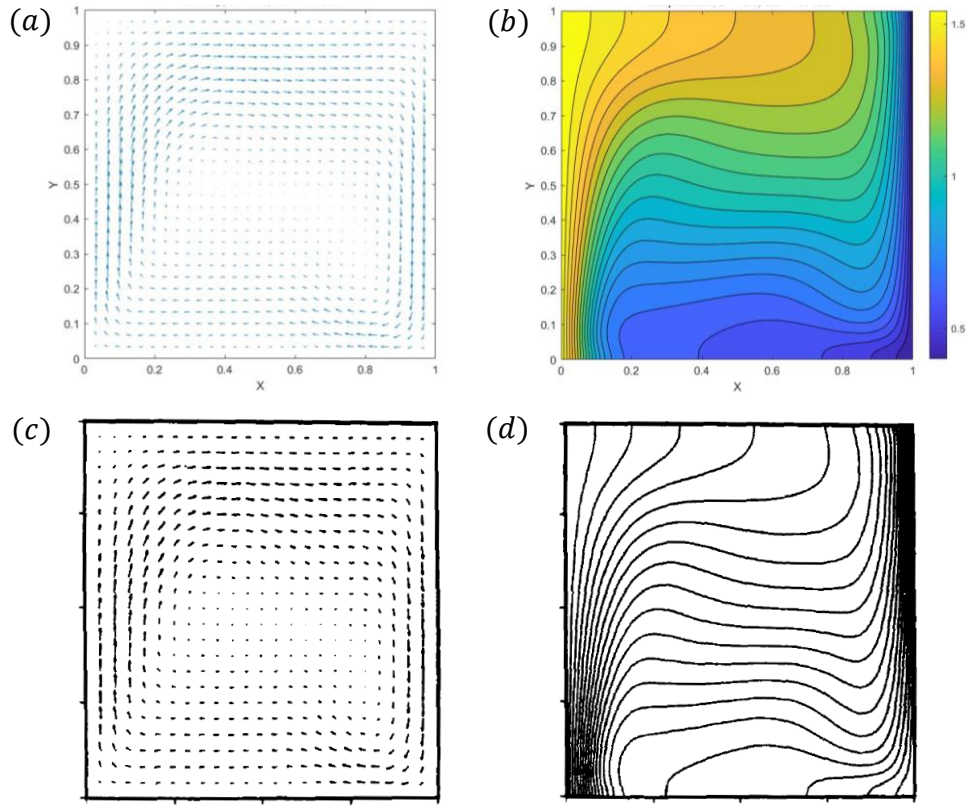

Figure 3.13: Comparison of contours for $\varepsilon = 0.6, Ra = 10^5$: (a) velocity, current study; (b) temperature, current study; (c) velocity, independent study [29]; (d) temperature, independent study [29]

Figs. 3.14-3.17 summarize the qualitative comparison of velocity and temperature fields for $\varepsilon = 0.6$ and the entire range of $Ra$ values. Good agreement is observed between the whole set of currently obtained and the independent results [29].

It can be clearly seen that for the highest value of $\varepsilon = 0.6$ the flow is characterized by broken skew-symmetry relative to the cavity center that is clearly recognized even at the lowest value of $Ra = 10^3$. For the higher $Ra$ flow regimes dominated by convective heat transfer, the skew-symmetry breaking becomes much more pronounced. This is confirmed when looking at the differences in the thicknesses of the boundary layers developing near the hot and the cold walls. In fact, the dynamic viscosity of the ideal gas increases with temperature, leading to the boundary layer thickening in the vicinity of the hot wall. At the same time, the boundary layer thickness decreases close to the cold vertical wall, which again is a consequence of a local decrease in the dynamic viscosity values as a result of lower temperatures prevailing in this region.

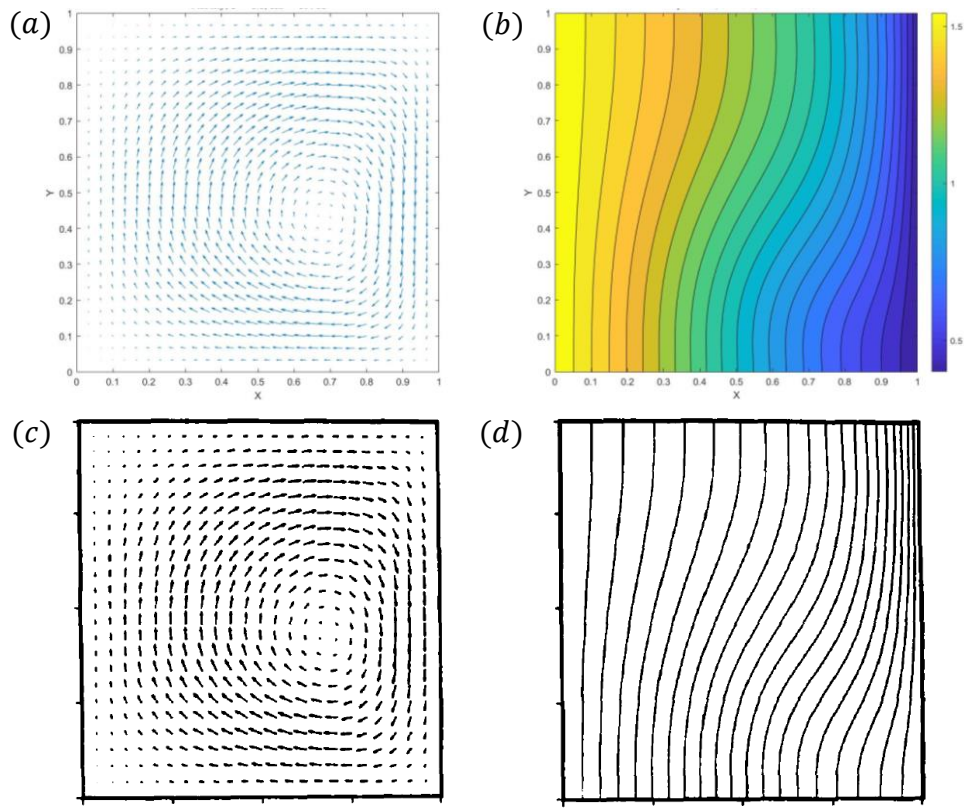

Figure 3.14: Comparison of contours for $\varepsilon = 0.6, Ra = 10^3$: (a) velocity, current study; (b) temperature, current study; (c) velocity, independent study [29]; (d) temperature, independent study [29]

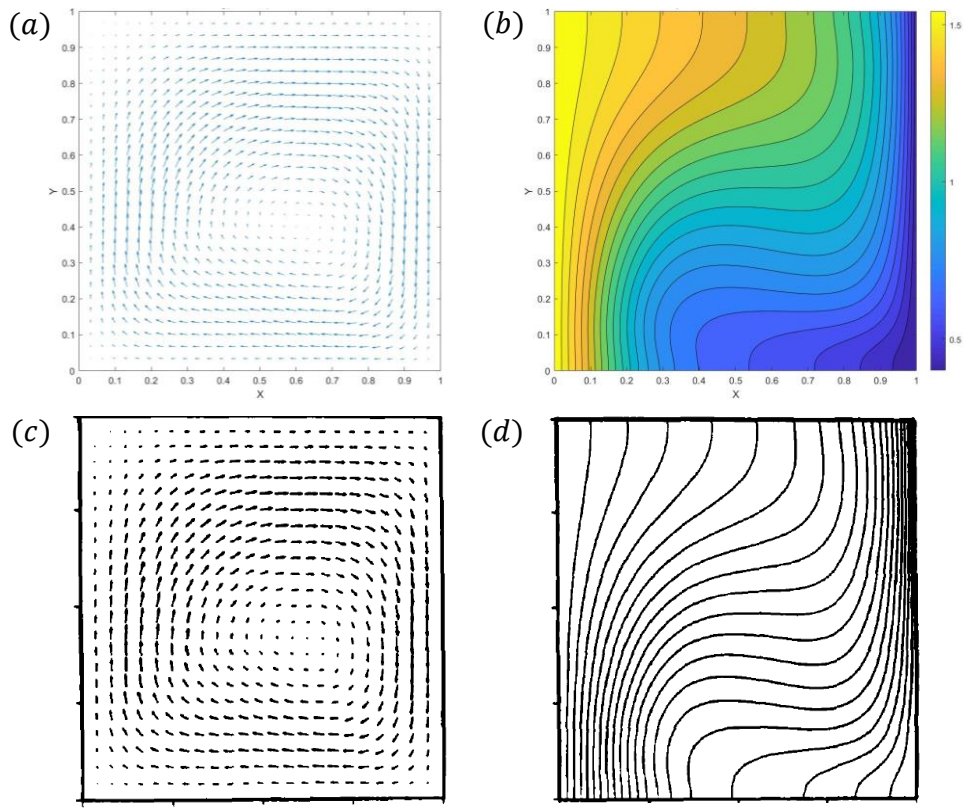

Figure 3.15: Comparison of contours for $\varepsilon = 0.6, Ra = 10^4$: (a) velocity, current study; (b) temperature, current study; (c) velocity, independent study [29]; (d) temperature, independent study [29]

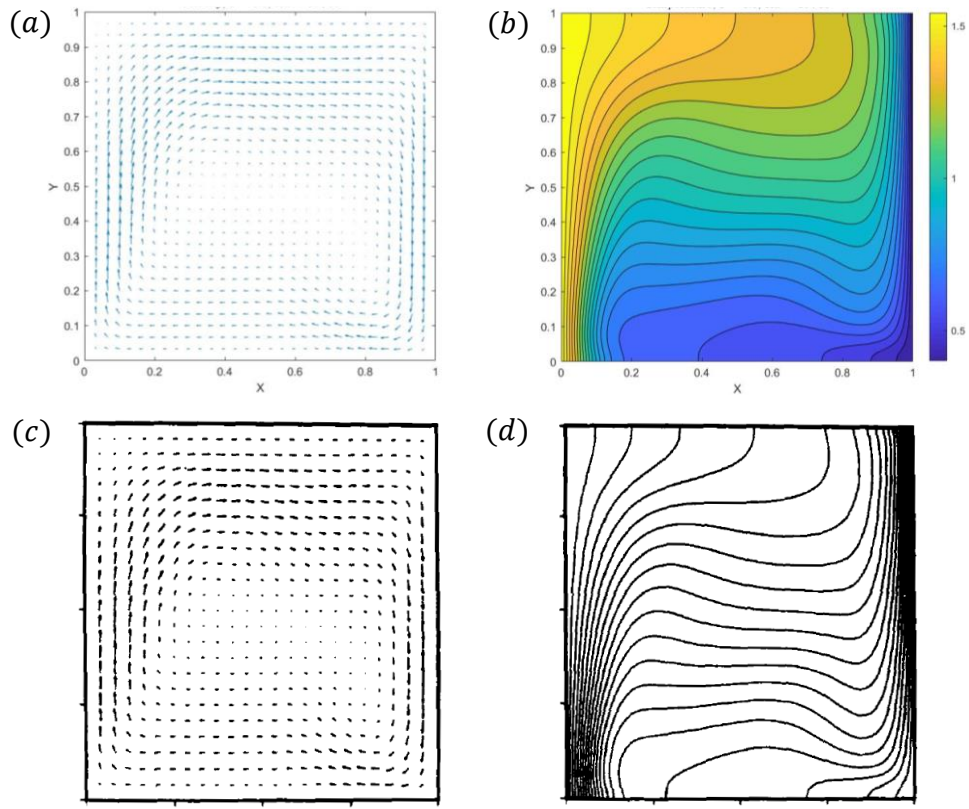

Figure 3.16: Comparison of contours for $\varepsilon = 0.6, Ra = 10^5$: (a) velocity, current study; (b) temperature, current study; (c) velocity, independent study [29]; (d) temperature, independent study [29]

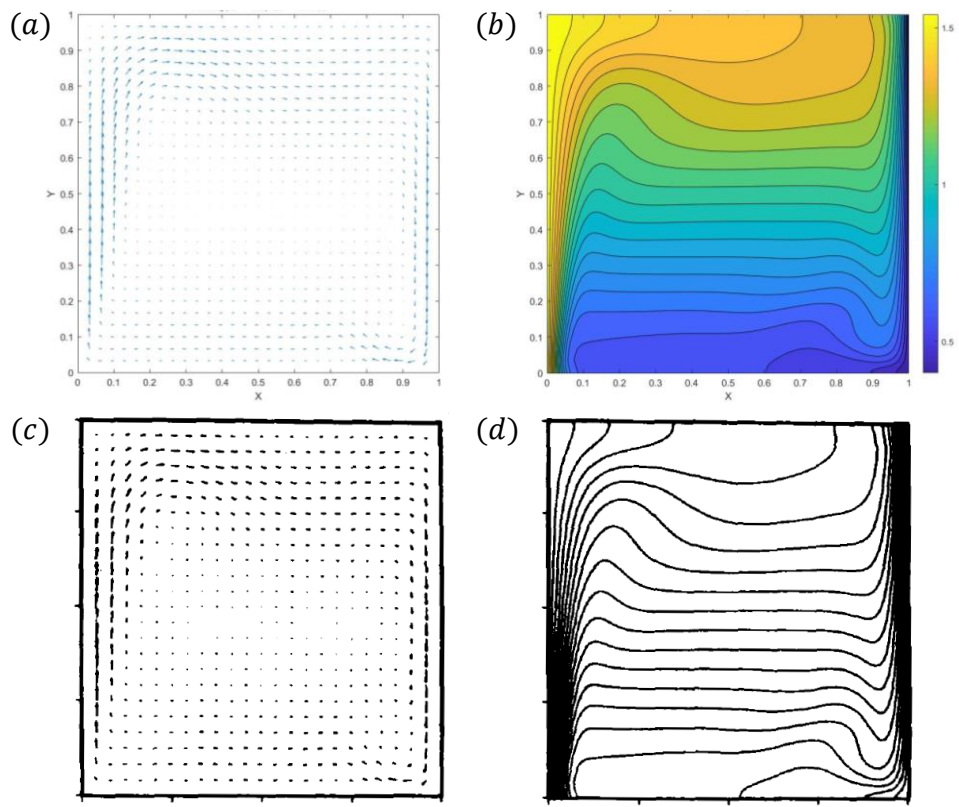

Figure 3.17: Comparison of contours for $\varepsilon = 0.6, Ra = 10^6$: (a) velocity, current study; (b) temperature, current study; (c) velocity, independent study [29]; (d) temperature, independent study [29]

Figs. 3.18-3.19 present quantitative comparisons of the vertical temperature stratification parameter and the averaged Nusselt number versus the $Ra$ values. The simulations were performed on three different grids: a uniform grid of $100 \times 100$ cells, a uniform grid of $200 \times 200$ cells and a non-uniform grid of $100 \times 100$ cells stretched towards the cavity walls to accurately resolve the thinnest boundary layers.

As can be seen from Fig. 3.18, there is good agreement between the currently obtained and independent results of the vertical temperature stratification parameter, except for two cases: (i) stretched grid, $\varepsilon = 0.005$, $Ra = 10^6$ and (ii) stretched grid, $\varepsilon = 0.2$, $Ra = 10^3$. This could indicate that stretching the grid towards the cavity boundaries is not an ultimate way to increase the results' accuracy, especially when the control parameters are acquired close to the cavity center. Fig. 3.19 demonstrates that the $Nu$ values obtained on the uniform and stretched grids built of $200 \times 200$ and $100 \times 100$ cells, respectively, agree with the independent results better than the corresponding $Nu$ values obtained on the uniform grid built of $100 \times 100$ cells, especially for high Rayleigh numbers. Thus, stretching the grid towards the cavity boundaries is preferable when analyzing characteristics based on the temperature gradients.

To sum up, an acceptable qualitative and quantitative agreement exists between the currently obtained and the independent results for the entire range of operating conditions and flow characteristics, which successfully verifies the currently developed numerical methodology when applied for the simulation of compressible natural convection confined flows.

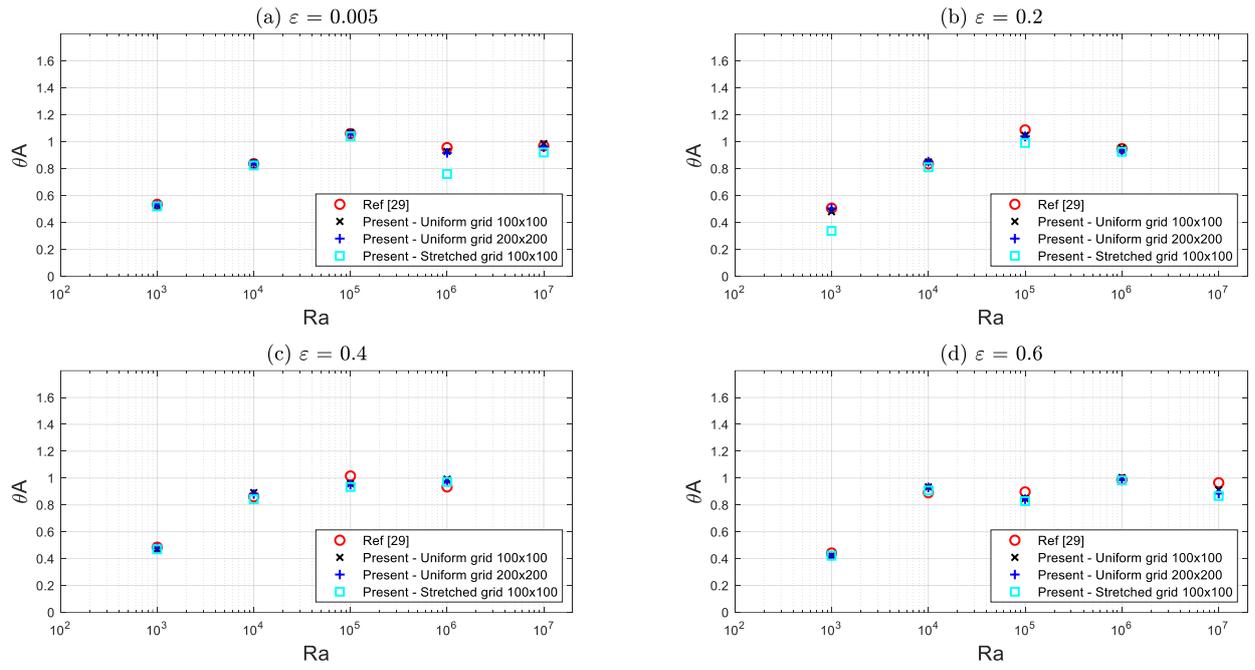

Figure 3.18: Vertical temperature stratification parameter vs Rayleigh number

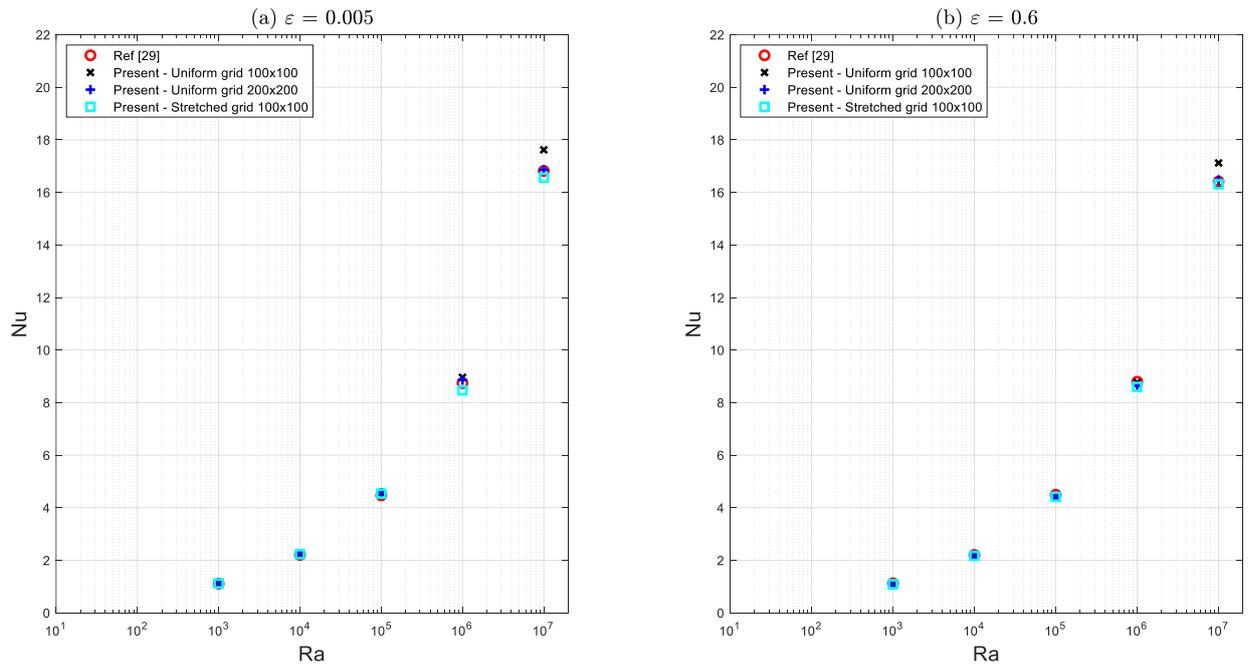

Figure 3.19: Nusselt number vs Rayleigh number

# 4. Results and discussion

## 4.1 System's overview

The results presented in this chapter were obtained by applying the developed methodology to the simulation of a two-dimensional compressible natural convection flow developing around a hot cylinder placed within a square cavity all of whose walls are held at a constant cold temperature. The flow is driven by the temperature difference between the cylinder and the cavity boundaries in the presence of gravity. The surface of the hot cylinder and the walls of the cold cavity are maintained at constant values of temperature equal to $T_h = 1 + \varepsilon$, and $T_c = 1 - \varepsilon$, respectively. No-slip and zero gradient boundary conditions were applied for all the velocity components and pressure, respectively, at all the cavity walls and the surface of the cylinder. A schematic description of the geometry and boundary conditions of the above flow configuration is shown in Fig. 4.1. The governing equations, characteristic values, and non-dimensional groups governing the flow within the considered configuration were defined in chapter 2.

The qualitative and quantitative results obtained for the wide range of parameters are next analyzed to attain deeper insight into the physics governing the considered non-Boussinesq natural convection confined flow.

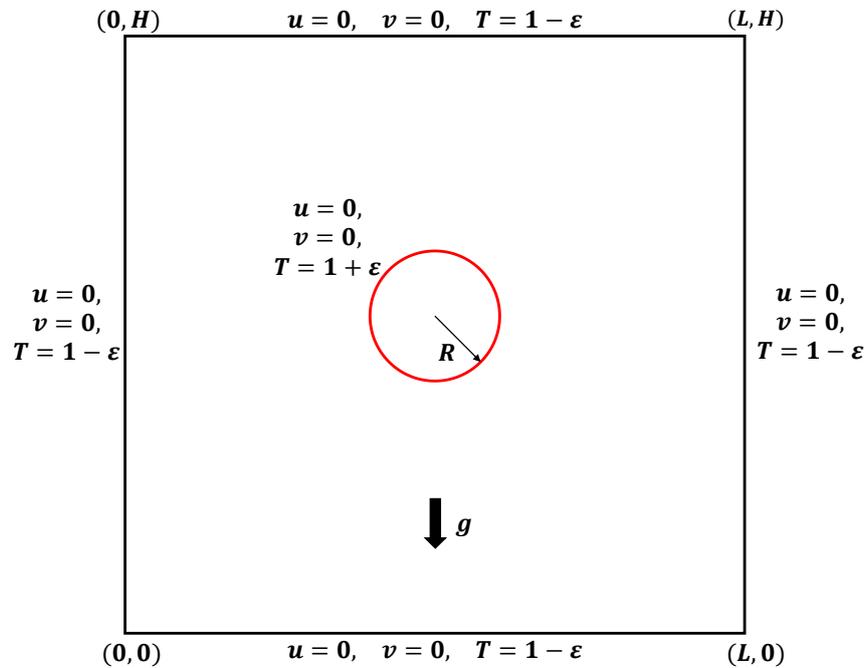

Figure 4.1: The cold cavity with a hot cylinder at the center: geometry and boundary conditions

## 4.2 Results and discussion

The results obtained in the current study were performed for a wide range of governing parameters: $Ra \in \{10^3, 10^4, 10^5, 10^6\}$, $\varepsilon \in \{0.005, 0.2, 0.4, 0.6\}$ and $R/L \in \{0.1, 0.2, 0.3, 0.4\}$. The simulations were performed on two uniform grids having 100 and 200 divisions in each direction, by utilizing time steps $\Delta t = 10^{-7}$ and $\Delta t = 10^{-8}$, respectively. The criterion for achieving the steady state was a value of $10^{-6}$ of the $L_2$ norm calculated for the relative difference between the two consecutive time steps for all the flow fields.

The obtained results emphasize the qualitative differences between the corresponding streamlines and temperature distributions typical of the lowest ($\varepsilon = 0.005$), and the highest ($\varepsilon = 0.6$) temperature-difference parameters obtained for the same $Ra$ and $R/L$ values. The quantitative results are presented in terms of the values of the $Nu - Ra$ functional relationship approximated to the power law, by employing the least-squares technique, where the Nusselt number $Nu$ was calculated as detailed further in subsection 4.2.2. Additionally, $Nu$ values obtained for the whole range of $Ra$ and for the lowest value temperature-difference parameter ($\varepsilon = 0.005$) were compared with the corresponding results available in the literature [72], [58] obtained by employing the Boussinesq approximation.

### 4.2.1 Qualitative observations

Figs. 4.2-4.17 summarize the qualitative results in terms of the spatial distribution of the streamline and the temperature fields obtained for the values of $\varepsilon = 0.005$ and $\varepsilon = 0.6$ for the entire range of $Ra$ and $R/L$ values on the uniform grid of 200 cells in each direction. The main purpose of this part of the study is to investigate qualitative similarities and differences between the flow characteristics typical of the lowest ($\varepsilon = 0.005$) and the highest ($\varepsilon = 0.6$) values of temperature-difference parameters.

As can be seen in the figures, every configuration necessarily contains two large (primary) convective cells, whereas several configurations also contain small (secondary) convective cells. The temperature gradients and shapes of the isotherms differ as a function of $Ra$, $\varepsilon$ and $R/L$ values.

As follows from Figs. 4.2-4.3 and 4.10-4.11, for the lowest Rayleigh number, $Ra = 10^3$, there are no significant differences between the spatial distributions of the streamlines and the temperature fields obtained for the lowest ($\varepsilon = 0.005$) and the highest ($\varepsilon = 0.6$) values of the temperature-difference parameter, regardless of the cylinder's diameter. The configurations with the lowest $Ra$ number do not contain secondary convective cells, and the temperature distribution is close to linear along the radial direction from the hot cylinder to the cold cavity walls, as expected from systems for which conduction constitutes the major heat transfer mechanism.

As the $Ra$ value increases, the differences become more visible. In Figs. 4.4-4.5 and 4.12-4.13 the streamline and temperature distributions are shown for $Ra = 10^4$. The differences in streamline distribution remain insignificant without any presence of secondary convective cells, while the temperature distribution for the highest value of $\varepsilon$ is slightly shifted up compared to that observed for the lowest $\varepsilon$ value.

For the two highest values of $Ra = 10^5$ and $Ra = 10^6$, the differences between the streamline and temperature distributions corresponding to the configurations characterized by the two edge values of $\varepsilon$ are quite significant, as expected for systems for which convection constitutes the major heat transfer mechanism (see Figs. 4.6-4.9 and 4.14-4.17). In this range of $Ra$ secondary convective cells may be generated, while the temperature distribution along the radial direction is non-linear with clearly recognizable single or multiple thermal plumes rising up from the top of the cylinder.

In summary, secondary convective cells never appear for $Ra \leq 10^4$ and/or $R/L \leq 0.1$. For $R/L = 0.2$ secondary convective cells appear only for the value of $\varepsilon = 0.6$ and the two values of $Ra = 10^5$ and $Ra = 10^6$. For the value of $R/L = 0.3$ secondary convective cells appear for the two values of $\varepsilon = 0.005$ and $\varepsilon = 0.6$ and the two values of $Ra = 10^5$ and $Ra = 10^6$. For $R/L = 0.4$ secondary convective cells appear for the two values of $\varepsilon$ and only for $Ra = 10^6$. Additionally, it was observed that the flows characterized by $\varepsilon = 0.6$ may contain more secondary convective cells compared to their counterparts characterized by $\varepsilon = 0.005$; this trend never takes place the other way around. As such, sufficient conditions for the flow separation to occur should apparently include simultaneously high $Ra$, $R/L$ and $\varepsilon$ values. It is also noteworthy that an increasing number of secondary convective cells with increasing values of $Ra$ follows the same trend observed in the Rayleigh-Benard convection with an increasing aspect ratio. That is, for high values of $R/L$ the cylinder curvature can be locally neglected and the flow resembles the Rayleigh-Benard configuration at the top region of the cavity.

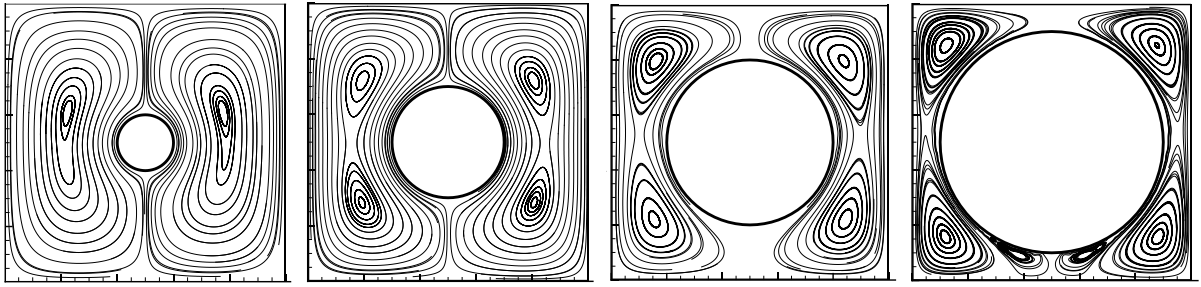
Figure 4.2: Streamlines distribution for $Ra = 10^3$, $\varepsilon = 0.005$ and $0.1 \leq R/L \leq 0.4$

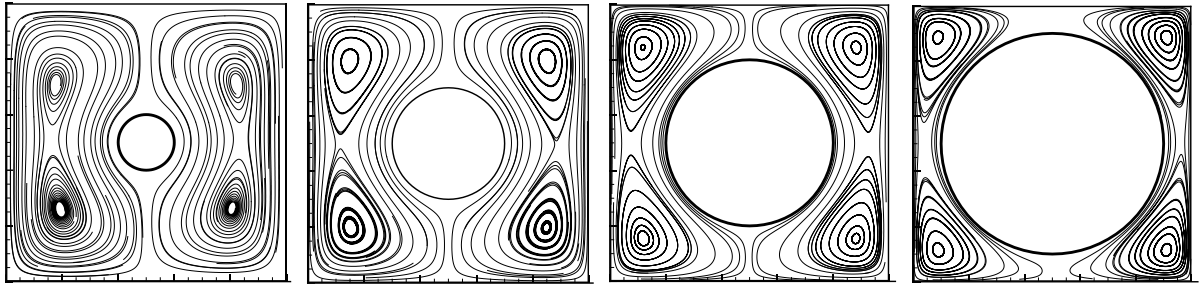
Figure 4.3: Streamlines distribution for $Ra = 10^3$, $\varepsilon = 0.6$ and $0.1 \leq R/L \leq 0.4$

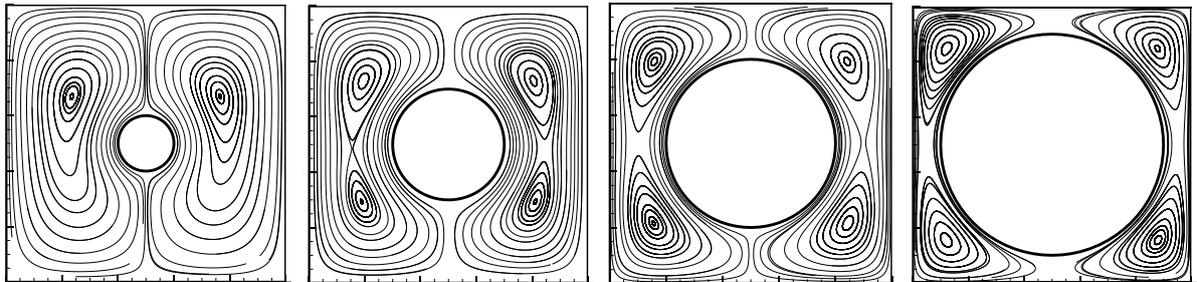
Figure 4.4: Streamlines distribution for $Ra = 10^4$, $\varepsilon = 0.005$ and $0.1 \leq R/L \leq 0.4$

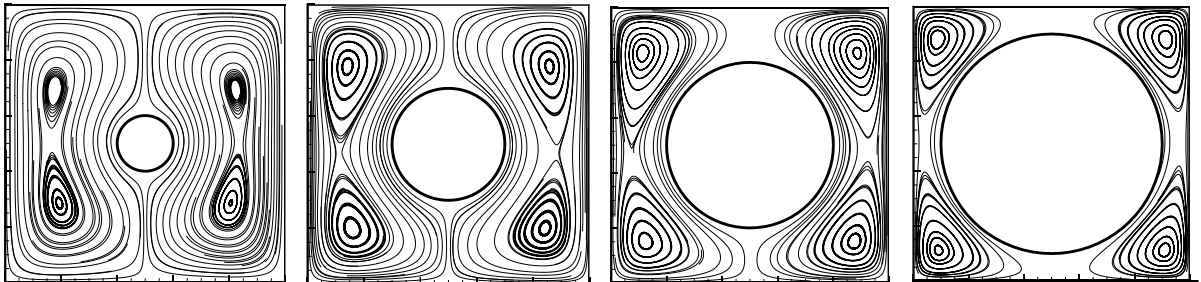
Figure 4.5: Streamlines distribution for $Ra = 10^4$, $\varepsilon = 0.6$ and $0.1 \leq R/L \leq 0.4$

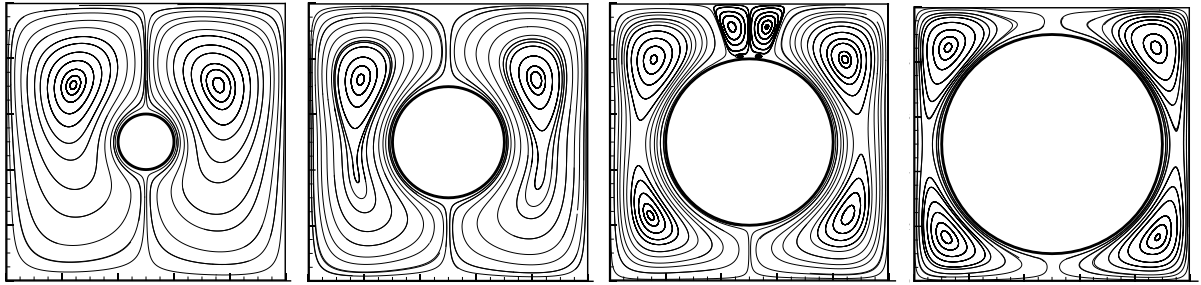

Figure 4.6: Streamlines distribution for $Ra = 10^5, \varepsilon = 0.005$ and $0.1 \leq R/L \leq 0.4$

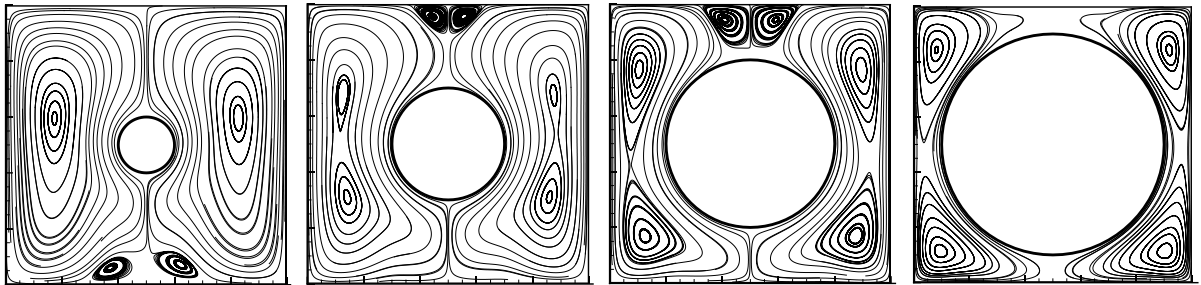

Figure 4.7: Streamlines distribution for $Ra = 10^5, \varepsilon = 0.6$ and $0.1 \leq R/L \leq 0.4$

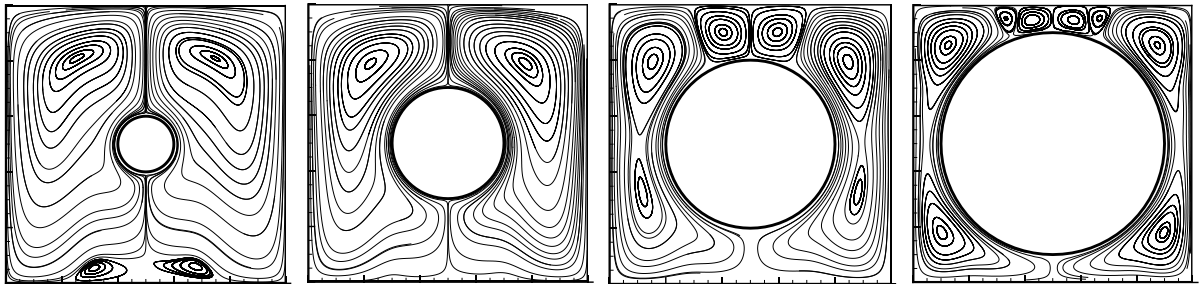

Figure 4.8: Streamlines distribution for $Ra = 10^6, \varepsilon = 0.005$ and $0.1 \leq R/L \leq 0.4$

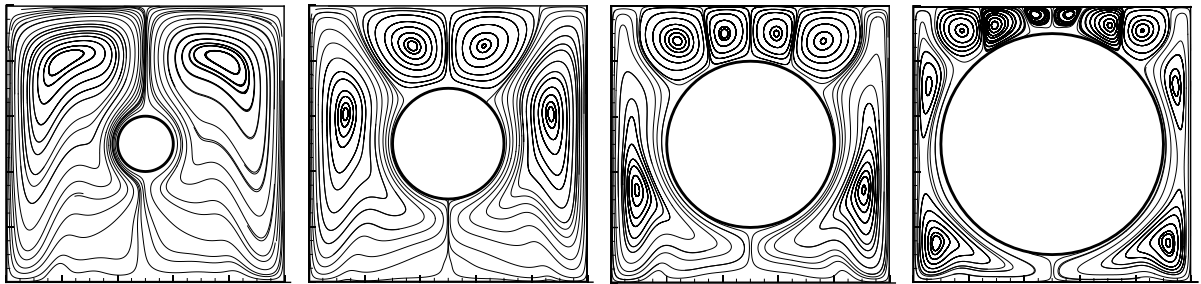

Figure 4.9: Streamlines distribution for $Ra = 10^6, \varepsilon = 0.6$ and $0.1 \leq R/L \leq 0.4$

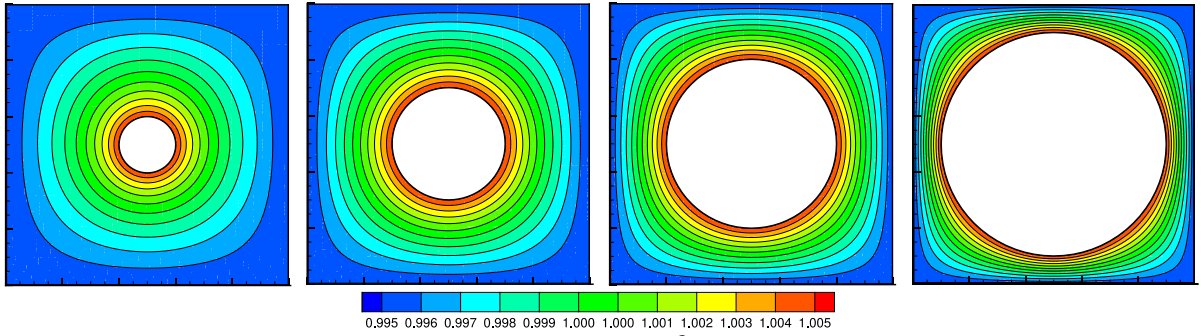

Figure 4.10: Temperature distribution for $Ra = 10^3, \varepsilon = 0.005$ and $0.1 \leq R/L \leq 0.4$

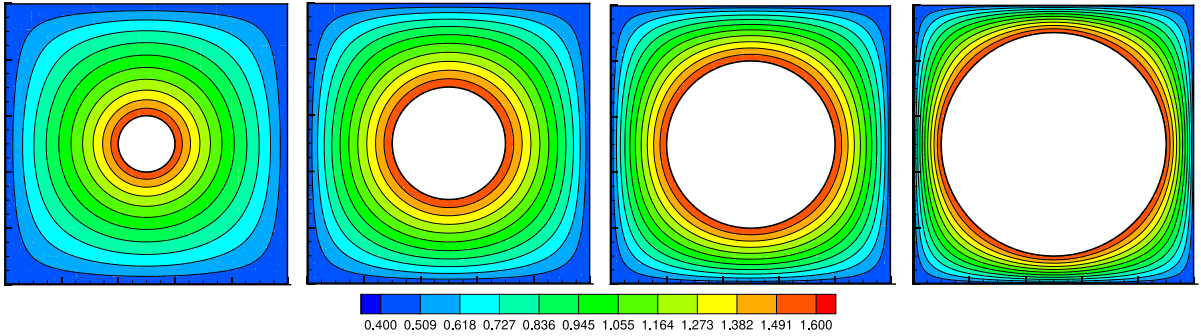

Figure 4.11: Temperature distribution for $Ra = 10^3, \varepsilon = 0.6$ and $0.1 \leq R/L \leq 0.4$

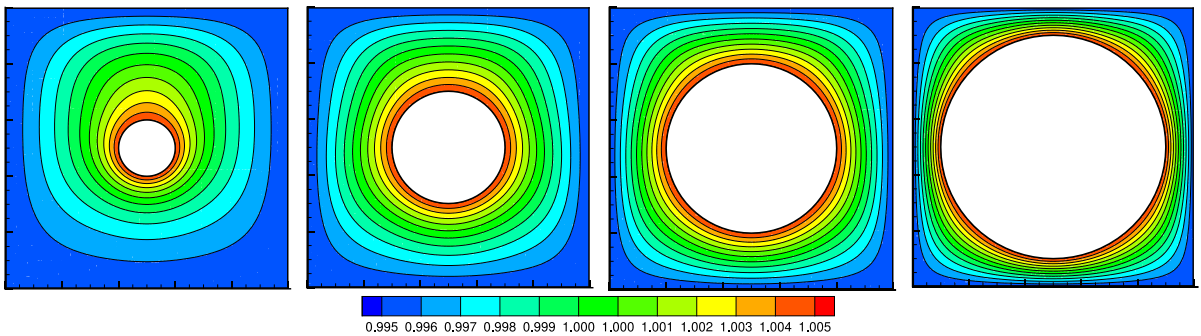

Figure 4.12: Temperature distribution for $Ra = 10^4, \varepsilon = 0.005$ and $0.1 \leq R/L \leq 0.4$

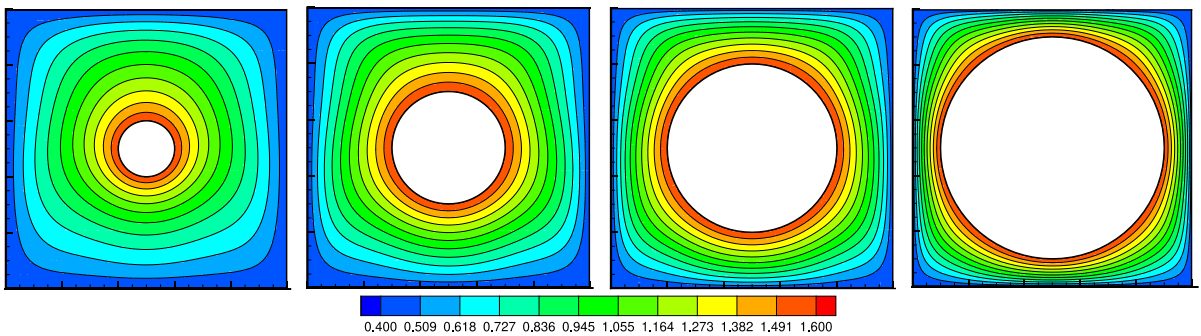

Figure 4.13: Temperature distribution for $Ra = 10^4, \varepsilon = 0.6$ and $0.1 \leq R/L \leq 0.4$

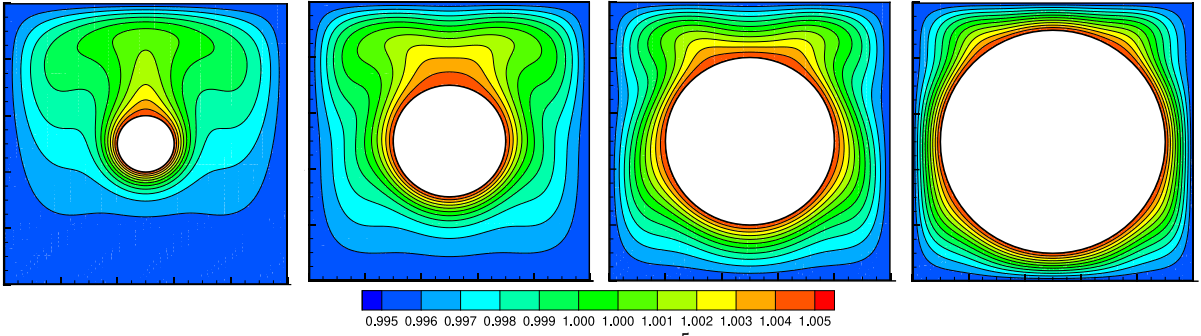

Figure 4.14: Temperature distribution for $Ra = 10^5, \varepsilon = 0.005$ and $0.1 \leq R/L \leq 0.4$

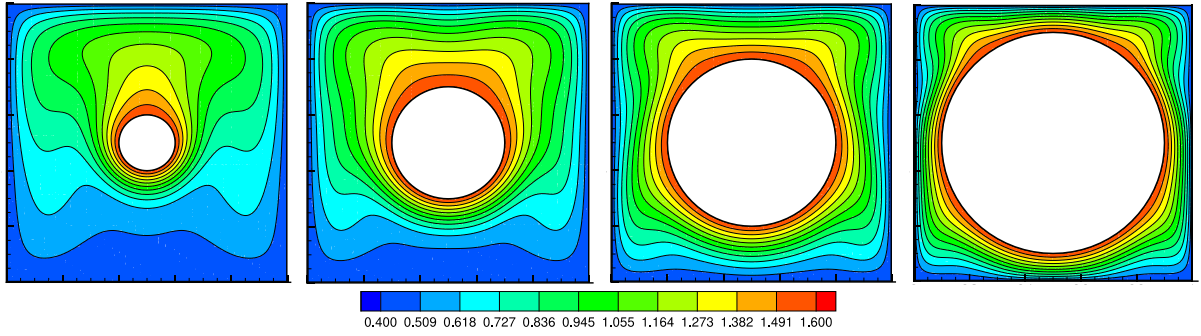

Figure 4.15: Temperature distribution for $Ra = 10^5, \varepsilon = 0.6$ and $0.1 \leq R/L \leq 0.4$

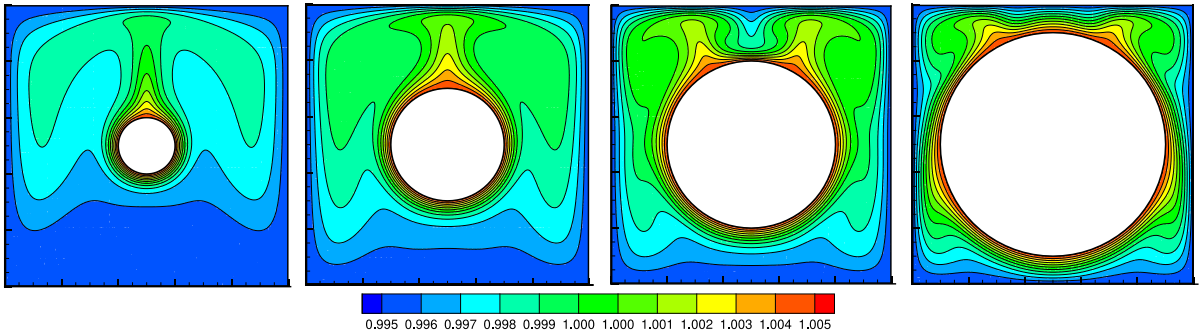

Figure 4.16: Temperature distribution for $Ra = 10^6, \varepsilon = 0.005$ and $0.1 \leq R/L \leq 0.4$

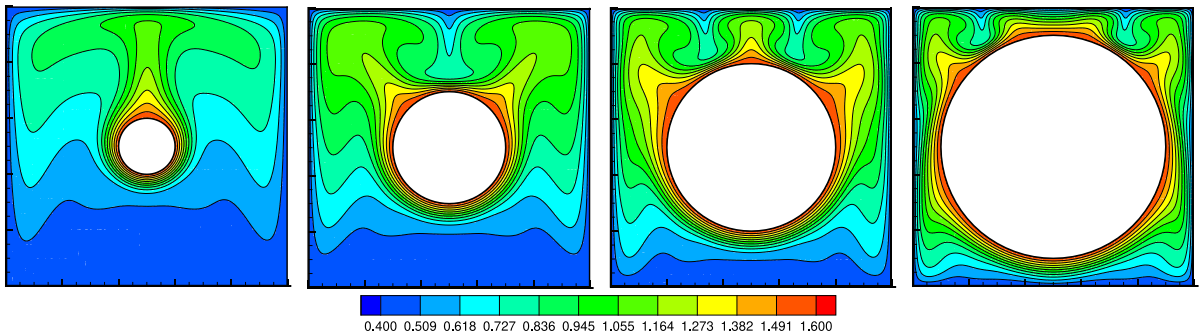

Figure 4.17: Temperature distribution for $Ra = 10^6, \varepsilon = 0.6$ and $0.1 \leq R/L \leq 0.4$

### 4.2.2 Calculation of the Nusselt number

In the present study calculation of the Nusselt number included calculation of the cold Nusselt number, $\overline{Nu}_c$, related to the surfaces of the cavity, and of the hot Nusselt number, $\overline{Nu}_h$, related the surface of the cylinder. The calculation of $\overline{Nu}_c$ is based on the arithmetic average of the Nusselt numbers at every wall of the cavity, each calculated by the method detailed in subsection 3.2.2. Calculation of $\overline{Nu}_h$ is obtained by accounting for the heat flux from the cylinder surface:

$$q_s = \tilde{\rho}\widetilde{C_p}\tilde{A}\widetilde{\Delta x}\frac{\partial \tilde{T}}{\partial \tilde{t}} = \tilde{h}\tilde{A}(\tilde{T}_h - \tilde{T}_c). \tag{4.1}$$

By utilizing the scaling determined in subsection 2.1.1, Eq. (4.1) can be written as:

$$\frac{\rho_0 C_{p_0} T_0 \alpha_0}{L_0^2} L_0 \Delta x \rho C_p \frac{\partial T}{\partial t} = \tilde{h} T_0 [(1+\varepsilon) - (1-\varepsilon)]. \tag{4.2}$$

Recalling that $\alpha_0 = k_0/\rho_0 C_{p_0}$, we next determine the local Nusselt number, $Nu_{L_0}$ as:

$$Nu_{h_{L_0}} = \frac{\tilde{h}L_0}{k_0} = \frac{1}{2\varepsilon}\rho C_p \frac{\partial T}{\partial t}, \tag{4.3}$$

and the average Nusselt number, $\overline{Nu}$ as:

$$\overline{Nu}_h = \frac{1}{2\varepsilon S}\int_S \rho C_p \frac{T_K^d - T_K}{\Delta t} dS. \tag{4.4}$$

In the current study the calculation of the Nusselt number was an integral part of the IBM applied for imposing the temperature constraint on the cylinder surface.

### 4.2.3 Comparison between the results of the lowest-temperature-difference cases and previous studies

The results of the current study obtained for the value of $\varepsilon = 0.005$ were compared with the corresponding independently reported data obtained by employing the Boussinesq approximation. In particular, we focus on the studies in [72] and [58] that simulate the natural convection flow from a hot cylinder placed within a three-dimensional cavity. The geometry and the boundary conditions of this specific configuration are shown in Fig. 4.18. Interestingly, there is an acceptable agreement between the current two-dimensional and the independent three-dimensional results in terms of $\overline{Nu}_h$ and $\overline{Nu}_c$ values over the whole range of $Ra$ and $R/L$, as summarized in Tables 4.1-4.2.

The maximal relative deviation between the currently obtained and the independent results is bounded by 19% and can be attributed to the impact of the lateral walls in the three-dimensional configuration suppressing the convective flow, and thus decreasing the total heat flux.

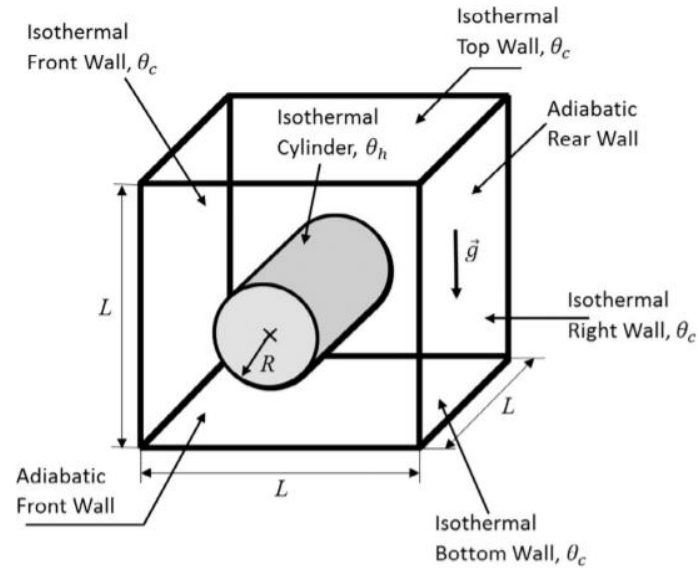

Figure 4.18: Physical model of a hot cylinder inside a cold tube in [72] and [58]

Table 4.1: Comparison between the present and the previously published $\overline{Nu}_h$ values averaged over the surface of a hot cylinder placed within a cold cube for $\varepsilon = 0.005$

| R/L      | 0.1     |          |          | 0.2     |          |          |
|----------|---------|----------|----------|---------|----------|----------|
| Ra       | Present | Ref [58] | Ref [72] | Present | Ref [58] | Ref [72] |
| 1.00E+04 | 6.4920  | 6.4880   | 6.2493   | 5.1990  | 5.1500   | 5.1184   |
| 1.00E+05 | 11.8700 | 11.6620  | 11.1380  | 7.7780  | 7.5800   | 7.2271   |
| 1.00E+06 | 18.1000 | 19.2500  | 18.3260  | 14.3500 | 13.3610  | 13.9370  |
| R/L      | 0.3     |          |          | 0.4     |          |          |
| Ra       | Present | Ref [58] | Ref [72] | Present | Ref [58] | Ref [72] |
| 1.00E+04 | 6.2630  | 5.7304   | 5.8084   | 8.8840  | 8.5544   | 8.7030   |
| 1.00E+05 | 7.3740  | 6.5169   | 6.4790   | 9.1240  | 8.7643   | 8.7030   |
| 1.00E+06 | 13.3600 | 11.4010  | 11.2720  | 11.9200 | 10.8320  | 10.7160  |

*Table 4.2: Comparison between the present and the previously published $\overline{Nu}_c$ values averaged over the surface of a hot cylinder placed within a cold cube for $\varepsilon = 0.005$*

| R/L      | 0.1     |          |          | 0.2     |          |          |
|----------|---------|----------|----------|---------|----------|----------|
| Ra       | Present | Ref [58] | Ref [72] | Present | Ref [58] | Ref [72] |
| 1.00E+04 | 1.0345  | 1.0208   | 1.0201   | 1.6662  | 1.6188   | 1.6161   |
| 1.00E+05 | 1.9112  | 1.8360   | 1.8099   | 2.5649  | 2.3814   | 2.3766   |
| 1.00E+06 | 2.8683  | 3.0348   | 2.9945   | 4.6134  | 4.3677   | 4.3985   |
| R/L      | 0.3     |          |          | 0.4     |          |          |
| Ra       | Present | Ref [58] | Ref [72] | Present | Ref [58] | Ref [72] |
| 1.00E+04 | 2.8655  | 2.9091   | 2.6216   | 5.4591  | 5.3928   | 5.1919   |
| 1.00E+05 | 3.3675  | 3.0702   | 2.9726   | 5.6192  | 5.5131   | 5.2651   |
| 1.00E+06 | 6.1671  | 5.3844   | 5.1956   | 7.2361  | 6.8313   | 6.6106   |

### 4.2.4 Analysis of the heat fluxes in the flow domain

As was mentioned in section 4.1, the walls of the cavity were maintained at a cold temperature $T_c = 1 - \varepsilon$, while the walls of the cylinder that was placed in the center of the cavity were maintained at a hot temperature $T_h = 1 + \varepsilon$. Therefore, the heat flux direction is from the cylinder surface towards the cavity surfaces. The value of heat flux at each cavity surface can differ, reflecting characteristics of the specific flow regime as can be quantified by a calculation of the average values of $Nu$ numbers on each wall. The $Nu_c$ values were calculated by formulas given in subsection 3.2.2.

As expected from symmetry considerations, the $Nu$ values obtained for the left and the right walls of the cavity are close to each other (see Tables 4.3-4.4) over the entire range of $Ra - R/L$ values. At the same time, significant differences are observed for the $Nu$ values obtained for the bottom and the top walls of the cavity (see Tables 4.5-4.6) for the entire range of $Ra - R/L$ values, whereas the $Nu$ values at the top are always higher than those at the bottom. These differences increase with $Ra$ values, and for $Ra = 10^6$ can reach up to one order higher.

To summarize, we can say that as the convective heat transfer becomes more pronounced with an increase in the $Ra$ values, the top of the cavity starts to play a more dominant role in removing heat from the system, which is clearly reflected in a gradual increase in the corresponding $Nu$ values.

*Table 4.3: $Nu_c$ on the left wall*

| ε | 0.005 | | | | 0.2 | | | |
|---|---|---|---|---|---|---|---|---|
| **R/L** | **0.1** | **0.2** | **0.3** | **0.4** | **0.1** | **0.2** | **0.3** | **0.4** |
| **Ra** | | | | | | | | |
| 1.00E+03 | 0.9389 | 1.5939 | 2.8186 | 5.4620 | 0.9877 | 1.5958 | 2.7040 | 5.4354 |
| 1.00E+04 | 0.9287 | 1.6347 | 2.8549 | 5.4550 | 1.0373 | 1.6550 | 2.7334 | 5.4341 |
| 1.00E+05 | 1.4945 | 2.3019 | 3.1750 | 5.5250 | 1.7238 | 2.2965 | 3.0942 | 5.5877 |
| 1.00E+06 | 2.2876 | 4.5233 | 4.5233 | 6.0957 | 2.3220 | 3.7539 | 5.3523 | 7.3707 |
| **ε** | **0.4** | | | | **0.6** | | | |
| **R/L** | **0.1** | **0.2** | **0.3** | **0.4** | **0.1** | **0.2** | **0.3** | **0.4** |
| **Ra** | | | | | | | | |
| 1.00E+03 | 0.9671 | 1.5791 | 2.6765 | 5.3313 | 0.9276 | 1.5618 | 2.6305 | 5.1832 |
| 1.00E+04 | 1.0173 | 1.6197 | 2.6896 | 5.3356 | 0.9774 | 1.5729 | 2.6453 | 5.1905 |
| 1.00E+05 | 1.7228 | 2.3632 | 3.1533 | 5.5551 | 1.7032 | 2.3394 | 3.1851 | 5.4981 |
| 1.00E+06 | 2.3817 | 3.7652 | 6.3741 | 7.0895 | 2.3033 | 4.6269 | 6.1530 | 7.9808 |

*Table 4.4: $Nu_c$ on the right wall*

| ε | 0.005 | | | | 0.2 | | | |
|---|---|---|---|---|---|---|---|---|
| **R/L** | **0.1** | **0.2** | **0.3** | **0.4** | **0.1** | **0.2** | **0.3** | **0.4** |
| **Ra** | | | | | | | | |
| 1.00E+03 | 0.9392 | 1.5945 | 2.8197 | 5.4640 | 0.9840 | 1.5976 | 2.7048 | 5.4374 |
| 1.00E+04 | 0.9291 | 1.6354 | 2.8559 | 5.4570 | 1.0337 | 1.6583 | 2.7340 | 5.4361 |
| 1.00E+05 | 1.4950 | 2.3030 | 3.1760 | 5.5270 | 1.7275 | 2.2982 | 3.0925 | 5.5897 |
| 1.00E+06 | 2.8784 | 4.5249 | 6.0970 | 7.5153 | 2.3751 | 3.6198 | 5.5117 | 7.3734 |
| **ε** | **0.4** | | | | **0.6** | | | |
| **R/L** | **0.1** | **0.2** | **0.3** | **0.4** | **0.1** | **0.2** | **0.3** | **0.4** |
| **Ra** | | | | | | | | |
| 1.00E+03 | 0.9612 | 1.5818 | 2.6775 | 5.3333 | 0.9235 | 1.5625 | 2.6315 | 5.1851 |
| 1.00E+04 | 1.0119 | 1.6241 | 2.6905 | 5.3375 | 0.9750 | 1.0880 | 2.2705 | 5.1924 |
| 1.00E+05 | 1.7259 | 2.3645 | 3.1544 | 5.5571 | 1.7048 | 2.3403 | 3.1862 | 5.5001 |
| 1.00E+06 | 2.4127 | 3.8251 | 6.3767 | 7.0920 | 2.2621 | 4.6286 | 6.1556 | 7.9837 |

*Table 4.5: $Nu_c$ on the bottom wall*

| ε | 0.005 | | | | 0.2 | | | |
|---|---|---|---|---|---|---|---|---|
| **R/L** | **0.1** | **0.2** | **0.3** | **0.4** | **0.1** | **0.2** | **0.3** | **0.4** |
| **Ra** | | | | | | | | |
| 1.00E+03 | 0.8819 | 1.5548 | 2.7986 | 5.4534 | 0.9575 | 1.5546 | 2.6772 | 5.4229 |
| 1.00E+04 | 0.5861 | 1.3616 | 2.6723 | 5.3717 | 0.8554 | 1.3812 | 2.4982 | 5.3157 |
| 1.00E+05 | 0.2150 | 0.7787 | 1.9192 | 4.8322 | 0.2756 | 0.6655 | 1.7119 | 4.6725 |
| 1.00E+06 | 0.2358 | 0.7924 | 1.3864 | 7.5153 | 0.2299 | 0.3993 | 0.7655 | 3.1495 |
| ε | 0.4 | | | | 0.6 | | | |
| **R/L** | **0.1** | **0.2** | **0.3** | **0.4** | **0.1** | **0.2** | **0.3** | **0.4** |
| **Ra** | | | | | | | | |
| 1.00E+03 | 0.9308 | 1.5322 | 2.6421 | 5.3133 | 0.8861 | 1.4995 | 2.5864 | 5.1582 |
| 1.00E+04 | 0.8119 | 1.2999 | 2.3834 | 5.1648 | 0.7617 | 1.0880 | 2.2705 | 4.9589 |
| 1.00E+05 | 0.2707 | 0.6223 | 1.2821 | 4.2901 | 0.2845 | 0.3204 | 1.1078 | 3.9579 |
| 1.00E+06 | 0.2032 | 0.4767 | 0.6584 | 2.6102 | 0.1486 | 0.1582 | 0.4593 | 2.0442 |

*Table 4.6: $Nu_c$ on the top wall*

| ε | 0.005 | | | | 0.2 | | | |
|---|---|---|---|---|---|---|---|---|
| **R/L** | **0.1** | **0.2** | **0.3** | **0.4** | **0.1** | **0.2** | **0.3** | **0.4** |
| **Ra** | | | | | | | | |
| 1.00E+03 | 1.0040 | 1.6374 | 2.8412 | 5.4732 | 1.0177 | 1.6432 | 2.7330 | 5.4502 |
| 1.00E+04 | 1.6941 | 2.0329 | 3.0787 | 5.5527 | 1.2933 | 2.0418 | 3.0288 | 5.5709 |
| 1.00E+05 | 4.4403 | 1.8762 | 5.1999 | 6.5926 | 4.2639 | 4.9223 | 5.3219 | 6.9002 |
| 1.00E+06 | 5.4828 | 8.6132 | 11.0893 | 10.2210 | 7.2042 | 9.7895 | 9.3328 | 11.2149 |
| ε | 0.4 | | | | 0.6 | | | |
| **R/L** | **0.1** | **0.2** | **0.3** | **0.4** | **0.1** | **0.2** | **0.3** | **0.4** |
| **Ra** | | | | | | | | |
| 1.00E+03 | 1.0012 | 1.6333 | 2.7134 | 5.3519 | 0.9687 | 1.6273 | 2.6768 | 5.2105 |
| 1.00E+04 | 1.3028 | 2.0593 | 3.0640 | 5.5315 | 1.2711 | 2.1806 | 3.0940 | 5.4540 |
| 1.00E+05 | 3.8549 | 4.5273 | 5.0862 | 7.0268 | 3.3704 | 3.9956 | 4.8861 | 6.9461 |
| 1.00E+06 | 6.9131 | 9.2124 | 10.5519 | 10.8403 | 6.3005 | 7.8221 | 9.9085 | 10.7687 |

### 4.2.5 Approximation to the $Nu - Ra$ power law

The qualitative results of the average Nusselt number for each configuration can be approximated to obtain the $Nu - Ra$ power law, by employing the least squares technique. Following the dimensional analysis stemming from the boundary layer theory, the $Nu - Ra$ relationship obeys the following power law [73]:

$$\overline{Nu_f} = C(Gr_f Pr_f)^m = C Ra_f^m, \tag{4.5}$$

where $Gr, Pr$ are the Grashof and the Prandtl numbers and $Ra = GrPr$, and $C$ and $m$ are specific constants, whose values depend on the $Ra$ number and on the geometric properties of the system under consideration.

Figs. 4.19-4.22 present the distribution of the $Nu$ number as a function of the $Ra$ number over the entire range of $\varepsilon \in \{0.005, 0.2, 0.4\ 0.6\}$, $Ra \in \{10^4, 10^5, 10^6\}$ and $R/L \in \{0.1, 0.2, 0.3, 0.4\}$. It is noteworthy that the $Nu$ values obtained for the lowest value of $Ra = 10^3$ were not taken into account because of the dominance of the conductive heat transfer mechanism (see also Figs. 4.2-4.3, 4.10-4.11).

All the $Nu$ values were approximated to the $Nu - Ra$ power law as formulated by Eq. (4.5). Generally, an acceptable accuracy was obtained when approximating the $Nu - Ra$ relationship by the power law over the entire range of $\varepsilon$ and $R/L$ values. More specifically, the results obtained for smaller cylinders ($R/L = 0.1, 0.2$) exhibit more precise power law fits, for which the smallest value of $R^2$ was equal to $R^2 = 0.975$. Remarkably, for all the configurations characterized by $R/L = 0.1, 0.2$ $Nu$ was approximately proportional to $Ra^{0.22}$, which is very close to the results reported with respect to the laminar natural convection in spherical shells [74], [75], [76].

Configurations with larger cylinders ($R/L = 0.3, 0.4$) exhibit a less pronounced power law fit for the $Nu - Ra$ relationship, characterized with the smallest value of $R^2$ equal to $R^2 = 0.84$. Note also that $Nu$ was approximately proportional to $Ra^{0.16}$ and to $Ra^{0.06}$ for $R/L = 0.3$ and $R/L = 0.4$ geometries, respectively. Such a considerable decrease in the heat flux rate compared with the geometries characterized by smaller cylinders can be attributed to the blocking effects of both cylinder and cavity boundaries, suppressing the momentum of the convective flow (see e.g., [72], [58]).

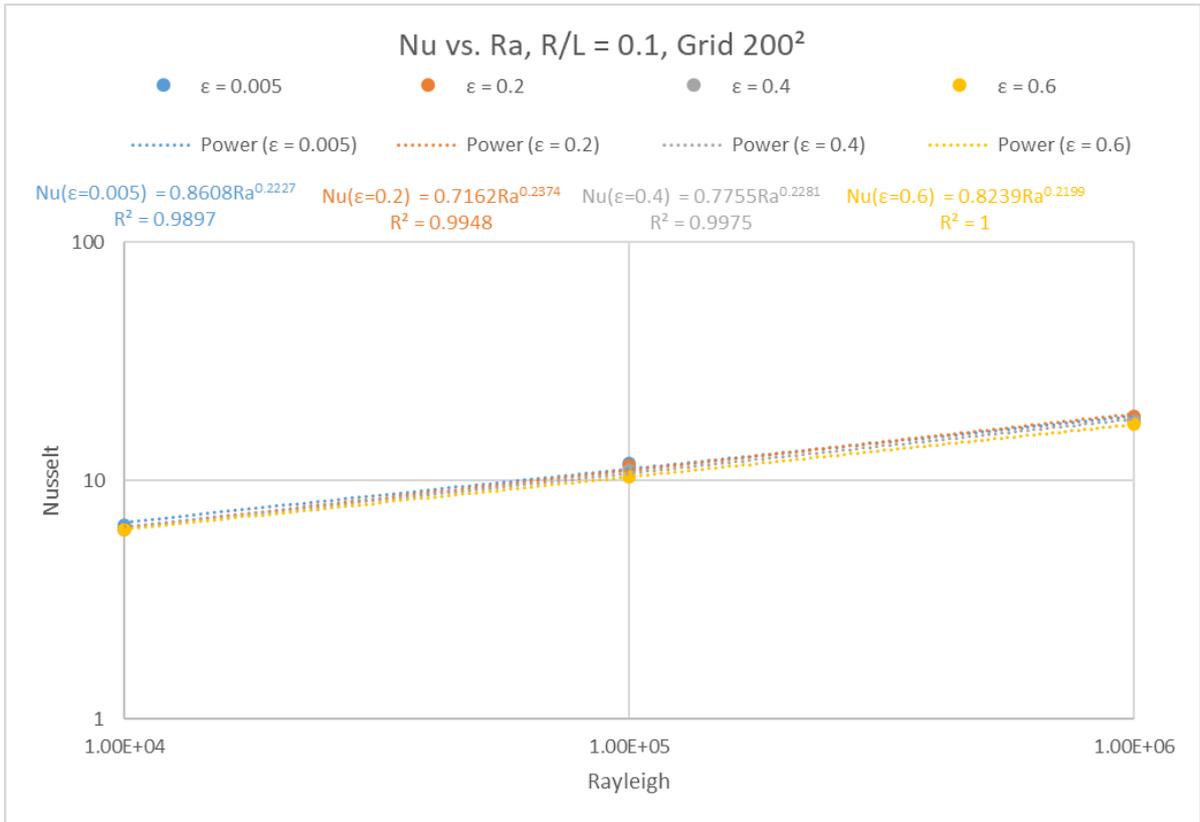

Figure 4.19: Nusselt vs Rayleigh for $R/L = 0.1$

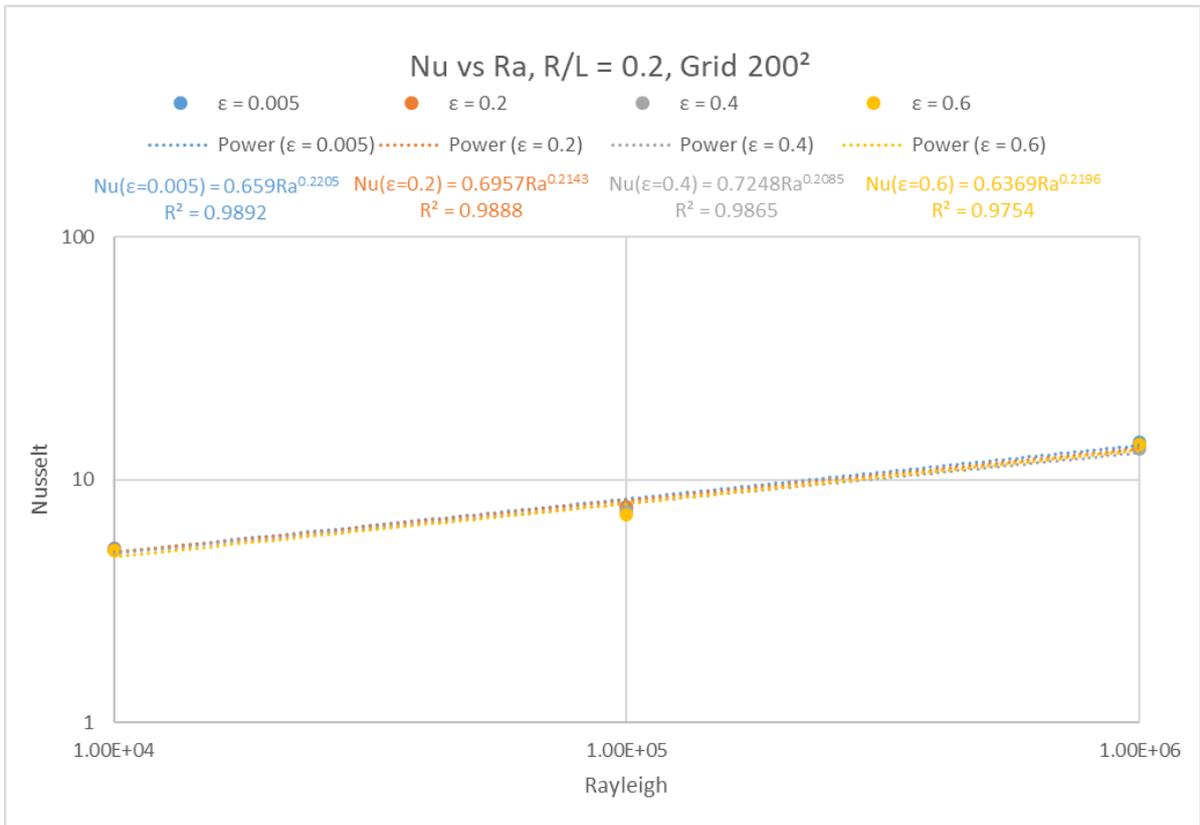

Figure 4.20: Nusselt vs Rayleigh for $R/L = 0.2$

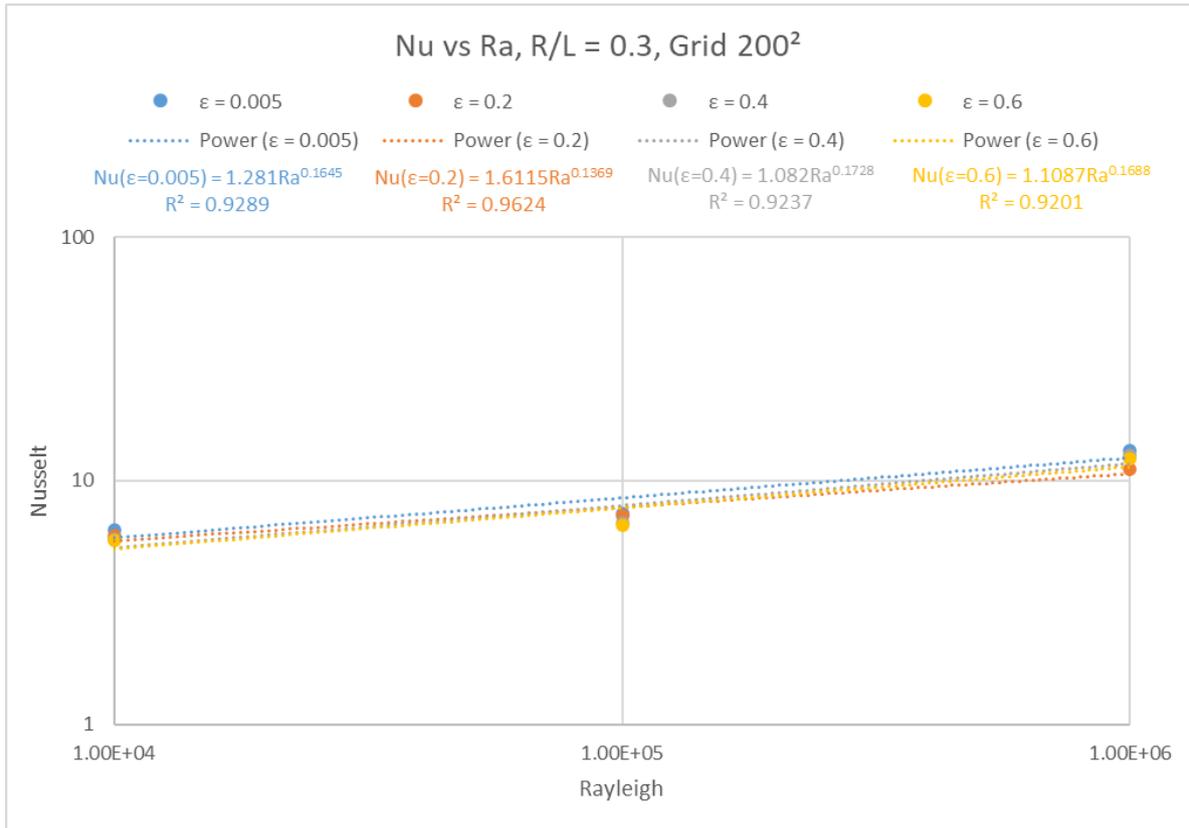

Figure 4.21: Nusselt vs Rayleigh for $R/L = 0.3$

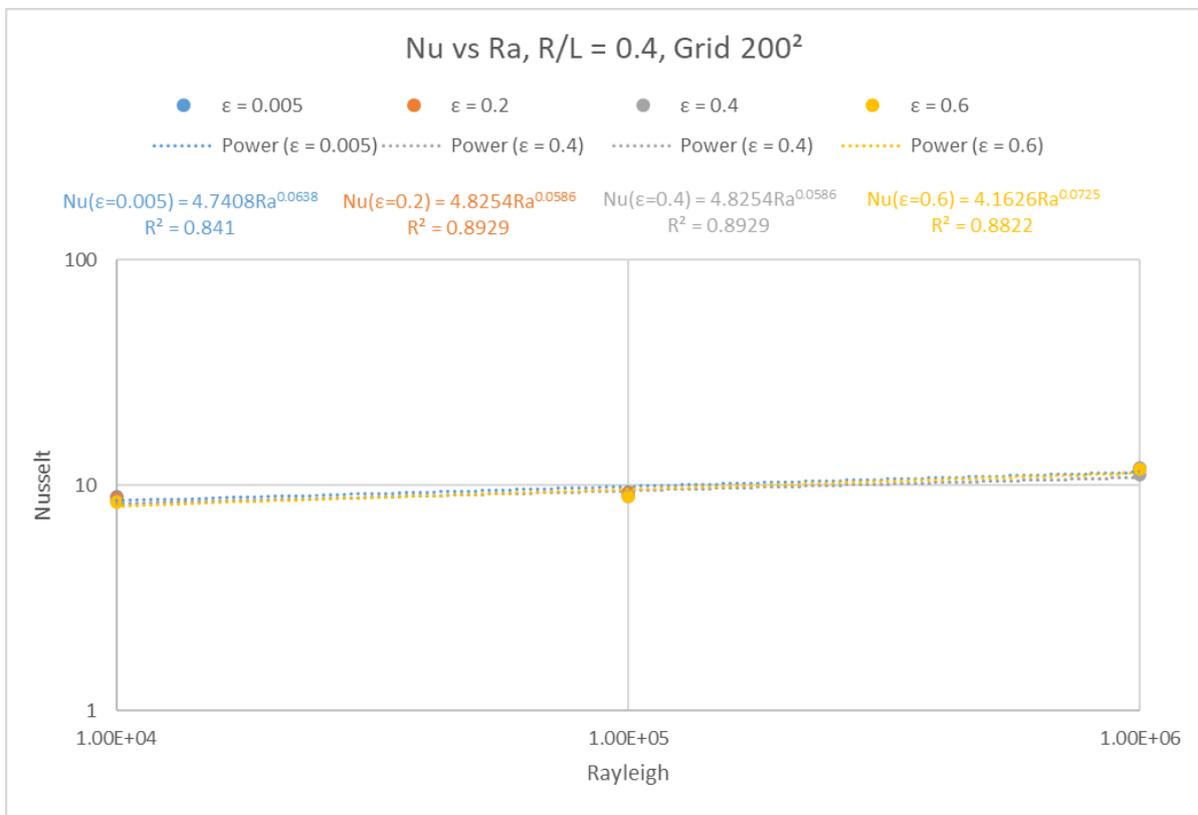

Figure 4.22: Nusselt vs Rayleigh for $R/L = 0.4$

### 4.2.6 Multiple steady-state regimes

Further numerical analysis revealed that steady non-Boussinesq natural convection flows can exhibit multiple steady state regimes. In particular, two independent steady state branches were found for the values of $\varepsilon = 0.4$, $R/L = \{0.2, 0.4\}$ and $Ra = 10^6$, as shown in Figs. 4.23-4.26. The stability of the revealed regimes was checked by randomly perturbing all the flow variables with values deviating by about 10% from the corresponding steady state values, and verifying that the flow further converges to the previously observed steady state. Remarkably, while the corresponding steady states differ by the number and the size of convective cells hosted within the flow domain, they are characterized by very close $Nu$ values, averaged over the cylinder and the cavity boundaries. It can thus be concluded that both steady state regimes obtained for the same values of governing parameters and belonging to the different branches are still characterized by the same averaged heat fluxes at all the flow boundaries.

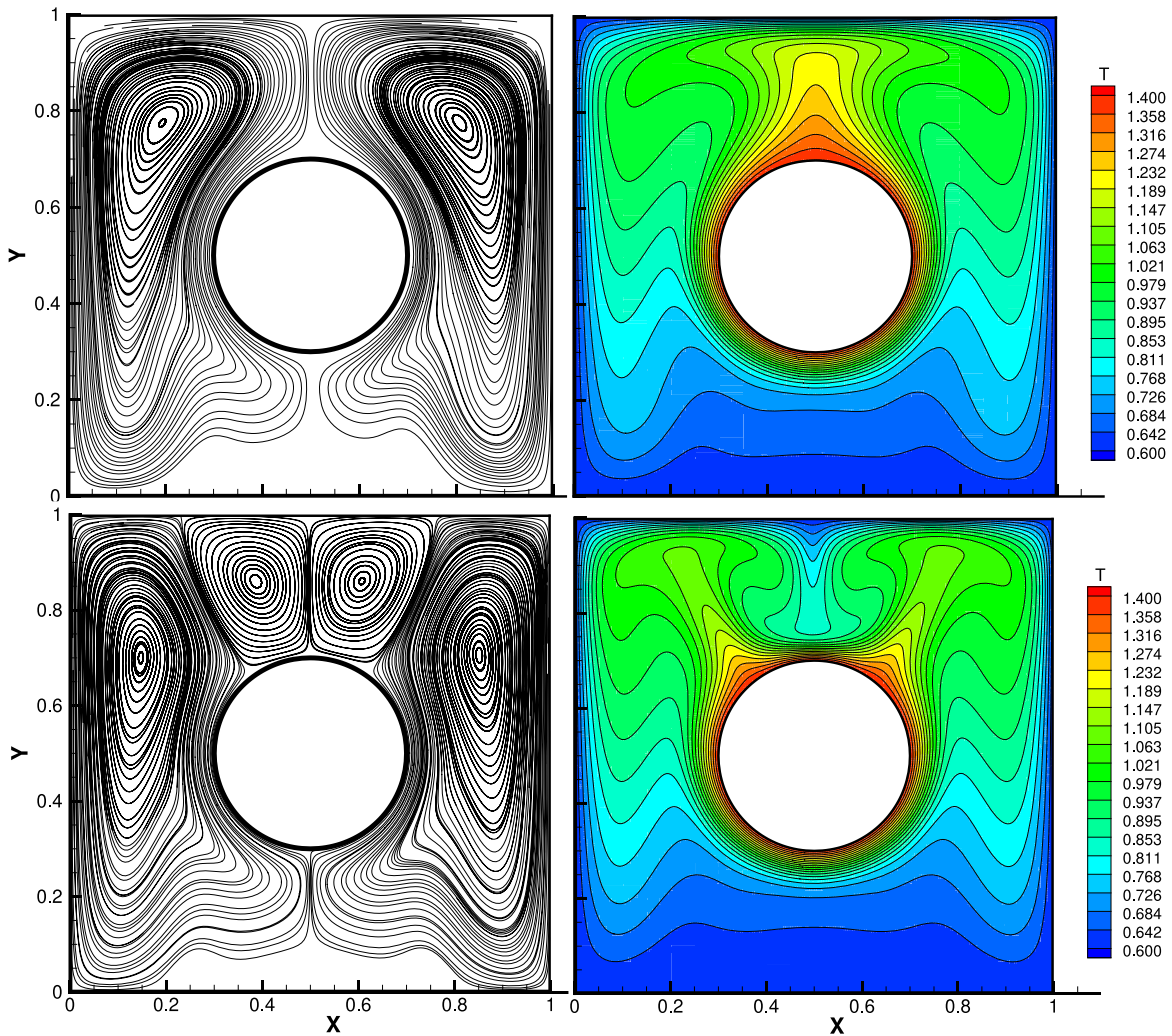

Figure 4.23: Flow and temperature patterns corresponding to two different steady state branches obtained for $\varepsilon = 0.4$, $Ra = 10^6$ and $R/L = 0.2$

To sum up, an acceptable quantitative agreement exists between the currently obtained lowest temperature-difference cases and the independent results that were computed by applying the Boussinesq approximation over the entire range of operating conditions and flow characteristics, which successfully verifies the currently developed numerical methodology with an incorporated IBM when applied for the simulation of compressible natural convection confined flows with complex geometry. The results of the high-temperature-gradient cases show good agreement with the boundary layer theory. In addition, multiple configurations of the steady state flow were discovered.

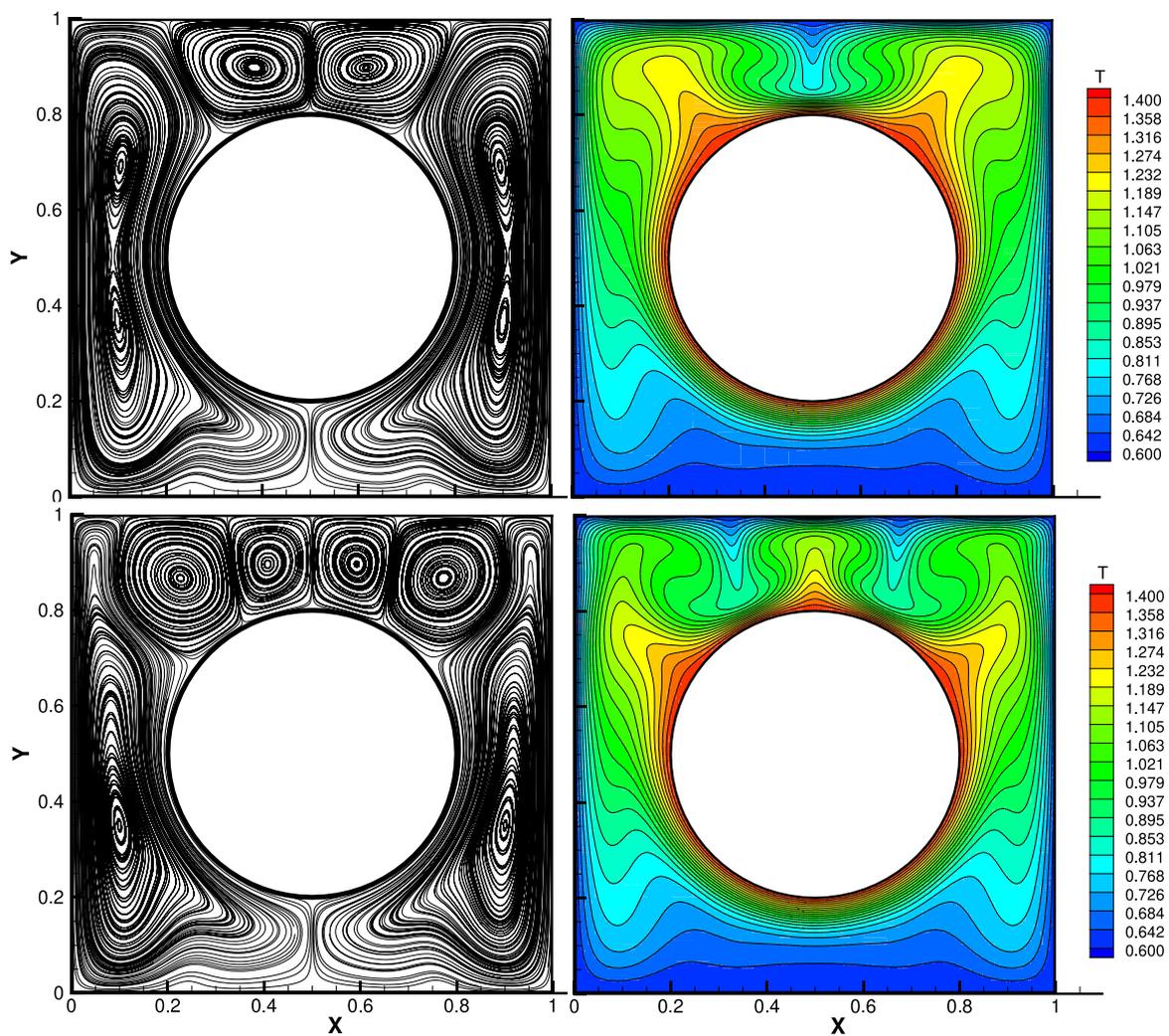

Figure 4.24: Flow and temperature patterns corresponding to two different steady state branches obtained for $\varepsilon = 0.4$, $Ra = 10^6$ and $R/L = 0.3$

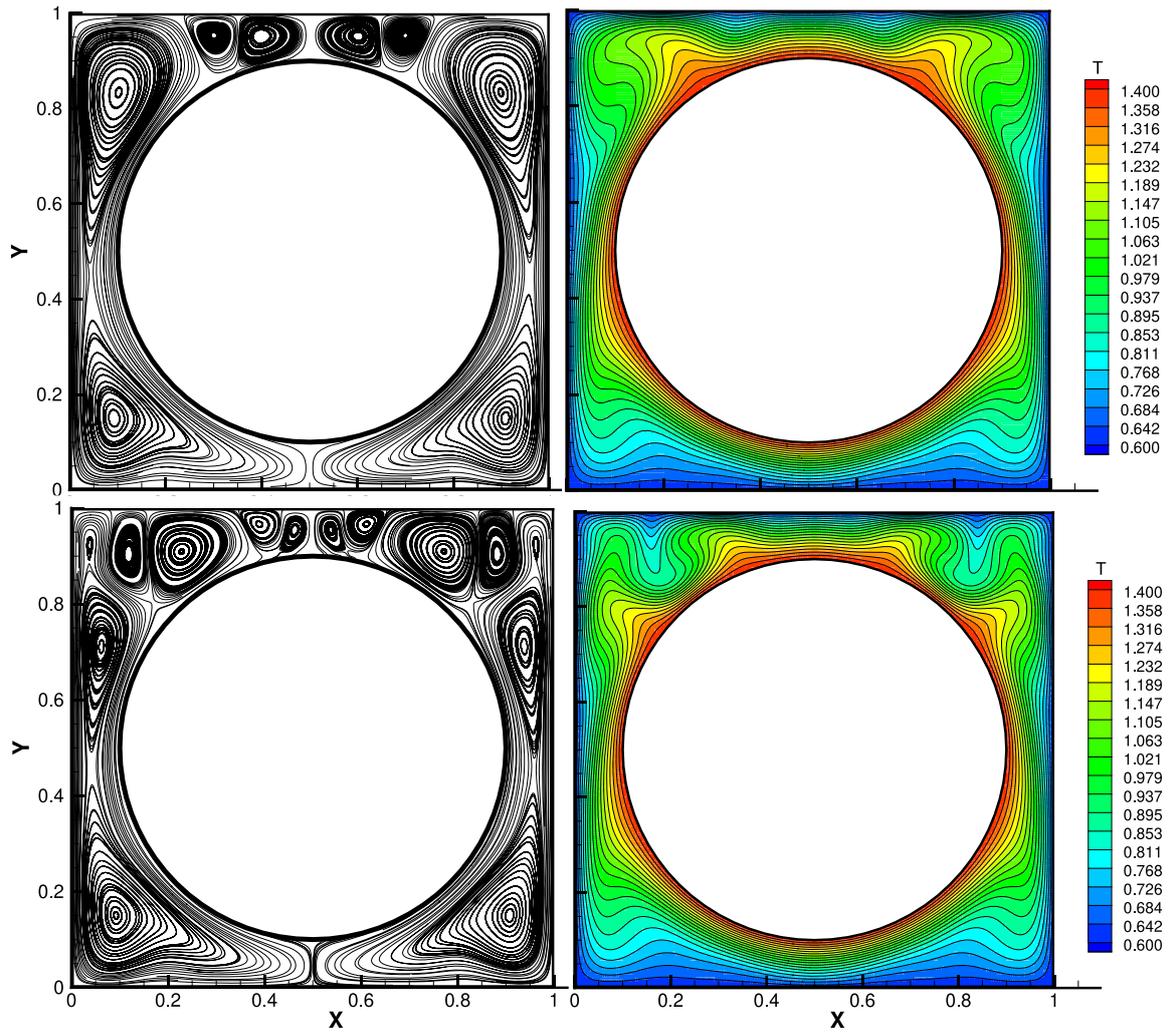

Figure 4.25: Flow and temperature patterns corresponding to two different steady state branches obtained for $\varepsilon = 0.4$, $Ra = 10^6$ and $R/L = 0.4$

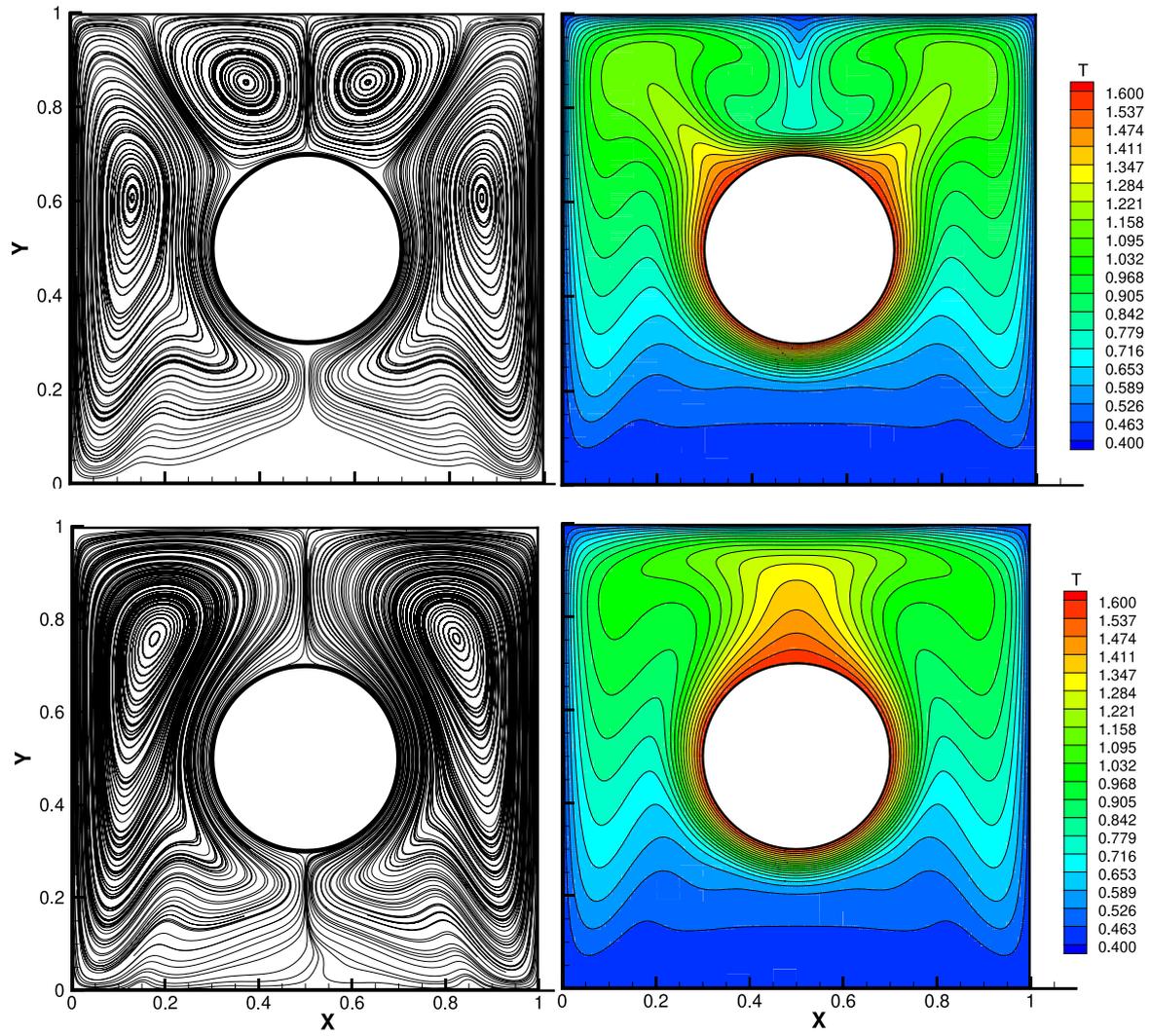

Figure 4.26: Flow and temperature patterns corresponding to two different steady state branches obtained for $\varepsilon = 0.6$, $Ra = 10^6$ and $R/L = 0.2$

# 5. Summary and conclusions

In the present work a pressure-based algorithm for the solution of the thermal compressible flow of an ideal gas was developed. This algorithm employs a semi-implicit fractional-step method, and a second order backward scheme for the temporal and standard second order finite volume method for the spatial discretization. In addition, a novel pressure-corrected direct forcing IBM, introduced by Riahi et al. [63], was adopted and extended in the framework of the current study to enforce the kinematic constraints of no-slip and the given temperature on the surface of the immersed body. The algorithm does not rely on the low Mach number assumption and does not employ the pressure splitting into hydrodynamic and thermodynamic terms, but rather, all the equations were solved in their original formulations. The viscous heating was neglected in the framework of the current study as is commonly done when simulating flows characterized by low values of shear stress.

The developed methodology was extensively verified by favorable comparison with corresponding independent numerical data available in the literature for incompressible [22], and non-Bossinesq compressible [29] flows. The pressure-corrected direct forcing IBM was implemented and further utilized for the simulation of natural convection non-Boussinesq flow developing within a square cold cavity with a centrally located cold cylinder over a broad range of governing parameters.

The results achieved were analyzed qualitatively and quantitatively. First, the spatial distribution of streamline and temperature fields was obtained and discussed. Second, the Nusselt numbers on the hot cylinder and the cold cavity surfaces were calculated. Third, the thermal fluxes were analyzed by comparing the corresponding values of the Nusselt numbers obtained for all the domain boundaries. Finally, multiple steady state solutions for several configurations were discovered and discussed.

The present study focused on the simulation of compressible flows characterized by high temperature differences. For this reason, a full set of compressible Naviers-Stokes equations, supplemented by equations of state, was solved without applying any simplifications, e.g., the Boussinesq approximation. As such, the currently developed approach can safely be considered to be an important milestone towards developing a comprehensive methodology capable of the simulation of steady state and transient multiphase flow regimes typical of the reactor core. To achieve this, future research will be focused on extending the developed solver to a general fluid (other than an ideal gas), by utilizing tables for obtaining thermodynamic properties of the fluid. Next, the currently developed methodology must be extended to be able to solve compressible two-phase flows, which would include various phase transition and phase interaction models.

## תקציר

מטרת מסמך זה הינה הצגת סיכום העבודה אשר בוצעה במסגרת הדרוש להשלמת תואר מגיסטר בהנדסה (M.Sc.) במחלקה להנדסת מכונות, אוניברסיטת בן-גוריון בנגב (BGU). הפרוייקט כולל מחקר תיאורטי של הסעה טבעית דחיסה עם הפרשי טמפרטורות גבוהים תוך כדי התייחסות לגיאומטריות מורכבות. המוטיבציה המחקרית מגיעה מהמחקר ארוך טווח שהוקם על-ידי מרכז למחקר גרעיני בנחל שורק המיועד לחקור ולסמלץ את משטרי הזרימה הרב-משטרית במצב מתמיד ומצבים טרנזיאנטים הקיימים בליבת הכור.

מטרתו העיקרית של פרויקט זה היא לפתח מתודולוגיה נומרית מקיפה המסוגלת למדל את הסעה טבעית דחיסה עם הפרשי טמפרטורות גבוהים תוך כדי התייחסות, באמצעות כלים סטנדרטיים של דינמיקת נוזלים חישובית (Computational Fluid Dynamics, CFD) – פתרנים עם אלגוריתמים מבוססי לחץ ושיטות גבול שקועות (Immersed Boundary methods, IBM).

דו"ח זה מכיל:

- רקע מודולרי אודות מרכיבי העניין שלמרכז למחקר גרעיני בנחל שורק.
- סקירת ספרות מקיפה של שיטות סימולציה של הסעה טבעית דחיסה ושיטות גבול שקועות.
- מטרות מורחבות של המחקר שבוצע.
- מודל פיזיקלי מקיף, הכולל את המשוואות השולטות, הגדרות, חוקים מכוננים וניתוח ממדי.
- מחקר אימות על ידי השוואה עם מחקרים נומריים עצמאיים מתאימים הקיימים בספרות עבור זרימה בלתי דחיסה והסעה טבעית דחיסה, ללא גיאומטריה מורכבת.
- השוואה עבור הפרשי טמפרטורות נמוכים עם גיאומטריה מורכבת, בין תוצאות ממחקרים קודמים לתוצאות שהושגו על ידי מחקר נוכחי.
- פתרון וניתוח עבור המקרים עם הפרשי טמפרטורות גבוהים עם גיאומטריה מורכבת.
- סיכום, מסקנות והמלצות לעבודה אפשרית בעתיד.

למילות מפתח בתזה ראה תקציר בשפה האנגלית המופיע בתחילת מסמך זה.

# אוניברסיטת בן-גוריון בנגב
## הפקולטה למדעי ההנדסה
### המחלקה להנדסת מכונות

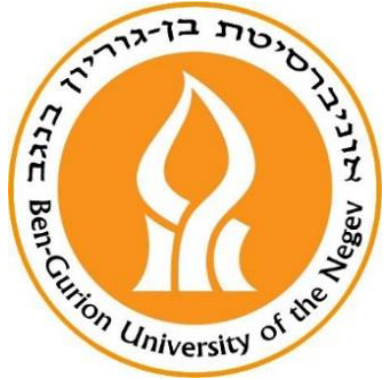

# פיתוח שיטות לסימולציה תרמו-הידראולית של כורים גרעיניים ומערכות דומות בתנאי עבודה נורמליים ובתהליכים טרנזיאנטיים

חיבור זה מהווה חלק מהדרישות לקבלת תואר מגיסטר בהנדסה

מאת : דמיטרי זביאגה

מנחה : ד״ר יורי פלדמן

| | | |
|---|---|---|
| חתימת המחבר : ................ | | תאריך : 30/09/2021 |
| אישור המנחים : ................ | | תאריך : 30/09/2021 |
| אישור יו״ר ועדת תואר שני מחלקתית : ................ | | תאריך : 30/09/2021 |

ספטמבר 2021